\documentclass[pdflatex,iicol,sn-standardnature]{sn-jnl}

\jyear{2022}%

\theoremstyle{thmstyleone}%
\theoremstyle{thmstyletwo}%

\theoremstyle{thmstylethree}%

\begin{document}

\title[Enhanced Excitation Energy Transfer]{Enhanced Excitation Energy Transfer under Strong Light-Matter Coupling: Insights from Multi-Scale Molecular Dynamics Simulations}


\author[1]{\fnm{Ilia} \sur{Sokolovskii}}\email{ilia.sokolovskii@jyu.fi}
\equalcont{These authors contributed equally to this work.}

\author[1]{\fnm{Ruth H.} \sur{Tichauer}}\email{ruth.tichauer@uam.es}
\equalcont{These authors contributed equally to this work.}

\author[1]{\fnm{Dmitry} \sur{Morozov}}\email{dmitry.morozov@jyu.fi}

\author[2]{\fnm{Johannes} \sur{Feist}}\email{johannes.feist@uam.es}

\author*[1]{\fnm{Gerrit} \sur{Groenhof}}\email{gerrit.x.groenhof@jyu.fi}

\affil[1]{\orgdiv{Nanoscience Center and
  Department of Chemistry}, \orgname{University of Jyv\"{a}skyl\"{a}}, \orgaddress{\street{P.O. Box 35}, \city{Jyv\"{a}skyl\"{a}}, \postcode{40014},  \country{Finland}}}

\affil[2]{\orgdiv{Departamento de F\'{i}sica Te\'{o}rica de la Materia Condensada and Condensed Matter Physics Center (IFIMAC)}, \orgname{Universidad Aut\'{o}noma de Madrid},
\orgaddress{ \city{Madrid}, \country{Spain}}}

\abstract{
Exciton transport can be enhanced in the strong coupling regime where excitons hybridise with confined light modes to form polaritons. Because polaritons have group velocity, their propagation should be ballistic and long-ranged. However, experiments indicate that organic polaritons propagate in a diffusive manner and more slowly than their group velocity. Here, we resolve this controversy by means of molecular dynamics simulations of Rhodamine molecules in a Fabry-Perot cavity. Our results suggest that polariton propagation is limited by the cavity lifetime and appears diffusive due to reversible population transfers between polaritonic states that propagate ballistically at their group velocity, and dark states that are stationary. Furthermore, because long-lived dark states transiently trap the excitation, propagation is observed on timescales beyond the intrinsic polariton lifetime. These insights not only help to better understand and interpret experimental observations, but also pave the way towards rational design of molecule-cavity systems for coherent exciton transport.

}

\keywords{Excitation energy transfer, Strong light-matter coupling, polariton, Fabry-P\'{e}rot cavity, QM/MM, molecular dynamics}



\maketitle

\section*{Introduction}\label{sec1}

Solar cells based on organic molecules are promising alternatives to the silicon-based technologies that dominate today's market,
mostly because organic photovoltaics (OPV) are cheaper to mass-produce, lighter, more flexible and easier to dispose of. A key step in light harvesting is transport of excitons from where photons are absorbed to where this energy is needed for initiating a photochemical process~\cite{Croce2014}, usually deeper inside the material of the solar cell. Because 
excitons in organic materials are predominantly localized onto single molecules, exciton transport proceeds via incoherent hops~\cite{Mikhnenko2015}. Such random-walk diffusion is, however, too slow to compete with ultra-fast deactivation processes of singlet excitons, such as radiative and non-radiative decay. As exciton diffusion is furthermore hindered by thermal disorder, propagation distances in organic materials typically remain below 10~nm~\cite{Mikhnenko2015}. Such short diffusion lengths limit the efficiency of solar energy harvesting and require complex morphologies of active layers into nanometer sized domains, {\it{e.g.}}, bulk heterojunctions in OPVs, which not only complicates device fabrication, but also reduces device stability~\cite{Cao2014,Rafique2018}.


Distances of hundreds of nanometers have been observed for the diffusion of longer-lived triplet states~\cite{Akselrod2014}, but because not all organic materials can undergo efficient inter-system crossing or singlet fission, it may be difficult to exploit triplet diffusion in general. Exciton mobility can also be increased through transient exciton delocalization~\cite{Sneyd2021,Kong2022, Sneyd2022}, but as the direct excitonic interactions are weak in most organic materials, molecules need to be ordered to reach this enhanced transport regime.

\begin{figure}[!htb]
\centering
\includegraphics[width=7cm]{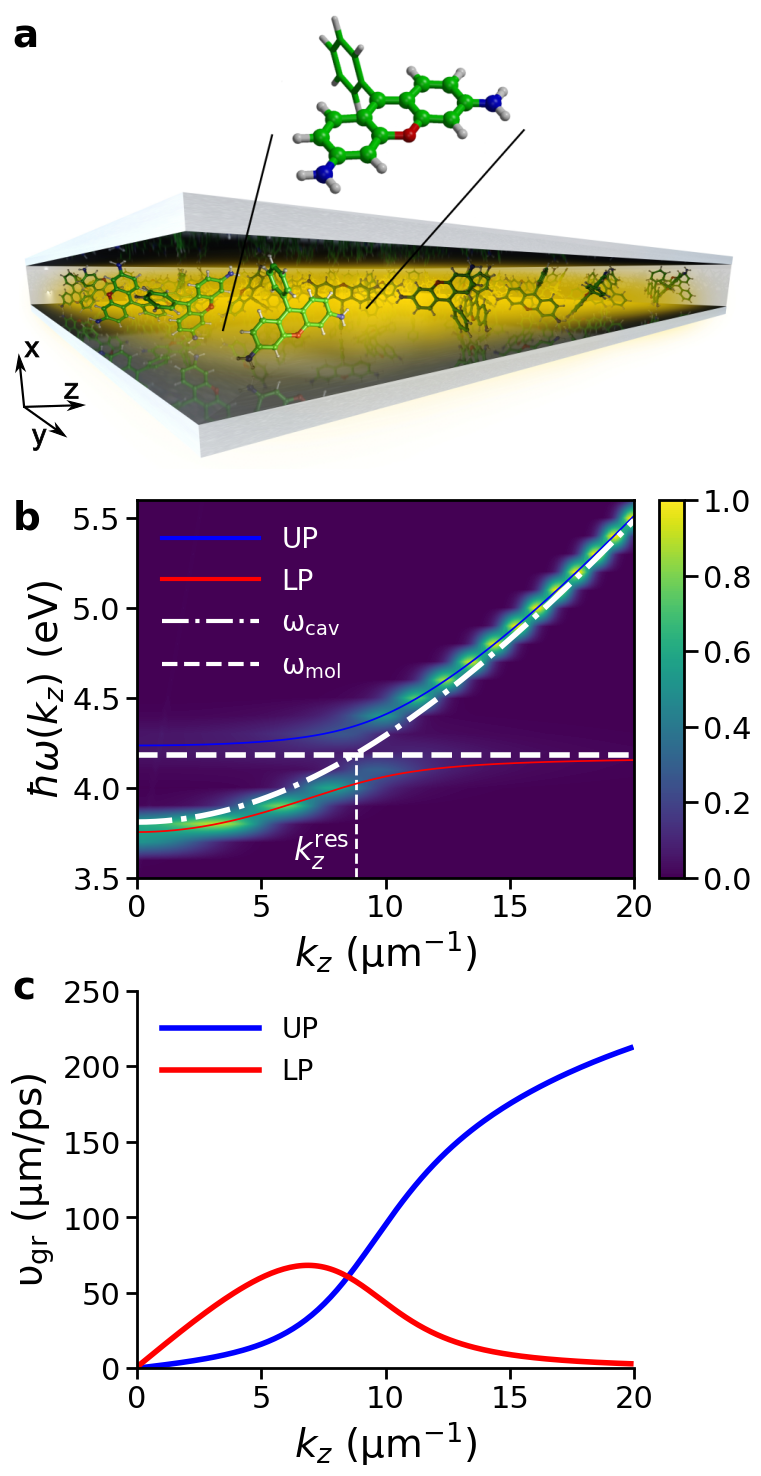}
  \caption{Panel {\bf{a}}: Schematic illustration of an optical Fabry-P\'erot microcavity filled with Rhodamine chromophores (not to scale). The QM subsystem, shown in ball-and-stick representation in the inset, is described at the HF/3-21G level of theory in the electronic ground state (S$_0$), and at the CIS/3-21G level of theory in the first singlet excited state (S$_1$). The MM subsystem, consisting of the atoms shown in stick representation and the water molecules (not shown), is modelled with the Amber03 force field. Panel {\bf b}: Normalised angle-resolved absorption spectrum of the cavity, showing Rabi splitting between lower polariton (LP, red line) and upper polariton (UP, blue line) branches. The cavity dispersion and excitation energy of the molecules (4.18~eV at the CIS/3-21G//Amber03 level of theory) are plotted by point-dashed and dashed lines, respectively. Panel {\bf c}: Group velocity of the LP (red) and UP (blue), defined as $\partial\omega(k_z)/\partial k_z$.}\label{fig:structure+dispersion}
\end{figure}

Alternatively, {\it permanent} delocalization over large numbers of molecules can be achieved by strongly coupling the excitons in the material to the confined light modes of optical cavities, such as Fabry-P\'{e}rot resonators (Figure~\ref{fig:structure+dispersion}{\bf{a}}) or nano-structured devices~\cite{Feist2015,Schachenmayer2015,Wellnitz2022}. In this strong light-matter coupling regime the rate of energy exchange between molecular excitons and confined light modes exceeds the intrinsic decay rates of both the excitons and the confined modes leading to the formation of new coherent light-matter states, called polaritons~\cite{Skolnick1998,Litinskaya2006,Torma2015,Ribeiro2018,Hertzog2019,Garcia-Vidal2021,Fregoni2022,Ruggenthaler2022,Rider2022}.


The majority of hybrid states in realistic molecule-cavity systems are dark~\cite{Agranovich2003,Litinskaya2004,delPino2015}, 
meaning that they have negligible contributions from the cavity modes.
In contrast, the few states with such contributions, are the bright polaritonic states that have dispersion and hence group velocity, defined as the derivative of the polariton energy with respect to in-plane momentum ({\it{i.e.}}, $k_z$ in Figure~\ref{fig:structure+dispersion}{\bf{b}}). 
In the out-of-plane cavity direction ({\it{i.e.}}, perpendicular to the mirrors), these states are delocalised over the molecules inside the mode volume, while in the in-plane direction ({\it{i.e.}}, parallel to the mirrors) they behave as quasi-particles with a low effective mass and large group velocity ({\it i.e.}, fractions of the speed of light). These polaritonic properties can be exploited for both out-of-plane~\cite{Coles2014,Feist2015,Schachenmayer2015,Zhong2016,Zhong2017,Georgiou2018,Groenhof2018,Xiang2020,Georgiou2021,Wellnitz2022,Son2022,Engelhardt2022,George2023}, and in-plane energy transport~\cite{Freixanet2000,Agranovich2007,Litinskaya2008,Michetti2008b,Lerario2017,Myers2018,Rozenman2018,Zakharko2018,Forrest2020,Pandya2021,Ostrovskaya2021,Ferreira2022,Berghuis2022,Pandya2022,Xu2022,Ribeiro2022,Allard2022,Balasubrahmaniyam2023,Aroeira2023,Engelhardt2023,Jin2023}.


Indeed, at cryogenic temperatures in-plane ballistic propagation at the group velocity of polaritons was observed for polariton wavepackets in a Fabry-P\'{e}rot microcavity containing an In$_{0.05}$Ga$_{0.95}$As quantum well~\cite{Freixanet2000}.
Ballistic propagation was also observed for polaritons formed between organic molecules and Bloch surface waves~\cite{Lerario2017,Forrest2020,Balasubrahmaniyam2023}, while a combination of ballistic transport on an ultrashort timescale (sub-50~fs) and diffusive motion on longer timescales was observed for cavity-free polaritons~\cite{Pandya2021}, for which strong coupling was achieved through a mismatch of the refractive indices between thin layers of densely-packed organic molecules and a host material~\cite{Daehne1998}. In contrast, experiments on strongly coupled organic J-aggregates in metallic micro-cavities suggest that molecular polaritons propagate in a diffusive manner and 
much more slowly than their group velocities~\cite{Rozenman2018}. Furthermore, despite a low cavity lifetime in the order of tens of femtoseconds in these experiments, propagation was observed over several picoseconds, which was attributed to a long lifetime of the lower polariton (LP)~\cite{Rozenman2018,Garcia-Vidal2021}.


To address these controversies and acquire atomistic insights into polariton propagation, we performed multi-scale molecular dynamics (MD) simulations~\cite{Luk2017,Tichauer2021} of solvated Rhodamine molecules strongly coupled to the confined light modes of a one-dimensional (1D) Fabry-P\'{e}rot microcavity, shown in Figure~\ref{fig:structure+dispersion}{\bf a}~\cite{Michetti2005}. As in previous work~\cite{Luk2017}, the electronic ground state (S$_0$) of the molecules was modeled at the hybrid Quantum Mechanics / Molecular Mechanics (QM/MM) level~\cite{Warshel1976b}, using the restricted Hartree-Fock (HF) method for the QM subsystem, which contains the fused rings, in combination with the 3-21G basis set~\cite{Ditchfield1971}. The MM subsystem, consisting of the rest of the Rhodamine molecule and the water, was modeled with the Amber03 force field~\cite{Duan2003}. The first electronic excited state (S$_1$) of the QM region was modelled with Configuration Interaction, truncated at single electron excitations (CIS/3-21G//Amber03). 
At this level of theory, the excitation energy of Rhodamine is 4.18~eV, which is significantly overestimated with respect to experiments. This discrepancy is due to the limited size of the basis set and the neglect of electron-electron correlation in the {\it{ab initio}} methods. While including electron-electron correlation into the description of the QM region improves the vertical excitation energy, we show in the Supporting Information (SI) that this does not significantly change the topology of the relevant potential energy surfaces, which determines the molecular dynamics (Figure S3).


We computed semi-classical Ehrenfest~\cite{Ehrenfest1927} MD trajectories of 
1024 Rhodamine molecules inside a 1D cavity of length $L_z = $~50~$\mu$m, with $z$ indicating the in-plane direction ($L_x = 163$~nm is the distance between the mirrors and $x$ thus indicates the out-of-plane direction). The cavity was red-detuned by 370~meV with respect to the molecular excitation energy (4.18~eV at the CIS/3-21G//Amber03 level of theory, dashed line in Figure~\ref{fig:structure+dispersion}{\bf b}), such that at wave vector $k_z = 0$, the cavity resonance is  $\hslash\omega_0 =$~3.81~eV. The dispersion of this cavity, $\omega_\text{cav}(k_{z}) = \sqrt{\omega_0^2+c^2k_{z}^2/n^2}$, was modelled with 160 modes ($0\le p\le 159$ for $k_{z}=2\pi p/L_z$, with $c$ the speed of light and $n$ the refractive index)~\cite{Agranovich2007}. With a cavity vacuum field strength of 0.26~MVcm$^{-1}$, the Rabi splitting, defined as the energy difference between the bright lower (LP) and upper polariton (UP) branches at the wave-vector $k^{\text{res}}_z$ where the cavity dispersion matches the molecular excitation energy (Figure~\ref{fig:structure+dispersion}{\bf b}), was $\sim$~325~meV. While the choice for a 1D cavity model with only positive $k_z$ vectors was motivated by the necessity to keep our simulations computationally tractable, it precludes the observation of elastic scattering events that would change the direction ({\it{i.e.}}, in-plane momentum, $\hslash{\bf{k}}$) of propagation. Furthermore, with only positive $k_z$ vectors, polariton motion is restricted to the $+z$ direction, but we show in the SI (Figure~S14) that this assumption does not affect our conclusions about the transport mechanism. 

Newton's equations of motion were integrated numerically with a 0.1~fs time step using forces derived on-the-fly from the mean-field potential energy surface provided by the total time-dependent polaritonic wave function, $\vert\Psi(t)\rangle$, which was expanded in the basis of the time-independent adiabatic eigenstates of the cavity-molecule Hamiltonian (SI)~\cite{Tavis1969,Michetti2005,Agranovich2007,Tichauer2021}. The total wavefunction was evolved along with the classical MD trajectory by unitary propagation in the local diabatic basis~\cite{Granucci2001}. A complete description of the methods employed in this work, including details of the multi-scale MD model for strongly coupled molecules, is provided as SI.


Experimentally, polariton propagation has been investigated by means of optical microscopy. While stationary microscopy measurements provides information on the distance over which polaritons  propagate~\cite{Lerario2017,Zakharko2018,Forrest2020,Berghuis2022}, transient microscopy also yields insight into the time evolution of the propagation~\cite{Freixanet2000,Rozenman2018,Pandya2021,Pandya2022,Xu2022,Balasubrahmaniyam2023,Jin2023}. In these experiments the strongly-coupled systems were excited either resonantly into the bright polaritonic states~\cite{Pandya2021,Pandya2022}, or off-resonantly into an uncoupled molecular electronic state~\cite{Lerario2017,Rozenman2018,Forrest2020,Berghuis2022,Balasubrahmaniyam2023}. To understand the effect of the excitation on polariton-mediated transport, we performed simulations for both initial conditions. 

Resonant excitation into the LP branch by a short broad-band laser pulse, typically used in time-resolved experiments~\cite{Freixanet2000,Pandya2021,Pandya2022} was modeled by preparing a Gaussian wavepacket of LP states centered at $\hslash\omega$ = 3.94~eV where the group velocity of the LP branch, defined as $v_{\text{gr}}^\text{LP}(k_z)=
\partial\omega_\text{LP}(k_z)/\partial k_z$, is highest, and with a bandwidth of $\sigma =$~ 0.707~$\mu$m$^{-1}$~\cite{Agranovich2007}. Off-resonant excitation in a molecule-cavity system is usually achieved by optically pumping a higher-energy electronic state of the molecules~\cite{Lerario2017,Rozenman2018,Forrest2020,Balasubrahmaniyam2023}, which then rapidly relaxes into the lowest energy excited state (S$_1$) according to Kasha's rule~\cite{Kasha1950}. We therefore modelled off-resonant photo-excitation by starting the simulations directly in the S$_1$ state of a single molecule, located at $z$ = 5~$\mu$m in the cavity (SI). 
We assume that the intensity of the excitation pulse in both cases is sufficiently weak for the system to remain within the single-excitation subspace. We thus exclude multi-photon absorption and model the interaction with the pump pulse as an instantaneous absorption of a \emph{single} photon.

Because the light-confining structures used in previous experiments ({\it{e.g.}}, Fabry-P\'{e}rot cavities~\cite{Freixanet2000,Rozenman2018,Pandya2022,Xu2022}, Bloch surface waves~\cite{Lerario2017,Forrest2020,Balasubrahmaniyam2023}, or plasmonic lattices~\cite{Zakharko2018,Berghuis2022,Jin2023}) span a wide range of quality factors (Q-factors), we also investigated the effect of the cavity mode lifetime on the transport by performing simulations 
in an ideal lossless cavity with no photon decay ({\it{i.e.}}, $\gamma_\text{cav}=$ 0~ps$^{-1}$), and a lossy cavity with decay rate of 66.7~ps$^{-1}$. This decay rate corresponds to a lifetime of 15~fs, which is in the same order of magnitude as the 2 - 15~fs lifetimes reported for metallic Fabry-P\'{e}rot cavities in experiments~\cite{Schwartz2013,George2015,Rozenman2018,Wu2022}. In addition to cavity loss, also internal conversion via the conical intersection seam between the S$_1$ and S$_0$ potential energy surfaces~\cite{Boggio-Pasqua2012}, can provide a decay channel for the excitation. However, because in our Rhodamine model, the minimum energy conical intersection is 1.3~eV higher in energy than the vertical excitation (SI), and is therefore unlikely to be reached on the timescale of our simulations, we neglect internal conversion processes altogether.


\section*{Results \& Discussion}\label{sec2}


\subsection*{Resonant excitation}

\begin{figure*}[!htb]
\centering
\includegraphics[width=\textwidth]{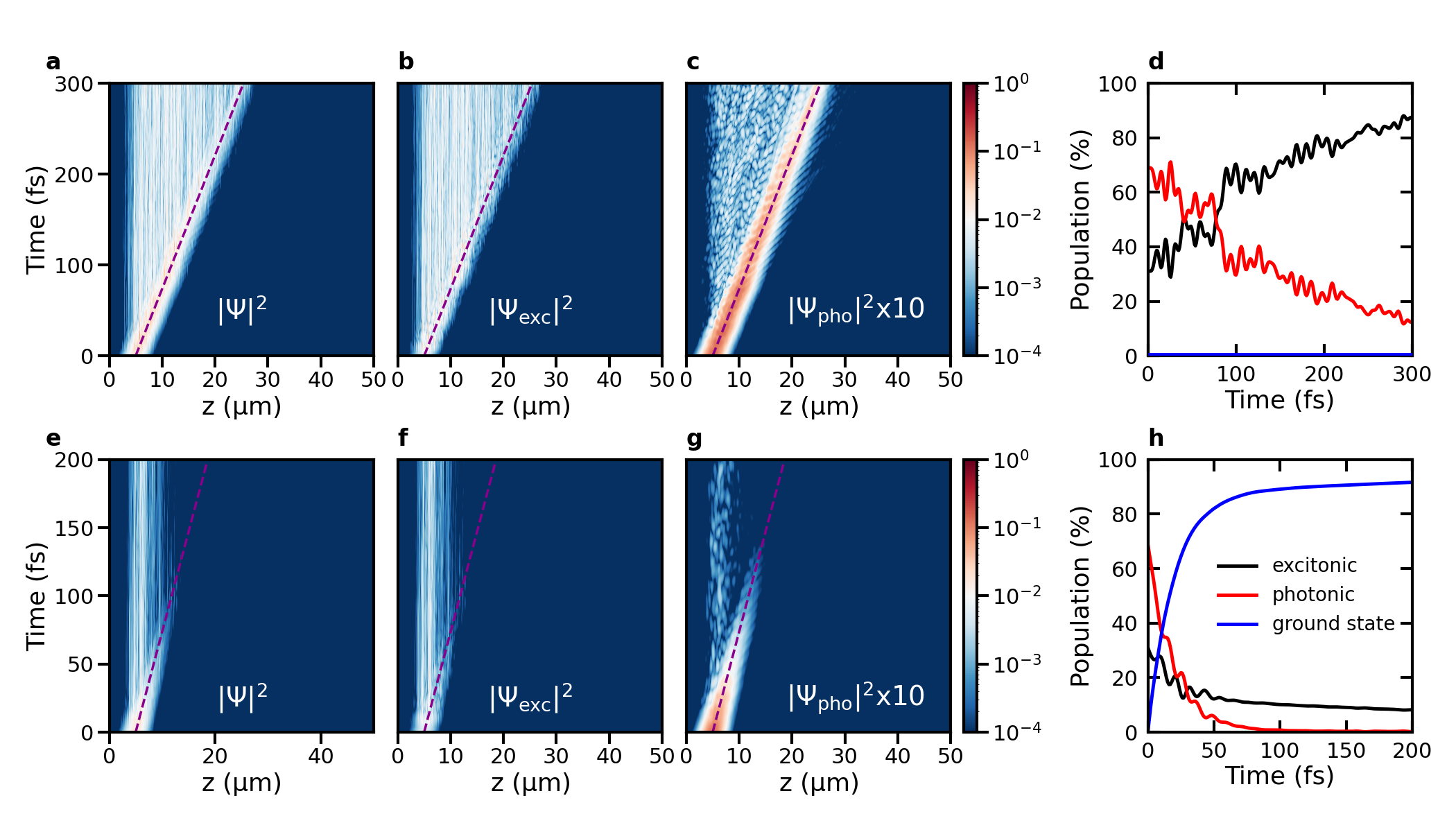}
  \caption{Polariton propagation after {\bf on-resonant} excitation of a wavepacket in the LP branch centered at $z=$~5~$\mu$m. Panels {\bf{a}}, {\bf{b}} and {\bf{c}}: total probability density $\vert\Psi(t)\vert^2$, probability density of the molecular excitons $\vert\Psi_{\text{exc}}(t)\vert^2$ and of the cavity mode excitations $\vert\Psi_{\text{pho}}(t)\vert^2$, respectively, as a function of distance (horizontal axis) and time (vertical axis), in a cavity with perfect mirrors ({\it i.e.}, $\gamma_{\text{cav}} = $~0~ps$^{-1}$). The magenta dashed line indicates propagation at the maximum group velocity of the LP (68~$\mu$mps$^{-1}$). Panel {\bf{d}}: Contributions of molecular excitons (black) and cavity mode excitations (red) to $\vert\Psi(t)\vert^2$ as a function of time in the perfect cavity. Without cavity losses, no ground state population (blue) can build up. Panels {\bf{e}}, {\bf{f}}, and {\bf{g}}: $\vert\Psi(t)\vert^2$, $\vert\Psi_{\text{exc}}(t)\vert^2$ and $\vert\Psi_{\text{pho}}(t)\vert^2$, respectively, as a function of distance (horizontal axis) and time (vertical axis), in a lossy cavity ($\gamma_{\text{cav}}=$~66.7~ps$^{-1}$). Panel {\bf{h}}: Contributions of the molecular excitons (black) and cavity mode excitations (red) to $\vert\Psi(t)\vert^2$ as a function of time. The population in the ground state, created by radiative decay through the imperfect mirrors, is plotted in blue.}\label{fig:wp_on_1024}
\end{figure*}

First, we explore how polaritons propagate after resonant excitation of a Gaussian wavepacket of LP states with a broad-band laser pulse. In Figure~\ref{fig:wp_on_1024}, we show the time evolution of the probability density of the polaritonic wave function, $\vert\Psi(t)\vert^2$ after such excitation in both a perfect lossless cavity with an infinite Q-factor ($\gamma_\text{cav}=$ 0~ps$^{-1}$, top panels) and a lossy cavity with a low Q-factor ($\gamma_\text{cav}=$ 66.7~ps$^{-1}$, bottom panels) containing 1024~Rhodamine molecules. Plots of wavepacket propagation in systems with 256 and 512~molecules are provided as SI (Figures S4-S5), as well as animations of the wavepackets for all system sizes. 


\subsubsection*{Lossless cavity}

\begin{figure}[!htb]
\centering
\includegraphics[width=0.45\textwidth]{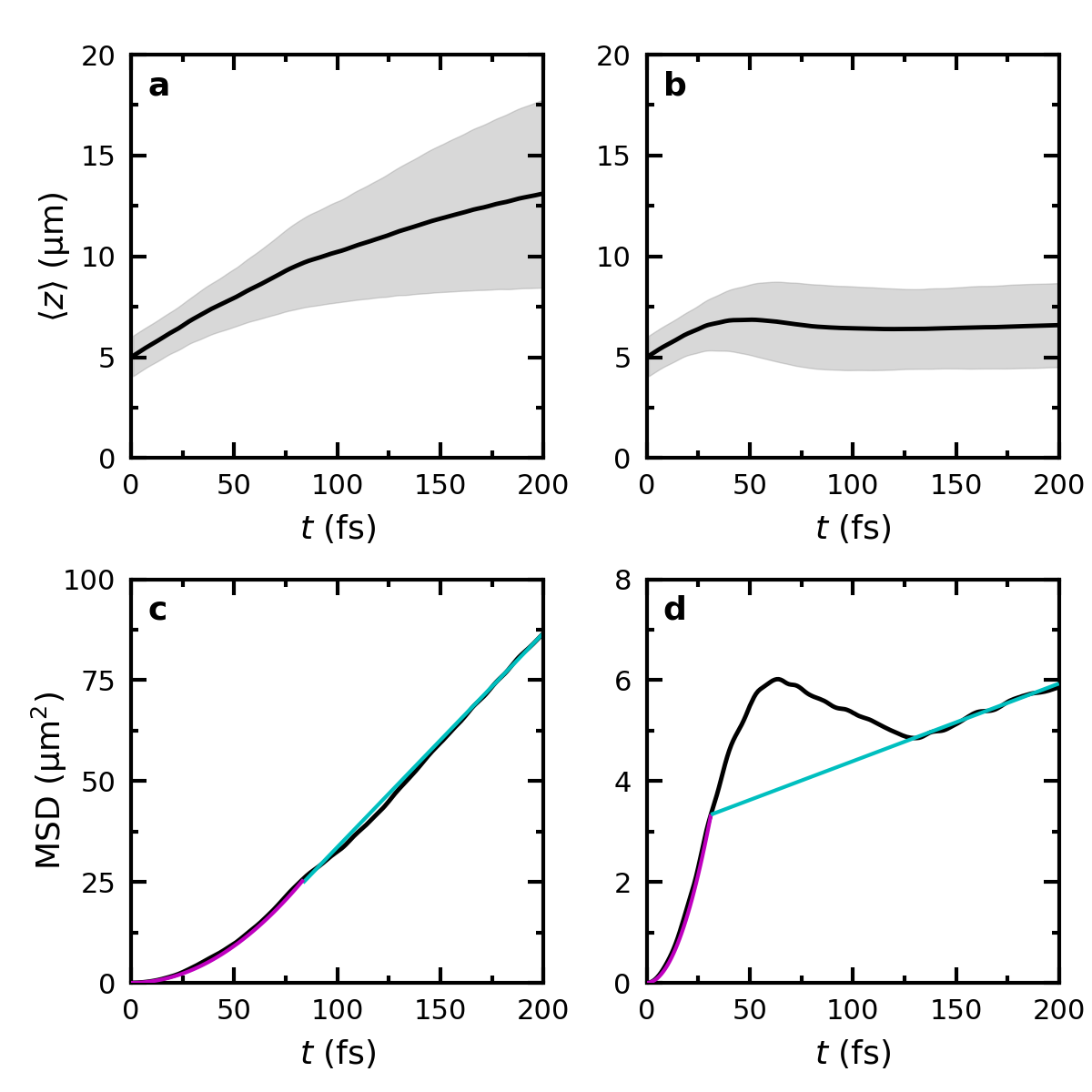}
  \caption{Top panels: Expectation value of the position of the total-time dependent wavefunction $\langle\Psi(t)\vert\hat{z}\vert\Psi(t)\rangle/\langle\Psi(t)\vert\Psi(t)\rangle$ after {\bf on-resonant} excitation in an ideal cavity ({\bf{a}}) and a lossy cavity ({\bf{b}}). The black lines represent $\langle z\rangle$ while the shaded area around the lines represents the root mean squared deviation (RMSD, {\it{i.e.}}, $\sqrt{\langle (z(t)-\langle z(t)\rangle)^2\rangle}$). Bottom panels: Mean square displacement (MSD, {\it{i.e.}}, $\sqrt{\langle (z(t)-\langle z(0)\rangle)^2\rangle}$) in the ideal lossless cavity ({\bf{c}}) and the lossy cavity ({\bf{d}}). Magenta lines are quadratic fits to the MSD and cyan lines are linear fits.
  }\label{fig:z_on}
\end{figure}

In the perfect lossless cavity the total wavepacket $\vert \Psi(t)\vert ^2$ initially propagates ballistically close to the maximum group velocity of the LP branch ($v^\text{LP,max}_\text{gr}=$ 68~$\mu$mps$^{-1}$, Figure~\ref{fig:structure+dispersion}{\bf c}), until around 100~fs (see animations in the SI), when it slows down as evidenced by a decrease in the slope of the expectation value of the position of the wavepacket $\langle z \rangle$ in Figure~\ref{fig:z_on}{\bf a}. The change from a quadratic to a linear time-dependence of the Mean Square Displacement (Figure~\ref{fig:z_on}{\bf c}) at $t=$ 100~fs furthermore suggests a transition from ballistic to diffusive motion. 

During propagation, the wavepacket broadens and sharp features appear, visible as vertical lines in both the total and molecular wavepackets in   Figure~\ref{fig:wp_on_1024}{\bf{a}-{\bf b}} and as peaks in the wavepacket animations provided as SI. These peaks coincide with the $z$ positions of molecules that contribute to the wavepacket with their excitations during propagation. 
Such peaks are not observed if there is no disorder and the molecular degrees of freedom are frozen (Figure~S15), but appear already at the start of the simulation when the initial configurations of the molecules are all different (Figure~S21). Similar observations were made by Agranovich and Gartstein~\cite{Agranovich2007}, who attributed these peaks to energetic disorder among the molecular excitons. We therefore also assign these peaks to a partial localization of the wavepacket at the molecules due to structural disorder that alters their contribution to the wavepacket.
In contrast, because the cavity modes are delocalized in space, the photonic wavepacket remains smooth throughout the propagation (Figure~\ref{fig:wp_on_1024}{\bf{c}}).


The transition from ballistic propagation to diffusion around 100~fs coincides with the onset of the molecular excitons dominating the polaritonic wavepacket, as shown in Figure~\ref{fig:wp_on_1024}{\bf d}, in which we plot the contributions of the molecular excitons (black line) and cavity mode excitations (red line) to the total wave function (see SI for details of this analysis). Because in the perfect cavity, photon leakage through the mirrors is absent ({\it i.e.}, $\gamma_{\text{cav}}=0~\text{ps}^{-1}$), the decrease of cavity mode excitations is due to population transfer from bright LP states into the dark state manifold (Figure~S19{\bf{b}})~\cite{Georgiou2018b,Groenhof2019,Takahashi2020}. Thus, while resonant excitation of LP states initially leads to ballistic motion with the central group velocity of the wavepacket, as evidenced by the quadratic dependence of the Mean Squared Displacement on time (Figure~\ref{fig:z_on}{\bf c}), population transfer into dark states turns the propagation into a diffusion process, as evidenced by a linear time-dependence of the Mean Square Displacement after $\sim$100~fs. 

Since dark states lack group velocity, and are therefore stationary, while excitonic couplings between molecules are neglected in our model (see SI), propagation in the diffusive regime must still involve bright polariton states. Our simulations therefore suggest that while, initially, molecular vibrations drive population transfer from the propagating bright states into the stationary dark states~\cite{Tichauer2022}, this process is reversible, causing new wavepackets to form continuously within the full range of LP group velocities. Likewise, the propagation of transiently occupied bright states is continuously interrupted by transfers into dark states, and re-started with different group velocities. This re-spawning process leads to the diffusive propagation of the excitation observed in Figure~\ref{fig:wp_on_1024}, with an increasing wavepacket width (Figure~\ref{fig:z_on}{\bf a}), in line with experimental observations~\cite{Rozenman2018,Forrest2020,Pandya2021,Balasubrahmaniyam2023}.


\subsubsection*{Lossy cavity}

Including a competing radiative decay channel by adding photon losses through the cavity mirrors at a rate of $\gamma_\text{cav}=$ 66.7~ps$^{-1}$, leads to a rapid depletion of the polariton population (Figure~\ref{fig:wp_on_1024}{\bf h}), but does not affect the overall transport mechanism: the wavepacket still propagates in two phases, with a fast ballistic regime followed by slower diffusion. However, in contrast to the propagation in the ideal lossless cavity, we observe that the wavepacket  temporarily contracts.
This contraction is visible as a reduction of both the expectation value of $\langle z\rangle$ and the Mean Squared Displacement between 60 to 130~fs in the right panels of Figure~\ref{fig:z_on}.

Initially the propagation of the wavepacket is dominated by ballistic motion of the population in the bright polaritonic states moving at the maximum group velocity of the LP branch. However, due to non-adiabatic coupling~\cite{Tichauer2022}, some of that population is transferred into dark states that are stationary. Because non-adiabatic population transfer is reversible, the wavepacket propagation undergoes a transition into a diffusion regime, which is significantly slower, as also observed in the ideal cavity (Figure~\ref{fig:z_on}{\bf{c}}). 

In addition to these non-adiabatic transitions, radiative decay further depletes population from the propagating bright polaritonic states. Because before decay, this population has moved much further than the population that got trapped in the dark states, the expectation value of $\langle z\rangle$, as well as the Mean Square displacement, which were dominated initially by the fast-moving population, decrease until the slower diffusion process catches up and reaches the same distance around 130~fs (right panels in Figure~\ref{fig:z_on}). Such contraction of the wavepacket in a lossy cavity is consistent with the measurements of Musser and co-workers, who also observe such contraction after on-resonant excitation of UP states~\cite{Pandya2022}.


Because of the contraction, it is difficult to  see where the transition between ballistic and diffusion regimes occurs in Figure~\ref{fig:z_on}{\bf{d}}. We therefore instead extrapolated the linear regime, and estimate the turn-over at 30~fs, where the quadratic fit to the ballistic regime intersects the extrapolated fit to the diffusion regime. As in the perfect lossless cavity, the transition between ballistic and diffusion regimes occurs when the population of molecular excitons exceeds the population of cavity mode excitations (Figure~\ref{fig:wp_on_1024}{\bf h}). However, due to the radiative decay of the latter, this turnover already happens around 30~fs in the lossy cavity simulations.

Owing to the short cavity mode lifetime (15~fs), most of the excitation has already decayed into the ground state at 100~fs, with a small remainder ``surviving" in dark states (Figure~\ref{fig:wp_on_1024}{\bf h}) that lack mobility. Because cavity losses restrict the lifetime of bright LP states, the distance a wavepacket can reach is limited due to (i) the shortening of the ballistic phase, and (ii) the reduction of the diffusion coefficient ({\it i.e.}, the slope of $\langle z\rangle$, Figure~\ref{fig:z_on}{\bf{c}}) in the second phase. Therefore, the overall velocity is significantly lower than in the perfect cavity, suggesting a connection between cavity Q-factor and propagation velocity~\cite{Pandya2022}, while also the broadening of the wavepacket is reduced (Figure~\ref{fig:z_on}{\bf b}). Furthermore, because the rate of population transfer is inversely proportional to the energy gap~\cite{Tichauer2022}, and hence highest when the LP and dark states overlap~\cite{Groenhof2019}, 
we speculate that the turn-over between the ballistic and diffusion regimes depends on the overlap between the absorption line width of the molecules and the polaritonic branches, and can hence be controlled by tuning the excitation energy to move the center of the initial polaritonic wavepacket along the LP branch. In addition, the direction of ballistic propagation can be controlled by varying the incidence angle of the on-resonant excitation pulse.



\subsubsection*{Comparison to experiments}

Our observations are in line with transient microscopy experiments, in which broad-band excitation pulses were used to initiate polariton propagation. At low temperatures Freixanet {\it{et al.}} observed ballistic wavepacket propagation for a strongly coupled quantum dot~\cite{Freixanet2000}. If we suppress  vibrations that drive population transfer by freezing the nuclear degrees of freedom, we also observe such purely ballistic motion (Figure~S15). In contrast, in room temperature experiments on cavity-free molecular polaritons, Pandya {\it{et al.}} identified two transport regimes: a short ballistic phase followed by diffusion~\cite{Pandya2021}. Based on the results of our simulations, we attribute the first phase to purely ballistic wavepacket propagation of photo-excited LP states. The slow-down of the transport in the second phase is attributed to reversible trapping of population inside the stationary dark state manifold. Owing to the reversible transfer of population between these dark states and the LP states, propagation continues diffusively at time scales exceeding the polariton lifetime, in line with experiment~\cite{Rozenman2018,Pandya2021}. 


\subsection*{Off-resonant excitation}

\begin{figure*}[!htb]
\centering
\includegraphics[width=16cm]
{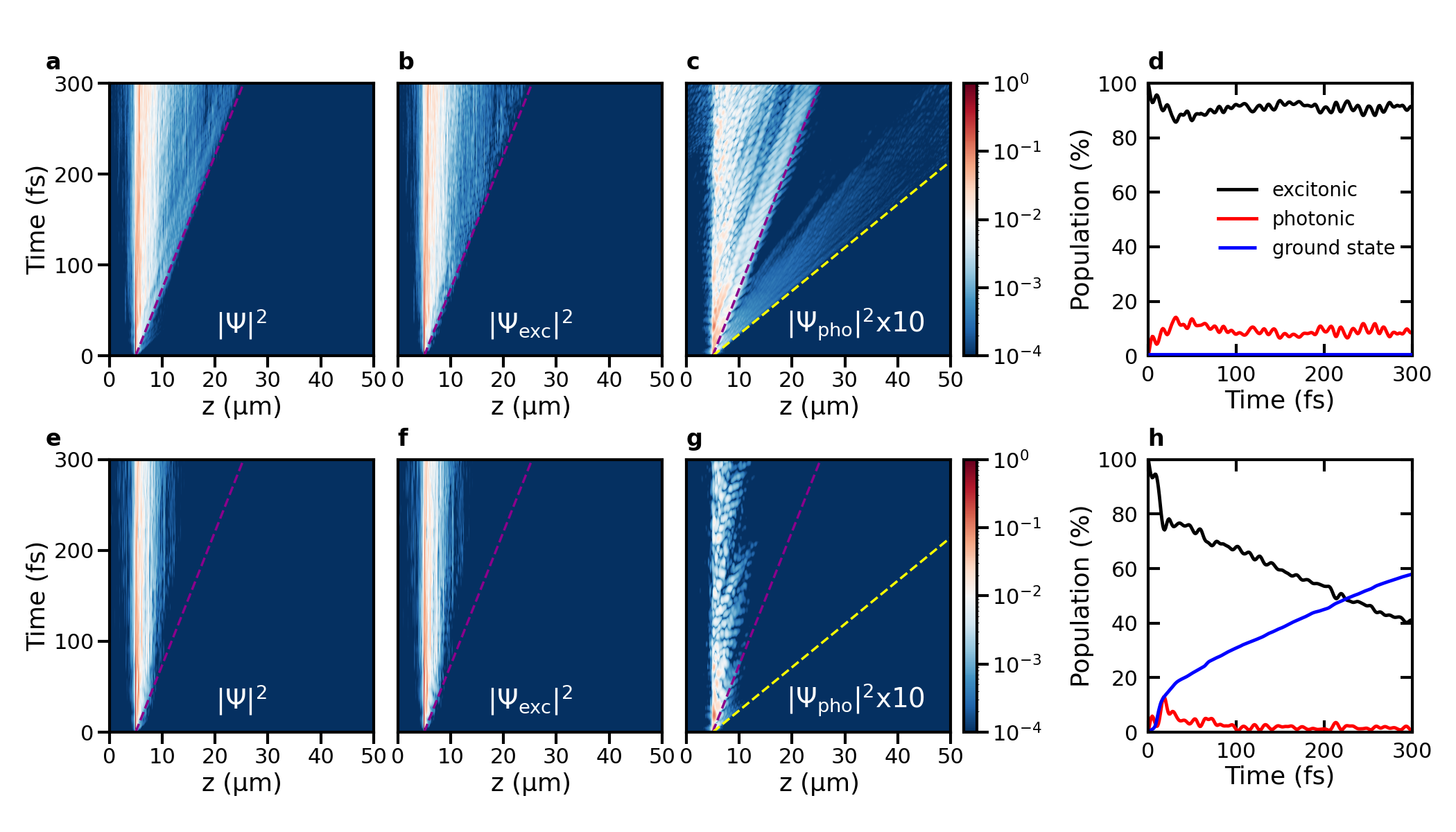}
  \caption{Polariton propagation after {\bf off-resonant} excitation of a single molecule located at $z=$~5~$\mu$m into S$_1$. Panels {\bf{a}}, {\bf{b}} and {\bf{c}}: total probability density $\vert\Psi(t)\vert^2$, probability density of the molecular excitons $\vert\Psi_{\text{exc}}(t)\vert^2$ and of the cavity mode excitations $\vert\Psi_{\text{pho}}(t)\vert^2$, respectively, as a function of distance (horizontal axis) and time (vertical axis), in a cavity with perfect mirrors ({\it i.e.}, $\gamma_{\text{cav}} = $~0~ps$^{-1}$). The magenta and yellow dashed lines indicate propagation at the maximum group velocity of the LP (68~$\mu$mps$^{-1}$) and UP (212~$\mu$mps$^{-1}$), respectively. Panel {\bf{d}}: Contributions of the  molecular excitons (black) and cavity mode excitations (red) to $\vert\Psi(t)\vert^2$ as a function of time in the perfect cavity. Without cavity decay, there is no build-up of ground state population (blue). Panels {\bf{e}}, {\bf{f}}, {\bf{g}}: $\vert\Psi(t)\vert^2$, $\vert\Psi_{\text{exc}}(t)\vert^2$ and $\vert\Psi_{\text{pho}}(t)\vert^2$, respectively, as a function of distance (horizontal axis) and time (vertical axis), in a lossy cavity ({\it i.e.}, $\gamma_{\text{cav}} = $~66.7~ps$^{-1}$). Panel {\bf{h}}: Contributions of the molecular excitons (black), and cavity mode excitations (red) to $\vert\Psi(t)\vert^2$ as a function of time in the lossy cavity. The population in the ground state, created by radiative decay through the imperfect mirrors, is plotted in blue.
  }
  \label{fig:wp_off_1024}
\end{figure*}

Next, we investigate polariton propagation after an off-resonant excitation into the S$_1$ electronic state of a single Rhodamine molecule, located at $z =$ 5 $\mu$m. In Figure~\ref{fig:wp_off_1024} we show the time evolution of the probability density of the total polaritonic wave function, $\vert\Psi(t)\vert^2$, after such excitation in both a perfect lossless cavity with an infinite Q-factor ($\gamma_\text{cav}=$ 0~ps$^{-1}$, top panels) and a lossy cavity with a low Q-factor ($\gamma_\text{cav}=$ 66.7~ps$^{-1}$, bottom panels) containing 1024~Rhodamine molecules. Plots of the wavepacket propagation in systems with 256 and 512~molecules are provided as SI (Figures S8-S9), as well as animations of the wavepackets for all system sizes. 


\subsubsection*{Lossless cavity}

\begin{figure}[!htb]
\centering
\includegraphics[width=0.45\textwidth]{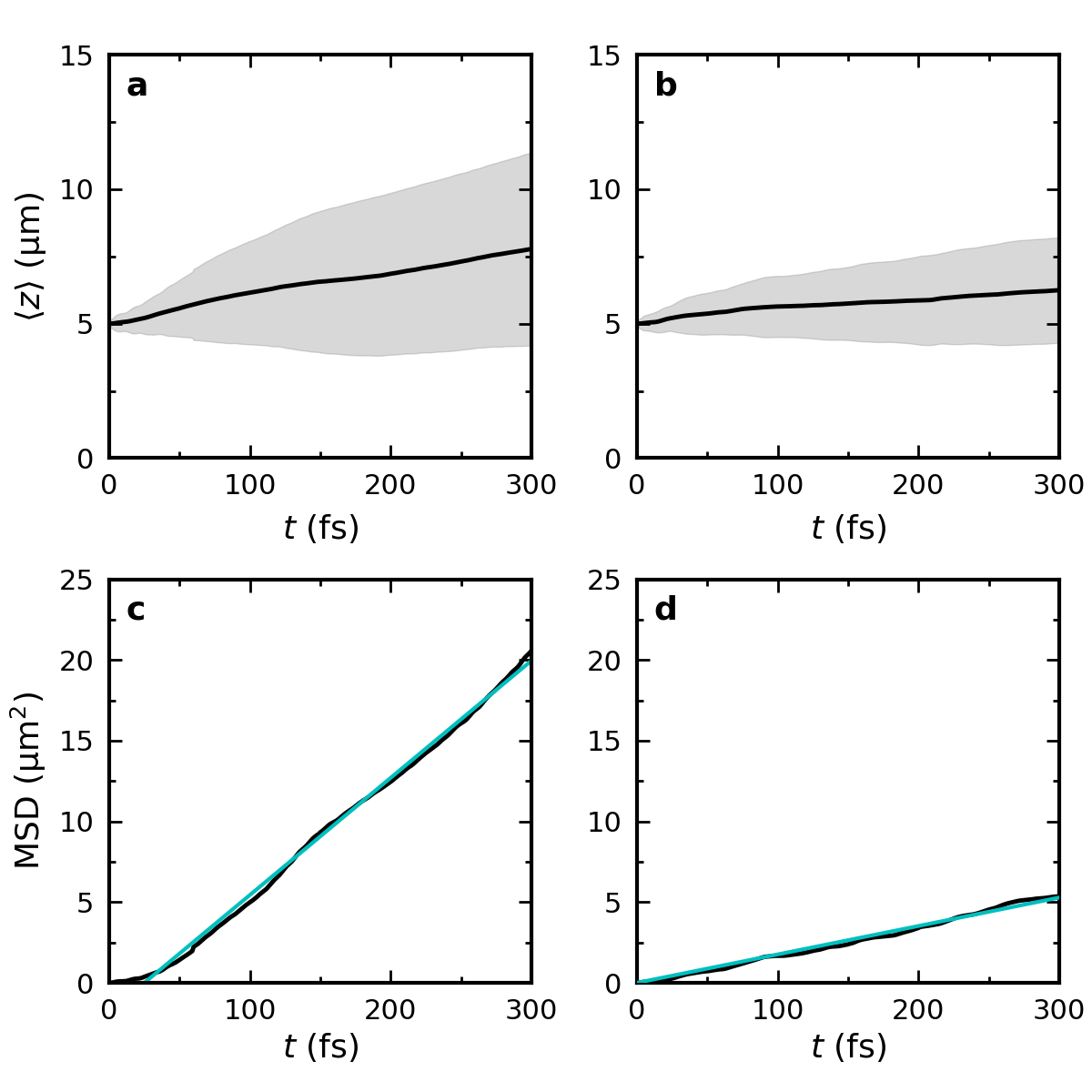}
  \caption{Top panels: Expectation value of the position of the total time-dependent wavefunction, $\langle z\rangle =  \langle\Psi(t)\vert \hat{z}\vert\Psi(t)\rangle/\langle\Psi(t)\vert\Psi(t)\rangle$, after {\bf off-resonant} excitation in an ideal cavity ({\bf{a}}, $\gamma_\text{cav}=0$~ps$^{-1}$) and a lossy cavity ({\bf{b}}, $\gamma_\text{cav}=66.7$~ps$^{-1}$). The black lines represent $\langle z\rangle$ while the shaded area around the lines represents the root mean squared deviation (RMSD, {\it{i.e.}}, $\sqrt{\langle (z(t)-\langle z(t)\rangle)^2\rangle}$). Bottom panels: Mean square displacement (MSD, {\it{i.e.}}, ($\langle (z(t)-z(0))^2\rangle$) in the ideal lossless cavity ({\bf{c}}) and the lossy cavity ({\bf{d}}). Cyan lines are linear fits to the MSD.
  }
  \label{fig:z_off}
\end{figure}

In the lossless cavity with perfect mirrors, the excitation, initially localised at a single molecule, rapidly spreads to other molecules (see animation in the SI). In contrast to the ballistic movement observed for on-resonant excitation, the wavepacket spreads out instead, with the front of the wavepacket propagating at a velocity that closely matches the maximum group velocity of the LP branch (68~$\mu$mps$^{-1}$, Figure~\ref{fig:structure+dispersion}{\bf c}), while the expectation value of the wavepacket position ($\langle z\rangle$, Figure~\ref{fig:z_off}{\bf a}) moves at a lower pace (
$\sim$10~$\mu$mps$^{-1}$). 

Because we do not include negative $k_z$-vectors in our cavity model, propagation can only occur in the positive $z$ direction. With negative $k_z$-vectors, propagation in the opposite direction cancels such motion leading to $\langle z\rangle\approx 0$ (Figure S14{\bf{a}}). Nevertheless, since the Mean Square Displacement is not affected by breaking the symmetry of the 1D cavity, and increases linearly with time in both uni- and bi-directional cavities (Figures~\ref{fig:z_off}{\bf{c}} and~S14{\bf{b}}), we consider it reasonable to assume that the mechanism underlying the propagation process is identical. 


Because the population of dark states dominates throughout these simulations (Figure~\ref{fig:wp_off_1024}{\bf d}), and direct excitonic couplings are not accounted for in our model (SI), the observed propagation must again involve bright polariton states. 
Since the initial state, with one molecule excited, is not an eigenstate of the molecule-cavity system, population exchange from this state into the propagating bright states is not only due to displacements along vibrational modes that are overlapping with the non-adiabatic coupling vector~\cite{Tichauer2022}, but also due to Rabi oscillations, in particular at the start of the simulation.

To quantify to what extent the overall propagation is driven by population transfers due to the molecular displacements, we performed additional simulations at 0~K with all nuclear degrees of freedom frozen. As shown in Figure~S17, the propagation is reduced at 0~K, and the wavepacket remains more localized on the molecule that was initially excited, than at 300~K. A quadratic time-dependence of the Mean Square Displacement of the cavity mode contributions to the wavepacket (Figure~S18{\bf f}) furthermore suggest that the mobility at 0~K is driven by the constructive and destructive interferences of the bright polaritonic states, which evolve with different phases ({\it{i.e.}}, $e^{-iE_mt/\hslash}$). 

The reduced mobility of the wavepacket at 0~K compared to 300~K (Figure~S18) confirms that thermally activated displacements of nuclear coordinates, which are absent at 0~K, are essential to drive population into the bright states and sustain the propagation of the polariton wavepacket. Thus, as during the diffusion phase observed for on-resonant excitation, ballistic motion of bright states is continuously interrupted and restarted with different group velocities, which makes the overall propagation appear diffusive with a Mean Square Displacement that depends linearly on time (Figure~\ref{fig:z_off}{\bf c}), in line with experimental observations~\cite{Rozenman2018,Forrest2020}. 

In the perfect cavity, propagation and broadening continue indefinitely due to the long-range ballistic motion of states with higher group velocities. Indeed, a small fraction at the front of the wavepacket, which moves even faster than the maximum group velocity of the LP (indicated by a magenta dashed line in Figure~\ref{fig:wp_off_1024}), is mostly composed of higher-energy UP states. These states not only have the highest in-plane momenta, but also decay most slowly into the dark state manifod of the perfect cavity due to the inverse dependence of the non-adiabatic coupling on the energy gap~\cite{Tichauer2021}. Momentum-resolved photo-luminenscence spectra at two distances from the initial excitation spot (Figure~S23, SI) confirm that the front of the wavepacket is indeed composed of UP states: at short distances ($z =$~10$\mu$m) from the excitation spot ($z =$ 5~$\mu$m), the emission spectrum, accumulated over 100~fs simulation time, closely matches the full polariton dispersion of Figure~\ref{fig:structure+dispersion}{\bf b}, displaying both the LP and UP branches. In contrast, further away from the excitation spot ($z =$ 20~$\mu$m), the  emission exclusively originates from the higher energy UP states, suggesting that only these states can reach the longer distance within 100~fs.

\subsubsection*{Lossy cavity}

Adding a radiative decay channel for the cavity mode excitations ($\gamma_\text{cav}$ = 66.7~ps$^{-1}$) restricts the distance over which polaritons propagate (Figure~\ref{fig:wp_off_1024}{\bf e}-{\bf g}), but does not affect the overall transport mechanism, as we also observe a linear increase of the MSD with time (Figure~\ref{fig:z_off}{\bf d}). While the propagation in the lossy cavity initially is very similar to that in the ideal lossless cavity, radiative decay selectively depletes population from the propagating bright states and the wavepacket slows down, as evidenced by the expectation value of the displacement, $\langle z\rangle$, levelling off in Figure~\ref{fig:z_off}{\bf b}. In addition, since the maximum distance a wavepacket can travel in lossy cavities is determined by the cavity lifetime in combination with the group velocity~\cite{Lerario2017}, the broadening of the wavepacket is also more limited when cavity losses are included (Figure~\ref{fig:z_off}{\bf b}). Furthermore, even if the dark states do not have a significant contribution from the cavity mode excitations, the reversible transfer of population between the dark state manifold and the decaying bright polaritonic states, also leads to a significant reduction of dark state population in the lossy cavity as compared to the ideal lossless cavity (Figure~\ref{fig:wp_off_1024}{\bf d} and {\bf f}). Nevertheless, dark states still provide ``protection" from cavity losses as the overall lifetime of the photo-excited molecule-cavity system ($>$ 150~fs) significantly exceeds that of the cavity modes (15~fs). 

\subsubsection*{Comparison to experiments}

In microscopy experiments relying on off-resonant optical pumping, polariton emission is typically observed between the excitation spot and a point several microns further away~\cite{Lerario2017, Rozenman2018,Forrest2020,Berghuis2022,Balasubrahmaniyam2023}. While such broad emission pattern is reminiscent of a diffusion process, the match between total distance over which that emission is detected on the one hand and the product of the maximum LP group velocity and cavity lifetime on the other hand, suggest ballistic propagation. The results of our simulations are thus in qualitative agreement with such observations as also our results suggest that, while polariton propagation appears diffusive under off-resonant excitation conditions, the front of the wave packet propagates close to the maximum group velocity of the LP branch. 

Based on the analysis of our MD trajectories we propose that on the experimentally accessible timescales, 
polariton propagation appears diffusive due to reversible population transfers between stationary dark states and propagating bright states. For lossy cavities, radiative decay of the cavity modes further slows down polariton transport such that the excitation reaches a maximum distance before decaying completely. Because a large fraction of the population resides in the non-decaying dark states, the lifetime of the molecule-cavity system is extended~\cite{Groenhof2019}, and polariton propagation can be observed on timescales far beyond the cavity lifetime, in line with experiment~\cite{Rozenman2018}.

Note that in our simulations we couple excitons only to the modes of the Fabry-P\'{e}rot cavity, whereas in experiments with micro-cavities constituted by metal mirrors, excitons can in principle also couple to the surface plasmon polaritons (SPP) below the light line that are supported by these metal surfaces. While their role will depend on the details of the set-up ({\it{e.g.}}, the materials used, energy of the relevant molecular excitations, {\it{etc.}}), we cannot rule out that reversible population transfer between the dark states and SPP-exciton polaritons also contributes to the effective diffusion constant observed in those experiments~\cite{Rozenman2018,Xu2022}. However, because the SPP decays exponentially away from the metal surface, and SPP-exciton polaritons also have group velocity, the qualitative behavior is not expected to change.

\subsection*{Size dependence}

Due to limitations on hard- and software, the number of molecules that we can include in our simulations is much smaller than in experiments~\cite{Houdre1996,Eizner2019,Martinez2019}. We therefore investigated how the number of molecules, $N$, coupled to the cavity affects the propagation by performing simulations for different $N$. To keep the Rabi splitting ($\sim$~325~meV) constant, and hence polariton dispersion the same, we scaled the cavity mode volume with the number of molecules $N$ (see SI for details). 

While the transport mechanism is not strongly affected by $N$ (Figures~S4-S12), the total population that resides in the bright states decreases when the number of molecules, and hence the number of dark states, increases, in particular in the diffusion phase. Such decrease in bright state population is due to the $1/N$ scaling of the rate at which population transfers between dark and bright states~\cite{Tichauer2022}. Because the number of dark states is proportional to $N$, while the number of bright states is constant for a fixed number of cavity modes, this dependency affects the ratio between the population in the dark and bright states, with the latter rapidly decreasing with increasing $N$. As the overall propagation velocity is determined by the population in bright states, also the velocity is inversely proportional to $N$ (Figure~S13). Therefore, in experiments, with 10$^5$-10$^8$ molecules inside the mode volume~\cite{Houdre1996,Eizner2019,Martinez2019}, the propagation velocity is much lower than in our simulations.

Nevertheless, because of the $1/N$ scaling, the effective polariton propagation velocity approaches the lower ``experimental limit" of 10$^5$ coupled molecules~\cite{Houdre1996} already around 1000 molecules. We therefore consider the results of the simulations with 1024 Rhodamines sufficiently representative for experiment and for providing qualitative insights into polariton propagation. Indeed, a propagation of 9.6~${\mu}$mps$^{-1}$ in the cavity containing 1024~molecules is about an order of magnitude below the maximum group velocity of the LP (68 ${\mu}$mps$^{-1}$) in line with experiments on organic microcavities~\cite{Rozenman2018}, and cavity-free polaritons~\cite{Pandya2021}


\section*{Conclusions}

We have investigated exciton transport in Rhodamine cavities by means of atomistic MD simulations that not only include the details of the cavity mode structure~\cite{Michetti2005,Tichauer2021}, but also the chemical details of the material~\cite{Luk2017}. 
The results of our simulations suggest that the transport is driven by an interplay between \emph{propagating} bright polaritonic states and \emph{stationary} dark states. Reversible population exchanges between these states interrupt  ballistic motion in bright states and make the overall propagation process appear diffusive.
While for off-resonant excitation of the molecule-cavity system, these exchanges are essential to transfer population from the initially excited molecule into the bright polaritonic branches and start the propagation process, the exchanges limit the duration of the initial ballistic phase for on-resonant excitation. As radiative decay of the cavity modes selectively depletes the population in bright states, ballistic propagation is restricted even further if the cavity is lossy. 
Because dark states lack in-plane momentum, the reversible population exchange between dark and bright states causes diffusion in \emph{all} directions. Therefore, under off-resonant excitation conditions, the propagation direction cannot be controlled. In contrast, because bright states carry momentum, the propagation direction in the ballistic phase can be controlled precisely by tuning the incidence angle and excitation wavelength under on-resonant excitation conditions.

The rate at which population transfers between bright and dark states depends on the non-adiabatic coupling vector, whose direction and magnitude are determined by the Huang-Rhys factor in combination with the frequency of the Franck-Condon active vibrations~\cite{Tichauer2022}, both of which are related to the molecular Stokes shift~\cite{DeJong2015}. In addition, because the non-adiabatic coupling is inversely proportional to the energy gap~\cite{Tichauer2022}, the Stokes shift in combination with the Rabi splitting, also determines the region on the LP branch into which population transfers after off-resonant excitation of a single molecule~\cite{Grant2016,Baieva2017,Luettgens2021,Hulkko2021}. We therefore speculate that the Stokes shift can be an important ``control knob" for tuning the coherent propagation of polaritons.

Because our Rhodamine model features the key photophysical characteristics of an organic dye molecule, we speculate that the propagation mechanism observed in our simulations is generally valid for exciton transport in strongly-coupled organic micro-cavities, in which the absorption line width of the material exceeds the Rabi splitting and there is  a significant overlap between bright and dark states.
To confirm this, we have also performed simulations of exciton transport in cavities containing Tetracene and Methylene Blue and observed that the propagation mechanism remains the same (SI).

Future work will be aimed at investigating how the propagation can be controlled by tuning molecular parameters, temperature, Rabi splitting, or cavity Q-factor~\cite{Tichauer2023}. 
Because we include the structural details of both cavity and molecules, our simulations, which are in qualitative agreement with experiments, pave the way to systematically optimize molecule-cavity systems for enhancing exciton energy transfer. 

\backmatter

\bmhead{Supplementary information}

The Supporting Information contains: (i) details of the multi-scale MD simulation model; (ii) details of simulation setups and analysis; (iii) results of simulations with different numbers of molecules, at 0~K, in symmetric cavities, with positional and energetic disorder, with different vacuum fields, as well as with Tetracene and Methylene Blue instead of Rhodamine; and (iv) animations of all wavepackets.

\bmhead{Acknowledgments}

We thank J. Jussi Toppari, A. M. Berghuis, J. G\'{o}mez Rivas, T. Schwartz, and M. Balusubrahmaniyam for fruitful discussions. We also thank the Center for Scientific Computing (CSC-IT Center for Science) for generous computational resources, and Nino Runenberg for his assistance in running the simulations on these resources.

\section*{Declarations}

\begin{itemize}
\item Funding: This work was supported by the Academy of Finland (Grant No. 323996 and 332743 to GG), the European Research Council (Grant No. ERC-2016-StG-714870 to JF), and by the Spanish Ministry for Science, Innovation, Universities-Agencia Estatal de Investigaci\'{o}n (AEI) through Grants (PID2021-125894NB-I00 and CEX2018-000805-M (through the Mar\'{i}a de Maeztu program for Units of Excellence in Research and Development).
\item Conflict of interest/Competing interests: The authors declare no competing financial interests.
\item Consent for publication: All authors consent to publication
\item Availability of data and materials: All data, including simulations models, input files, trajectories and structures, analysis scripts and programs, inlcuding raw data, are available for download from Fairdata IDA.
\item Code availability: The GROMACS-4.5.3 fork with the multi-scale Tavis-Cummings model is available for download from: https://github.com/rhti/gromacs$\_$qed
\item Authors' contributions: IS, RHT and GG conceptualized the project, IS and RHT performed the simulations, IS, RT and GG analysed the data, and DM performed higher-level {\it{ab initio}} computations. All authors contributed to the interpretation of the data and participated in writing the manuscript. IS and RHT contributed equally to this work. 
\end{itemize}

\bibliography{main}

\begin{thebibliography}{10}
\expandafter\ifx\csname url\endcsname\relax
  \def\url#1{\burl{#1}}\fi
\expandafter\ifx\csname urlprefix\endcsname\relax\def\urlprefix{URL }\fi
\providecommand{\bibinfo}[2]{#2}
\providecommand{\eprint}[2][]{\url{#2}}
\providecommand{\doi}[1]{\url{https://doi.org/#1}}
\bibcommenthead

\bibitem{Croce2014}
\bibinfo{author}{Croce, R.} \& \bibinfo{author}{Amerongen, H.~V.}
\newblock \bibinfo{title}{Natural strategies for photosynthetic light
  harvesting}.
\newblock \emph{\bibinfo{journal}{Nat. Chem. Biol.}}
  \textbf{\bibinfo{volume}{10}}, \bibinfo{pages}{492--501}
  (\bibinfo{year}{2014}) .

\bibitem{Mikhnenko2015}
\bibinfo{author}{Mikhnenko, O.~V.}, \bibinfo{author}{Blom, P. W.~M.} \&
  \bibinfo{author}{Nguyen, T.-Q.}
\newblock \bibinfo{title}{Exciton diffusion in organic semiconductors}.
\newblock \emph{\bibinfo{journal}{Energy Environ. Sci.}}
  \textbf{\bibinfo{volume}{8}}, \bibinfo{pages}{1867--1888}
  (\bibinfo{year}{2015}) .

\bibitem{Cao2014}
\bibinfo{author}{Cao, H.} \emph{et~al.}
\newblock \bibinfo{title}{{Recent progress in degradation and stabilization of
  organic solar cells}}.
\newblock \emph{\bibinfo{journal}{Journal of Power Sources}}
  \textbf{\bibinfo{volume}{264}}, \bibinfo{pages}{168--183}
  (\bibinfo{year}{2014}) .

\bibitem{Rafique2018}
\bibinfo{author}{Rafique, S.}, \bibinfo{author}{Abdullah, S.~M.},
  \bibinfo{author}{Sulaiman, K.} \& \bibinfo{author}{Iwamoto, M.}
\newblock \bibinfo{title}{Fundamentals of bulk heterojunction organic solar
  cells: An overview of stability/degradation issues and strategies for
  improvement}.
\newblock \emph{\bibinfo{journal}{Renew. Sust. Energ. Rev.}}
  \textbf{\bibinfo{volume}{84}}, \bibinfo{pages}{43--53} (\bibinfo{year}{2018})
  .

\bibitem{Akselrod2014}
\bibinfo{author}{Akselrod, G.~M.} \emph{et~al.}
\newblock \bibinfo{title}{Visualization of exciton transport in ordered and
  disordered molecular solids}.
\newblock \emph{\bibinfo{journal}{Nat. Comm.}} \textbf{\bibinfo{volume}{5}},
  \bibinfo{pages}{3646} (\bibinfo{year}{2014}) .

\bibitem{Sneyd2021}
\bibinfo{author}{Sneyd, A.} \emph{et~al.}
\newblock \bibinfo{title}{Efficient energy transport in an organic
  semiconductor mediated by transient exciton delocalization}.
\newblock \emph{\bibinfo{journal}{Sci. Adv.}} \textbf{\bibinfo{volume}{7}},
  \bibinfo{pages}{eabh4232} (\bibinfo{year}{2021}) .

\bibitem{Kong2022}
\bibinfo{author}{Kong, F.~F.} \emph{et~al.}
\newblock \bibinfo{title}{Wavelike electronic energy transfer in
  donor–acceptor molecular systems through quantum coherence}.
\newblock \emph{\bibinfo{journal}{Nat. Nanotechnol.}}
  \textbf{\bibinfo{volume}{17}}, \bibinfo{pages}{729--736}
  (\bibinfo{year}{2022}) .

\bibitem{Sneyd2022}
\bibinfo{author}{Sneyd, A.~J.}, \bibinfo{author}{Beljonne, D.} \&
  \bibinfo{author}{Rao, A.}
\newblock \bibinfo{title}{A new frontier in exciton transport: Transient
  delocalization}.
\newblock \emph{\bibinfo{journal}{J. Chem. Phys. Lett.}}
  \textbf{\bibinfo{volume}{13}}, \bibinfo{pages}{6820--6830}
  (\bibinfo{year}{2022}) .

\bibitem{Feist2015}
\bibinfo{author}{Feist, J.} \& \bibinfo{author}{Garcia-Vidal, F.~J.}
\newblock \bibinfo{title}{Extraordinary exciton conductance induced by strong
  coupling}.
\newblock \emph{\bibinfo{journal}{Phys. Rev. Lett.}}
  \textbf{\bibinfo{volume}{114}}, \bibinfo{pages}{196402}
  (\bibinfo{year}{2015}) .

\bibitem{Schachenmayer2015}
\bibinfo{author}{Schachenmayer, J.}, \bibinfo{author}{Genes, C.},
  \bibinfo{author}{Tignone, E.} \& \bibinfo{author}{Pupillo, G.}
\newblock \bibinfo{title}{{Cavity enhanced transport of excitons}}.
\newblock \emph{\bibinfo{journal}{Phys. Rev. Lett.}}
  \textbf{\bibinfo{volume}{114}}~(May), \bibinfo{pages}{196403}
  (\bibinfo{year}{2015}) .

\bibitem{Wellnitz2022}
\bibinfo{author}{Wellnitz, D.}, \bibinfo{author}{Pupillo, G.} \&
  \bibinfo{author}{Schachenmayer, J.}
\newblock \bibinfo{title}{Disorder enhanced vibrational entanglement and
  dynamics in polaritonic chemistry}.
\newblock \emph{\bibinfo{journal}{Comm. Phys.}} \textbf{\bibinfo{volume}{5}},
  \bibinfo{pages}{120} (\bibinfo{year}{2022}) .

\bibitem{Skolnick1998}
\bibinfo{author}{Skolnick, M.~S.}, \bibinfo{author}{Fisher, T.~A.} \&
  \bibinfo{author}{Whittaker, D.~M.}
\newblock \bibinfo{title}{Strong coupling phenomena in quantum microcavity
  structures}.
\newblock \emph{\bibinfo{journal}{Semicond. Sci. Technol}}
  \textbf{\bibinfo{volume}{13}}, \bibinfo{pages}{645--669}
  (\bibinfo{year}{1998}) .

\bibitem{Litinskaya2006}
\bibinfo{author}{Litinskaya, M.}, \bibinfo{author}{Reineker, P.} \&
  \bibinfo{author}{Agranovich, V.~M.}
\newblock \bibinfo{title}{Exciton-polaritons in organic microcavities}.
\newblock \emph{\bibinfo{journal}{J. Lumin.}} \textbf{\bibinfo{volume}{119}},
  \bibinfo{pages}{277--282} (\bibinfo{year}{2006}) .

\bibitem{Torma2015}
\bibinfo{author}{T{\"{o}}rm{\"{a}}, P.} \& \bibinfo{author}{Barnes, W.~L.}
\newblock \bibinfo{title}{Strong coupling between surface plasmon polaritons
  and emitters: a review}.
\newblock \emph{\bibinfo{journal}{Rep. Prog. Phys.}}
  \textbf{\bibinfo{volume}{78}}, \bibinfo{pages}{013901} (\bibinfo{year}{2015})
  .

\bibitem{Ribeiro2018}
\bibinfo{author}{Ribeiro, R.~F.}, \bibinfo{author}{Martin\'{e}z-Martin\'{e}z,
  L.~A.}, \bibinfo{author}{Du, M.}, \bibinfo{author}{Campos-Gonzalez-Angule,
  J.} \& \bibinfo{author}{Yuen-Zhou, J.}
\newblock \bibinfo{title}{Polariton chemistry: controlling molecular dynamics
  with optical cavities}.
\newblock \emph{\bibinfo{journal}{Chem. Sci.}} \textbf{\bibinfo{volume}{9}},
  \bibinfo{pages}{6325--6339} (\bibinfo{year}{2018}) .

\bibitem{Hertzog2019}
\bibinfo{author}{Hertzog, M.}, \bibinfo{author}{Wang, M.},
  \bibinfo{author}{Mony, J.} \& \bibinfo{author}{B{\"o}rjesson, K.}
\newblock \bibinfo{title}{Strong light--matter interactions: a new direction
  within chemistry}.
\newblock \emph{\bibinfo{journal}{Chem. Soc. Rev.}}
  \textbf{\bibinfo{volume}{48}}, \bibinfo{pages}{937--961}
  (\bibinfo{year}{2019}) .

\bibitem{Garcia-Vidal2021}
\bibinfo{author}{Garcia-Vidal, F.~J.}, \bibinfo{author}{Ciuti, C.} \&
  \bibinfo{author}{Ebbessen, T.~W.}
\newblock \bibinfo{title}{Manipulating matter by strong coupling to vacuum
  fields}.
\newblock \emph{\bibinfo{journal}{Science}} \textbf{\bibinfo{volume}{373}},
  \bibinfo{pages}{336} (\bibinfo{year}{2021}) .

\bibitem{Fregoni2022}
\bibinfo{author}{Fregoni, J.}, \bibinfo{author}{Garcia-Vidal, F.~J.} \&
  \bibinfo{author}{Feist, J.}
\newblock \bibinfo{title}{Theoretical challenges in polaritonic chemistry}.
\newblock \emph{\bibinfo{journal}{ACS Photonics}}
  \textbf{\bibinfo{volume}{9}}~(4), \bibinfo{pages}{1096--1107}
  (\bibinfo{year}{2022}) .

\bibitem{Ruggenthaler2022}
\bibinfo{author}{Ruggenthaler, M.}, \bibinfo{author}{Sidler, D.} \&
  \bibinfo{author}{Rubio, A.}
\newblock \bibinfo{title}{Understanding polaritonic chemistry from ab initio
  quantum electrodynamics}.
\newblock \emph{\bibinfo{journal}{ArXiv}}
  \textbf{\bibinfo{volume}{2211.04241v1}} (\bibinfo{year}{2022}) .

\bibitem{Rider2022}
\bibinfo{author}{Rider, M.~S.} \& \bibinfo{author}{Barnes, W.~L.}
\newblock \bibinfo{title}{Something from nothing: linking molecules with
  virtual light}.
\newblock \emph{\bibinfo{journal}{Contemporary Physics}}
  \textbf{\bibinfo{volume}{62}}~(4), \bibinfo{pages}{217--232}
  (\bibinfo{year}{2022}) .

\bibitem{Agranovich2003}
\bibinfo{author}{Agranovich, V.~M.}, \bibinfo{author}{Litinskaia, M.} \&
  \bibinfo{author}{Lidzey, D.~G.}
\newblock \bibinfo{title}{Cavity polaritons in microcavities containing
  disordered organic semiconductors}.
\newblock \emph{\bibinfo{journal}{Phys. Rev. B}} \textbf{\bibinfo{volume}{67}},
  \bibinfo{pages}{085311} (\bibinfo{year}{2003}) .

\bibitem{Litinskaya2004}
\bibinfo{author}{Litinskaya, M.}, \bibinfo{author}{Reineker, P.} \&
  \bibinfo{author}{Agranovich, V.~M.}
\newblock \bibinfo{title}{Fast polariton relaxation in strongly coupled organic
  microcavities}.
\newblock \emph{\bibinfo{journal}{J. Lumin.}} \textbf{\bibinfo{volume}{110}},
  \bibinfo{pages}{364--372} (\bibinfo{year}{2004}) .

\bibitem{delPino2015}
\bibinfo{author}{{del Pino}, J.}, \bibinfo{author}{Feist, J.} \&
  \bibinfo{author}{{Garcia-Vidal}, F.~J.}
\newblock \bibinfo{title}{{Quantum Theory of Collective Strong Coupling of
  Molecular Vibrations with a Microcavity Mode}}.
\newblock \emph{\bibinfo{journal}{New J. Phys.}}
  \textbf{\bibinfo{volume}{17}}~(5), \bibinfo{pages}{053040}
  (\bibinfo{year}{2015}) .

\bibitem{Coles2014}
\bibinfo{author}{Coles, D.~M.} \emph{et~al.}
\newblock \bibinfo{title}{Polariton-mediated energy transfer between organic
  dyes in a strongly coupled optical microcavity}.
\newblock \emph{\bibinfo{journal}{Nature Materials}}
  \textbf{\bibinfo{volume}{13}}, \bibinfo{pages}{712--719}
  (\bibinfo{year}{2014}) .

\bibitem{Zhong2016}
\bibinfo{author}{Zhong, X.} \emph{et~al.}
\newblock \bibinfo{title}{Non-radiative energy transfer mediated by hybrid
  light-matter states}.
\newblock \emph{\bibinfo{journal}{Angew. Chem. Int. Ed.}}
  \textbf{\bibinfo{volume}{55}}, \bibinfo{pages}{6202--6206}
  (\bibinfo{year}{2016}) .

\bibitem{Zhong2017}
\bibinfo{author}{Zhong, X.} \emph{et~al.}
\newblock \bibinfo{title}{Energy transfer between spatially separated entangled
  molecules}.
\newblock \emph{\bibinfo{journal}{Angew. Chem. Int. Ed.}}
  \textbf{\bibinfo{volume}{56}}, \bibinfo{pages}{9034--9038}
  (\bibinfo{year}{2017}) .

\bibitem{Georgiou2018}
\bibinfo{author}{Georgiou, K.} \emph{et~al.}
\newblock \bibinfo{title}{Control over energy transfer between fluorescent
  bodipy dyes in a strongly coupled microcavity}.
\newblock \emph{\bibinfo{journal}{ACS Photonics}} \textbf{\bibinfo{volume}{5}},
  \bibinfo{pages}{258--266} (\bibinfo{year}{2018}) .

\bibitem{Groenhof2018}
\bibinfo{author}{Groenhof, G.} \& \bibinfo{author}{Toppari, J.~J.}
\newblock \bibinfo{title}{Coherent light harvesting through strong coupling to
  confined light}.
\newblock \emph{\bibinfo{journal}{J. Phys. Chem. Lett.}}
  \textbf{\bibinfo{volume}{9}}, \bibinfo{pages}{4848--4851}
  (\bibinfo{year}{2018}) .

\bibitem{Xiang2020}
\bibinfo{author}{Xiang, B.} \emph{et~al.}
\newblock \bibinfo{title}{Intermolecular vibrational energy transfer enabled by
  microcavity strong light–matter coupling}.
\newblock \emph{\bibinfo{journal}{Science}} \textbf{\bibinfo{volume}{368}},
  \bibinfo{pages}{665--667} (\bibinfo{year}{2020}) .

\bibitem{Georgiou2021}
\bibinfo{author}{Georgiou, K.}, \bibinfo{author}{Jayaprakash, R.},
  \bibinfo{author}{Othonos, A.} \& \bibinfo{author}{Lidzey, D.~G.}
\newblock \bibinfo{title}{Ultralong-range polariton-assisted energy transfer in
  organic microcavities}.
\newblock \emph{\bibinfo{journal}{Angew. Chem. Int. Ed.}}
  \textbf{\bibinfo{volume}{60}}, \bibinfo{pages}{16661--16667}
  (\bibinfo{year}{2021}) .

\bibitem{Son2022}
\bibinfo{author}{Son, M.} \emph{et~al.}
\newblock \bibinfo{title}{Energy cascades in donor-acceptor exciton- polaritons
  observed by ultrafast two- dimensional white-light spectroscopy}.
\newblock \emph{\bibinfo{journal}{Nat. Comm.}} \textbf{\bibinfo{volume}{13}},
  \bibinfo{pages}{7305} (\bibinfo{year}{2022}) .

\bibitem{Engelhardt2022}
\bibinfo{author}{Engelhardt, G.} \& \bibinfo{author}{Cao, J.}
\newblock \bibinfo{title}{Unusual dynamical properties of disordered polaritons
  in microcavities}.
\newblock \emph{\bibinfo{journal}{Phys. Rev. B}}
  \textbf{\bibinfo{volume}{105}}, \bibinfo{pages}{064205}
  (\bibinfo{year}{2022}) .

\bibitem{George2023}
\bibinfo{author}{Bhatt, P.}, \bibinfo{author}{Dutta, J.},
  \bibinfo{author}{Kaur, K.} \& \bibinfo{author}{George, J.}
\newblock \bibinfo{title}{Long-range energy transfer in strongly coupled
  donor–acceptor phototransistors}.
\newblock \emph{\bibinfo{journal}{Nano Lett.}} \textbf{\bibinfo{volume}{XXXX}},
  \bibinfo{pages}{XXX--XXX} (\bibinfo{year}{2023}) .

\bibitem{Freixanet2000}
\bibinfo{author}{Freixanet, T.}, \bibinfo{author}{Sermage, B.},
  \bibinfo{author}{Tiberj, A.} \& \bibinfo{author}{Planel, R.}
\newblock \bibinfo{title}{In-plane propagation of excitonic cavity polaritons}.
\newblock \emph{\bibinfo{journal}{Phys. Rev. B}} \textbf{\bibinfo{volume}{61}},
  \bibinfo{pages}{7233} (\bibinfo{year}{2000}) .

\bibitem{Agranovich2007}
\bibinfo{author}{Agranovich, V.~M.} \& \bibinfo{author}{Gartstein, Y.~N.}
\newblock \bibinfo{title}{Nature and dynamics of low-energy exciton polaritons
  in semiconductor microcavities}.
\newblock \emph{\bibinfo{journal}{Phys. Rev. B}} \textbf{\bibinfo{volume}{75}},
  \bibinfo{pages}{075302} (\bibinfo{year}{2007}) .

\bibitem{Litinskaya2008}
\bibinfo{author}{Litinskaya, M.}
\newblock \bibinfo{title}{Propagation and localization of polaritons in
  disordered organic microcavities}.
\newblock \emph{\bibinfo{journal}{Phys. Lett. A}}
  \textbf{\bibinfo{volume}{372}}, \bibinfo{pages}{3898--3903}
  (\bibinfo{year}{2008}) .

\bibitem{Michetti2008b}
\bibinfo{author}{Michetti, P.} \& \bibinfo{author}{Rocca, G. C.~L.}
\newblock \bibinfo{title}{Polariton dynamics in disordered microcavities}.
\newblock \emph{\bibinfo{journal}{Physica E}} \textbf{\bibinfo{volume}{40}},
  \bibinfo{pages}{1926--1929} (\bibinfo{year}{2008}) .

\bibitem{Lerario2017}
\bibinfo{author}{Lerario, G.} \emph{et~al.}
\newblock \bibinfo{title}{High-speed flow of interacting organic polaritons}.
\newblock \emph{\bibinfo{journal}{Light Sci. Appl.}}
  \textbf{\bibinfo{volume}{6}}, \bibinfo{pages}{e16212} (\bibinfo{year}{2017})
  .

\bibitem{Myers2018}
\bibinfo{author}{Myers, D.~M.}, \bibinfo{author}{Mukherjee, S.},
  \bibinfo{author}{Beaumariage, J.} \& \bibinfo{author}{Snoke, D.~W.}
\newblock \bibinfo{title}{Polariton-enhanced exciton transport}.
\newblock \emph{\bibinfo{journal}{Phys. Rev. B}} \textbf{\bibinfo{volume}{98}},
  \bibinfo{pages}{235302} (\bibinfo{year}{2018}) .

\bibitem{Rozenman2018}
\bibinfo{author}{Rozenman, G.~G.}, \bibinfo{author}{Akulov, K.},
  \bibinfo{author}{Golombek, A.} \& \bibinfo{author}{Schwartz, T.}
\newblock \bibinfo{title}{Long-range transport of organic exciton-polaritons
  revealed by ultrafast microscopy}.
\newblock \emph{\bibinfo{journal}{ACS Photonics}} \textbf{\bibinfo{volume}{5}},
  \bibinfo{pages}{105--110} (\bibinfo{year}{2018}) .

\bibitem{Zakharko2018}
\bibinfo{author}{Zakharko, Y.} \emph{et~al.}
\newblock \bibinfo{title}{Radiative pumping and propagation of plexcitons in
  diffractive plasmonic crystals}.
\newblock \emph{\bibinfo{journal}{Nano Lett.}} \textbf{\bibinfo{volume}{18}},
  \bibinfo{pages}{4927--4933} (\bibinfo{year}{2018}) .

\bibitem{Forrest2020}
\bibinfo{author}{Hou, S.} \emph{et~al.}
\newblock \bibinfo{title}{Ultralong-range energy transport in a disordered
  organic semiconductor at room temperature via coherent exciton-polariton
  propagation}.
\newblock \emph{\bibinfo{journal}{Adv. Mater.}}
  \textbf{\bibinfo{volume}{32(28)}}, \bibinfo{pages}{2002127}
  (\bibinfo{year}{2020}) .

\bibitem{Pandya2021}
\bibinfo{author}{Pandya, R.} \emph{et~al.}
\newblock \bibinfo{title}{Microcavity-like exciton-polaritons can be the
  primary photoexcitation in bare organic semiconductors}.
\newblock \emph{\bibinfo{journal}{Nat. Commun.}} \textbf{\bibinfo{volume}{12}},
  \bibinfo{pages}{6519} (\bibinfo{year}{2021}) .

\bibitem{Ostrovskaya2021}
\bibinfo{author}{Wurdack, M.} \emph{et~al.}
\newblock \bibinfo{title}{Motional narrowing, ballistic transport, and trapping
  of room-temperature exciton polaritons in an atomically-thin semiconductor}.
\newblock \emph{\bibinfo{journal}{Nat. Comm.}} \textbf{\bibinfo{volume}{12}},
  \bibinfo{pages}{5366} (\bibinfo{year}{2021}) .

\bibitem{Ferreira2022}
\bibinfo{author}{Ferreira, B.}, \bibinfo{author}{Rosati, R.} \&
  \bibinfo{author}{Malic, E.}
\newblock \bibinfo{title}{Microscopic modeling of exciton-polariton diffusion
  coefficients in atomically thin semiconductors}.
\newblock \emph{\bibinfo{journal}{Phys. Rev. Mat.}}
  \textbf{\bibinfo{volume}{6}}, \bibinfo{pages}{034008} (\bibinfo{year}{2022})
  .

\bibitem{Berghuis2022}
\bibinfo{author}{Berghuis, M.~A.} \emph{et~al.}
\newblock \bibinfo{title}{Controlling exciton propagation in organic crystals
  through strong coupling to plasmonic nanoparticle arrays}.
\newblock \emph{\bibinfo{journal}{ACS Photonics}} \textbf{\bibinfo{volume}{9}},
  \bibinfo{pages}{123} (\bibinfo{year}{2022}) .

\bibitem{Pandya2022}
\bibinfo{author}{Pandya, R.} \emph{et~al.}
\newblock \bibinfo{title}{Tuning the coherent propagation of organic
  exciton-polaritons through dark state delocalization}.
\newblock \emph{\bibinfo{journal}{Adv. Sci.}} \textbf{\bibinfo{volume}{9}},
  \bibinfo{pages}{2105569} (\bibinfo{year}{2022}) .

\bibitem{Xu2022}
\bibinfo{author}{Xu, D.} \emph{et~al.}
\newblock \bibinfo{title}{Ultrafast imaging of coherent polariton propagation
  and interactions}.
\newblock \emph{\bibinfo{journal}{ArXiv}} \textbf{\bibinfo{volume}{2205.01176}}
  (\bibinfo{year}{2022}) .

\bibitem{Ribeiro2022}
\bibinfo{author}{Ribeiro, R.~F.}
\newblock \bibinfo{title}{Multimode polariton effects on molecular energy
  transport and spectral fluctuations}.
\newblock \emph{\bibinfo{journal}{Comm. Chem.}} \textbf{\bibinfo{volume}{5}},
  \bibinfo{pages}{48} (\bibinfo{year}{2022}) .

\bibitem{Allard2022}
\bibinfo{author}{Allard, T.~F.} \& \bibinfo{author}{Weick, G.}
\newblock \bibinfo{title}{Disorder-enhanced transport in a chain of lossy
  dipoles strongly coupled to cavity photons}.
\newblock \emph{\bibinfo{journal}{Phys. Rev. B}}
  \textbf{\bibinfo{volume}{106}}, \bibinfo{pages}{245424}
  (\bibinfo{year}{2022}) .

\bibitem{Balasubrahmaniyam2023}
\bibinfo{author}{Balasubrahmaniyam, M.}, \bibinfo{author}{Simkovich, A.},
  \bibinfo{author}{Golombek, A.}, \bibinfo{author}{Ankonina, G.} \&
  \bibinfo{author}{Schwartz, T.}
\newblock \bibinfo{title}{Unveiling the mixed nature of polaritonic transport:
  From enhanced diffusion to ballistic motion approaching the speed of light}.
\newblock \emph{\bibinfo{journal}{Nat. Mater.}} \textbf{\bibinfo{volume}{22}},
  \bibinfo{pages}{pages 338–344} (\bibinfo{year}{2023}) .

\bibitem{Aroeira2023}
\bibinfo{author}{Aroeira, G. J.~R.}, \bibinfo{author}{Kairys, K.} \&
  \bibinfo{author}{Ribeiro, R.~F.}
\newblock \bibinfo{title}{Theoretical analysis of exciton wave packet dynamics
  in polaritonic wires}.
\newblock \emph{\bibinfo{journal}{J. Phys. Chem. Lett.}}
  \textbf{\bibinfo{volume}{14}}, \bibinfo{pages}{5681--5691}
  (\bibinfo{year}{2023}) .

\bibitem{Engelhardt2023}
\bibinfo{author}{Engelhardt, G.} \& \bibinfo{author}{Cao, J.}
\newblock \bibinfo{title}{Polariton localization and dispersion properties of
  disordered quantum emitters in multimode microcavities}.
\newblock \emph{\bibinfo{journal}{Phys. Rev. Lett.}}
  \textbf{\bibinfo{volume}{130}}, \bibinfo{pages}{213602}
  (\bibinfo{year}{2023}) .

\bibitem{Jin2023}
\bibinfo{author}{Jin, L.} \emph{et~al.}
\newblock \bibinfo{title}{Enhanced two-dimensional exciton propagation via
  strong light–matter coupling with surface lattice plasmons}.
\newblock \emph{\bibinfo{journal}{ACS Photonics}}
  \textbf{\bibinfo{volume}{XXXX}}, \bibinfo{pages}{XXX--XXX}
  (\bibinfo{year}{2023}) .

\bibitem{Daehne1998}
\bibinfo{author}{D\"{a}hne, L.}, \bibinfo{author}{Biller, E.} \&
  \bibinfo{author}{Baumg\"{a}rtel, H.}
\newblock \bibinfo{title}{Polariton-induced color tuning of thin dye layers}.
\newblock \emph{\bibinfo{journal}{Angew. Chem. Int. Ed.}}
  \textbf{\bibinfo{volume}{37}}, \bibinfo{pages}{646--649}
  (\bibinfo{year}{1998}) .

\bibitem{Luk2017}
\bibinfo{author}{Luk, H.~L.}, \bibinfo{author}{Feist, J.},
  \bibinfo{author}{Toppari, J.~J.} \& \bibinfo{author}{Groenhof, G.}
\newblock \bibinfo{title}{Multiscale molecular dynamics simulations of
  polaritonic chemistry}.
\newblock \emph{\bibinfo{journal}{J. Chem. Theory Comput.}}
  \textbf{\bibinfo{volume}{13}}, \bibinfo{pages}{4324--4335}
  (\bibinfo{year}{2017}) .

\bibitem{Tichauer2021}
\bibinfo{author}{Tichauer, R.~H.}, \bibinfo{author}{Feist, J.} \&
  \bibinfo{author}{Groenhof, G.}
\newblock \bibinfo{title}{Multi-scale dynamics simulations of molecular
  polaritons: the effect of multiple cavity modes on polariton relaxation}.
\newblock \emph{\bibinfo{journal}{J. Chem. Phys.}}
  \textbf{\bibinfo{volume}{154}}, \bibinfo{pages}{104112}
  (\bibinfo{year}{2021}) .

\bibitem{Michetti2005}
\bibinfo{author}{Michetti, P.} \& \bibinfo{author}{Rocca, G. C.~L.}
\newblock \bibinfo{title}{Polariton states in disordered organic
  microcavities}.
\newblock \emph{\bibinfo{journal}{Phys. Rev. B.}}
  \textbf{\bibinfo{volume}{71}}, \bibinfo{pages}{115320} (\bibinfo{year}{2005})
  .

\bibitem{Warshel1976b}
\bibinfo{author}{Warshel, A.} \& \bibinfo{author}{Levitt, M.}
\newblock \bibinfo{title}{Theoretical studies of enzymatic reactions:
  Dielectric, electrostatic and steric stabilization of carbonium ion in the
  reaction of lysozyme}.
\newblock \emph{\bibinfo{journal}{J. Mol. Biol.}}
  \textbf{\bibinfo{volume}{103}}, \bibinfo{pages}{227--249}
  (\bibinfo{year}{1976}) .

\bibitem{Ditchfield1971}
\bibinfo{author}{Ditchfield, R.}, \bibinfo{author}{Hehre, W.~J.} \&
  \bibinfo{author}{Pople, J.~A.}
\newblock \bibinfo{title}{Self-consistent molecular-orbital methods. ix. an
  extended gaussian-type basis for molecular-orbital studies of organic
  molecules}.
\newblock \emph{\bibinfo{journal}{J. Chem. Phys.}}
  \textbf{\bibinfo{volume}{54}}, \bibinfo{pages}{724--728}
  (\bibinfo{year}{1971}) .

\bibitem{Duan2003}
\bibinfo{author}{Duan, Y.} \emph{et~al.}
\newblock \bibinfo{title}{{A point-charge force field for molecular mechanics
  simulations of proteins based on condensed-phase quantum mechanical
  calculations}}.
\newblock \emph{\bibinfo{journal}{J. Comput. Chem.}}
  \textbf{\bibinfo{volume}{{24}}}~({16}), \bibinfo{pages}{{1999--2012}}
  (\bibinfo{year}{{2003}}) .

\bibitem{Ehrenfest1927}
\bibinfo{author}{Ehrenfest, P.}
\newblock \bibinfo{title}{Bemerkung über die angen\"{a}herte g\"{u}ltigkeit
  der klassischen mechanik innerhalb der quantenmechanik}.
\newblock \emph{\bibinfo{journal}{Z. Phys.}} \textbf{\bibinfo{volume}{45}},
  \bibinfo{pages}{445--457} (\bibinfo{year}{1927}) .

\bibitem{Tavis1969}
\bibinfo{author}{Tavis, M.} \& \bibinfo{author}{Cummings, F.~W.}
\newblock \bibinfo{title}{Approximate solutions for an n-molecule
  radiation-field hamiltonian}.
\newblock \emph{\bibinfo{journal}{Phys. Rev.}} \textbf{\bibinfo{volume}{188}},
  \bibinfo{pages}{692--695} (\bibinfo{year}{1969}) .

\bibitem{Granucci2001}
\bibinfo{author}{Granucci, G.}, \bibinfo{author}{Persico, M.} \&
  \bibinfo{author}{Toniolo, A.}
\newblock \bibinfo{title}{Direct semiclassical simulation of photochemical
  processes with semiempirical wave functions}.
\newblock \emph{\bibinfo{journal}{J. Chem. Phys.}}
  \textbf{\bibinfo{volume}{114}}, \bibinfo{pages}{10608--10615}
  (\bibinfo{year}{2001}) .

\bibitem{Kasha1950}
\bibinfo{author}{Kasha, M.}
\newblock \bibinfo{title}{Characterization of electronic transitions in complex
  molecules}.
\newblock \emph{\bibinfo{journal}{Disc. Faraday Soc.}}
  \textbf{\bibinfo{volume}{9}}, \bibinfo{pages}{14--19} (\bibinfo{year}{1950})
  .

\bibitem{Schwartz2013}
\bibinfo{author}{Schwartz, T.} \emph{et~al.}
\newblock \bibinfo{title}{Polariton dynamics under strong light-molecule
  coupling}.
\newblock \emph{\bibinfo{journal}{ChemPhysChem}} \textbf{\bibinfo{volume}{14}},
  \bibinfo{pages}{125--131} (\bibinfo{year}{2013}) .

\bibitem{George2015}
\bibinfo{author}{George, J.} \emph{et~al.}
\newblock \bibinfo{title}{Ultra-strong coupling of molecular materials:
  spectroscopy and dynamics}.
\newblock \emph{\bibinfo{journal}{Faraday Discuss.}}
  \textbf{\bibinfo{volume}{178}}, \bibinfo{pages}{281--294}
  (\bibinfo{year}{2015}) .

\bibitem{Wu2022}
\bibinfo{author}{Wu, F.} \emph{et~al.}
\newblock \bibinfo{title}{Optical cavity-mediated exciton dynamics in
  photosynthetic light harvesting 2 complexes}.
\newblock \emph{\bibinfo{journal}{Nat. Comm.}} \textbf{\bibinfo{volume}{13}},
  \bibinfo{pages}{6864} (\bibinfo{year}{2022}) .

\bibitem{Boggio-Pasqua2012}
\bibinfo{author}{Boggio-Pasqua, M.}, \bibinfo{author}{Burmeister, C.~F.},
  \bibinfo{author}{Robb, M.~A.} \& \bibinfo{author}{Groenhof, G.}
\newblock \bibinfo{title}{Photochemical reactions in biological systems:
  probing the effect of the environment by means of hybrid quantum
  chemistry/molecular mechanics simulations}.
\newblock \emph{\bibinfo{journal}{Phys. Chem. Chem. Phys.}}
  \textbf{\bibinfo{volume}{14}}, \bibinfo{pages}{7912--7928}
  (\bibinfo{year}{2012}) .

\bibitem{Georgiou2018b}
\bibinfo{author}{Georgiou, K.} \emph{et~al.}
\newblock \bibinfo{title}{{Generation of Anti-Stokes Fluorescence in a Strongly
  Coupled Organic Semiconductor Microcavity}}.
\newblock \emph{\bibinfo{journal}{ACS Photonics}} \textbf{\bibinfo{volume}{5}},
  \bibinfo{pages}{4343--4351} (\bibinfo{year}{2018}) .

\bibitem{Groenhof2019}
\bibinfo{author}{Groenhof, G.}, \bibinfo{author}{Climent, C.},
  \bibinfo{author}{Feist, J.}, \bibinfo{author}{Morozov, D.} \&
  \bibinfo{author}{Toppari, J.~J.}
\newblock \bibinfo{title}{Tracking polariton relaxation with multiscale
  molecular dynamics simulations}.
\newblock \emph{\bibinfo{journal}{J. Chem. Phys. Lett.}}
  \textbf{\bibinfo{volume}{10}}, \bibinfo{pages}{5476--5483}
  (\bibinfo{year}{2019}) .

\bibitem{Takahashi2020}
\bibinfo{author}{Takahashi, S.} \& \bibinfo{author}{Watanabe, K.}
\newblock \bibinfo{title}{Decoupling from a thermal bath via molecular
  polariton formation}.
\newblock \emph{\bibinfo{journal}{J. Phys. Chem. Lett.}}
  \textbf{\bibinfo{volume}{11}}, \bibinfo{pages}{1349--1356}
  (\bibinfo{year}{2020}) .

\bibitem{Tichauer2022}
\bibinfo{author}{Tichauer, R.~H.}, \bibinfo{author}{Morozov, D.},
  \bibinfo{author}{Sokolovskii, I.}, \bibinfo{author}{Toppari, J.~J.} \&
  \bibinfo{author}{Groenhof, G.}
\newblock \bibinfo{title}{Identifying vibrations that control non-adiabatic
  relaxation of polaritons in strongly coupled molecule-cavity systems}.
\newblock \emph{\bibinfo{journal}{J. Phys. Chem. Lett.}}
  \textbf{\bibinfo{volume}{13}}, \bibinfo{pages}{6259--6267}
  (\bibinfo{year}{2022}) .

\bibitem{Houdre1996}
\bibinfo{author}{Houdr{\'{e}}, R.}, \bibinfo{author}{Stanley, R.~P.} \&
  \bibinfo{author}{Ilegems, M.}
\newblock \bibinfo{title}{Vacuum-field rabi splitting in the presence of
  inhomogeneous broadening: Resolution of a homogeneous linewidth in an
  inhomogeneously broadened system}.
\newblock \emph{\bibinfo{journal}{Phys. Rev. A}} \textbf{\bibinfo{volume}{53}},
  \bibinfo{pages}{2711--2715} (\bibinfo{year}{1996}) .

\bibitem{Eizner2019}
\bibinfo{author}{Eizner, E.}, \bibinfo{author}{Mart\'{i}nez-Mart\'{i}nez,
  L.~A.}, \bibinfo{author}{{Yuen-Shou}, J.} \& \bibinfo{author}{K\'{e}na-Cohen,
  S.}
\newblock \bibinfo{title}{Inverting singlet and triplet excited states using
  strong light-matter coupling}.
\newblock \emph{\bibinfo{journal}{Arxiv}}
  \textbf{\bibinfo{volume}{1903.09251v1}}, \bibinfo{pages}{1--29}
  (\bibinfo{year}{2016}) .

\bibitem{Martinez2019}
\bibinfo{author}{Mart\'{i}nez-Mart\'{i}nez, L.~A.}, \bibinfo{author}{Eizner,
  E.}, \bibinfo{author}{K\'{e}na-Cohen, S.} \& \bibinfo{author}{Yuemn-Zhou, K.}
\newblock \bibinfo{title}{Triplet harvesting in the polaritonic regime: A
  variational polaron approach}.
\newblock \emph{\bibinfo{journal}{J. Chem. Phys.}}
  \textbf{\bibinfo{volume}{151}}, \bibinfo{pages}{054106}
  (\bibinfo{year}{2019}) .

\bibitem{DeJong2015}
\bibinfo{author}{de~Jong, M.}, \bibinfo{author}{Seijo, L.},
  \bibinfo{author}{Meijerinka, A.} \& \bibinfo{author}{Rabouw, F.~T.}
\newblock \bibinfo{title}{Resolving the ambiguity in the relation between
  stokes shift and huang–rhys parameter}.
\newblock \emph{\bibinfo{journal}{Phys. Chem. Chem. Phys}}
  \textbf{\bibinfo{volume}{17}}, \bibinfo{pages}{16959--16969}
  (\bibinfo{year}{2015}) .

\bibitem{Grant2016}
\bibinfo{author}{Grant, R.~T.} \emph{et~al.}
\newblock \bibinfo{title}{Efficient radiative pumping of polaritons in a
  strongly coupled microcavity by a fluorescent molecular dye}.
\newblock \emph{\bibinfo{journal}{Adv. Optical Mater.}}
  \textbf{\bibinfo{volume}{4}}, \bibinfo{pages}{1615--1623}
  (\bibinfo{year}{2016}) .

\bibitem{Baieva2017}
\bibinfo{author}{Baieva, S.}, \bibinfo{author}{Hakamaa, O.},
  \bibinfo{author}{Groenhof, G.}, \bibinfo{author}{Heikkil{\"{a}}, T.~T.} \&
  \bibinfo{author}{Toppari, J.~J.}
\newblock \bibinfo{title}{Dynamics of strongly coupled modes between surface
  plasmon polaritons and photoactive molecules: the effect of the stokes
  shift}.
\newblock \emph{\bibinfo{journal}{ACS Photonics}} \textbf{\bibinfo{volume}{4}},
  \bibinfo{pages}{28--37} (\bibinfo{year}{2017}) .

\bibitem{Luettgens2021}
\bibinfo{author}{L\"{u}ttgens, J.~M.}, \bibinfo{author}{Berger, F.~J.} \&
  \bibinfo{author}{Zaumseil, J.}
\newblock \bibinfo{title}{Population of exciton-polaritons via luminescent
  sp$^3$ defects in single-walled carbon nanotubes}.
\newblock \emph{\bibinfo{journal}{ACS Photonics}} \textbf{\bibinfo{volume}{8}},
  \bibinfo{pages}{182--193} (\bibinfo{year}{2021}) .

\bibitem{Hulkko2021}
\bibinfo{author}{Hulkko, E.} \emph{et~al.}
\newblock \bibinfo{title}{Effect of molecular stokes shift on polariton
  dynamics}.
\newblock \emph{\bibinfo{journal}{J. Chem. Phys.}}
  \textbf{\bibinfo{volume}{154}}, \bibinfo{pages}{154303}
  (\bibinfo{year}{2021}) .

\bibitem{Tichauer2023}
\bibinfo{author}{Tichauer, R.~H.}, \bibinfo{author}{Sokolovskii, I.} \&
  \bibinfo{author}{Groenhof, G.}
\newblock \bibinfo{title}{Tuning coherent propagation of organic
  exciton-polaritons through the cavity q-factor}.
\newblock \emph{\bibinfo{journal}{arXiv}} \textbf{\bibinfo{volume}{2304.13123}}
  (\bibinfo{year}{2023}) .

\end{thebibliography}


\providecommand{\latin}[1]{#1}
\makeatletter
\providecommand{\doi}
  {\begingroup\let\do\@makeother\dospecials
  \catcode`\{=1 \catcode`\}=2 \doi@aux}
\providecommand{\doi@aux}[1]{\endgroup\texttt{#1}}
\makeatother
\providecommand*\mcitethebibliography{\thebibliography}
\csname @ifundefined\endcsname{endmcitethebibliography}
  {\let\endmcitethebibliography\endthebibliography}{}
\begin{mcitethebibliography}{48}
\providecommand*\natexlab[1]{#1}
\providecommand*\mciteSetBstSublistMode[1]{}
\providecommand*\mciteSetBstMaxWidthForm[2]{}
\providecommand*\mciteBstWouldAddEndPuncttrue
  {\def\EndOfBibitem{\unskip.}}
\providecommand*\mciteBstWouldAddEndPunctfalse
  {\let\EndOfBibitem\relax}
\providecommand*\mciteSetBstMidEndSepPunct[3]{}
\providecommand*\mciteSetBstSublistLabelBeginEnd[3]{}
\providecommand*\EndOfBibitem{}
\mciteSetBstSublistMode{f}
\mciteSetBstMaxWidthForm{subitem}{(\alph{mcitesubitemcount})}
\mciteSetBstSublistLabelBeginEnd
  {\mcitemaxwidthsubitemform\space}
  {\relax}
  {\relax}

\bibitem[Forn-D\'{i}az \latin{et~al.}(2019)Forn-D\'{i}az, Lamata, Rico, Kono,
  and Solano]{Forn-Diaz2019}
Forn-D\'{i}az,~P.; Lamata,~L.; Rico,~E.; Kono,~J.; Solano,~E. Ultrastrong
  coupling regimes of light-matter interaction. \emph{Rev. Mod. Phys.}
  \textbf{2019}, \emph{91}, 025005\relax
\mciteBstWouldAddEndPuncttrue
\mciteSetBstMidEndSepPunct{\mcitedefaultmidpunct}
{\mcitedefaultendpunct}{\mcitedefaultseppunct}\relax
\EndOfBibitem
\bibitem[Jaynes and Cummings(1963)Jaynes, and Cummings]{Jaynes1963}
Jaynes,~E.~T.; Cummings,~F.~W. Comparison of quantum and semiclassical
  radiation theories with application to the beam maser. \emph{Proc. IEEE}
  \textbf{1963}, \emph{51}, 89--109\relax
\mciteBstWouldAddEndPuncttrue
\mciteSetBstMidEndSepPunct{\mcitedefaultmidpunct}
{\mcitedefaultendpunct}{\mcitedefaultseppunct}\relax
\EndOfBibitem
\bibitem[Tavis and Cummings(1969)Tavis, and Cummings]{Tavis1969}
Tavis,~M.; Cummings,~F.~W. Approximate solutions for an N-molecule
  radiation-field Hamiltonian. \emph{Phys. Rev.} \textbf{1969}, \emph{188},
  692--695\relax
\mciteBstWouldAddEndPuncttrue
\mciteSetBstMidEndSepPunct{\mcitedefaultmidpunct}
{\mcitedefaultendpunct}{\mcitedefaultseppunct}\relax
\EndOfBibitem
\bibitem[Michetti and Rocca(2005)Michetti, and Rocca]{Michetti2005}
Michetti,~P.; Rocca,~G. C.~L. Polariton states in disordered organic
  microcavities. \emph{Phys. Rev. B.} \textbf{2005}, \emph{71}, 115320\relax
\mciteBstWouldAddEndPuncttrue
\mciteSetBstMidEndSepPunct{\mcitedefaultmidpunct}
{\mcitedefaultendpunct}{\mcitedefaultseppunct}\relax
\EndOfBibitem
\bibitem[Tichauer \latin{et~al.}(2021)Tichauer, Feist, and
  Groenhof]{Tichauer2021}
Tichauer,~R.; Feist,~J.; Groenhof,~G. Multi-scale Dynamics Simulations of
  Molecular Polaritons: the Effect of Multiple Cavity Modes on Polariton
  Relaxation. \emph{J. Chem. Phys.} \textbf{2021}, \emph{154}, 104112\relax
\mciteBstWouldAddEndPuncttrue
\mciteSetBstMidEndSepPunct{\mcitedefaultmidpunct}
{\mcitedefaultendpunct}{\mcitedefaultseppunct}\relax
\EndOfBibitem
\bibitem[Warshel and Levitt(1976)Warshel, and Levitt]{Warshel1976b}
Warshel,~A.; Levitt,~M. Theoretical studies of enzymatic reactions: Dielectric,
  electrostatic and steric stabilization of carbonium ion in the reaction of
  lysozyme. \emph{J. Mol. Biol.} \textbf{1976}, \emph{103}, 227--249\relax
\mciteBstWouldAddEndPuncttrue
\mciteSetBstMidEndSepPunct{\mcitedefaultmidpunct}
{\mcitedefaultendpunct}{\mcitedefaultseppunct}\relax
\EndOfBibitem
\bibitem[Boggio-Pasqua \latin{et~al.}(2012)Boggio-Pasqua, Burmeister, Robb, and
  Groenhof]{Boggio-Pasqua2012}
Boggio-Pasqua,~M.; Burmeister,~C.~F.; Robb,~M.~A.; Groenhof,~G. Photochemical
  reactions in biological systems: probing the effect of the environment by
  means of hybrid quantum chemistry/molecular mechanics simulations.
  \emph{Phys. Chem. Chem. Phys.} \textbf{2012}, \emph{14}, 7912--7928\relax
\mciteBstWouldAddEndPuncttrue
\mciteSetBstMidEndSepPunct{\mcitedefaultmidpunct}
{\mcitedefaultendpunct}{\mcitedefaultseppunct}\relax
\EndOfBibitem
\bibitem[Agranovich and Gartstein(2007)Agranovich, and
  Gartstein]{Agranovich2007}
Agranovich,~V.; Gartstein,~Y. Nature and Dynamics of Low-Energy Exciton
  Polaritons in Semiconductor Microcavities. \emph{Phys. Rev. B} \textbf{2007},
  \emph{75}, 075302\relax
\mciteBstWouldAddEndPuncttrue
\mciteSetBstMidEndSepPunct{\mcitedefaultmidpunct}
{\mcitedefaultendpunct}{\mcitedefaultseppunct}\relax
\EndOfBibitem
\bibitem[Granucci \latin{et~al.}(2001)Granucci, Persico, and
  Toniolo]{Granucci2001}
Granucci,~G.; Persico,~M.; Toniolo,~A. Direct semiclassical simulation of
  photochemical processes with semiempirical wave functions. \emph{J. Chem.
  Phys.} \textbf{2001}, \emph{114}, 10608--10615\relax
\mciteBstWouldAddEndPuncttrue
\mciteSetBstMidEndSepPunct{\mcitedefaultmidpunct}
{\mcitedefaultendpunct}{\mcitedefaultseppunct}\relax
\EndOfBibitem
\bibitem[Yarkony(2012)]{Yarkony2012}
Yarkony,~D.~R. Nonadiabatic Quantum Chemistry—Past, Present, and Future.
  \emph{Chem. Rev.} \textbf{2012}, \emph{112}, 481--498\relax
\mciteBstWouldAddEndPuncttrue
\mciteSetBstMidEndSepPunct{\mcitedefaultmidpunct}
{\mcitedefaultendpunct}{\mcitedefaultseppunct}\relax
\EndOfBibitem
\bibitem[Worth and Cederbaum(2004)Worth, and Cederbaum]{Worth2004}
Worth,~G.~A.; Cederbaum,~L.~A. Beyond Born-Oppenheimer: Molecular Dynamics
  Through a Conical Intersection. \emph{Annu. Rev. Phys. Chem.} \textbf{2004},
  \emph{55}, 127--158\relax
\mciteBstWouldAddEndPuncttrue
\mciteSetBstMidEndSepPunct{\mcitedefaultmidpunct}
{\mcitedefaultendpunct}{\mcitedefaultseppunct}\relax
\EndOfBibitem
\bibitem[Azumi and Matsuzaki(1977)Azumi, and Matsuzaki]{Azumi1977}
Azumi,~T.; Matsuzaki,~K. What does the term "vibronic coupling" mean?
  \emph{Photochem. Photobiol.} \textbf{1977}, \emph{25}, 315--326\relax
\mciteBstWouldAddEndPuncttrue
\mciteSetBstMidEndSepPunct{\mcitedefaultmidpunct}
{\mcitedefaultendpunct}{\mcitedefaultseppunct}\relax
\EndOfBibitem
\bibitem[Crespo-Otero and Barbatti(2018)Crespo-Otero, and Barbatti]{Crespo2018}
Crespo-Otero,~R.; Barbatti,~M. Recent Advances and Perspectives on Nonadiabatic
  Mixed Quantum-Classical Dynamics. \emph{Chem. Rev.} \textbf{2018},
  \emph{118}, 7026--7068\relax
\mciteBstWouldAddEndPuncttrue
\mciteSetBstMidEndSepPunct{\mcitedefaultmidpunct}
{\mcitedefaultendpunct}{\mcitedefaultseppunct}\relax
\EndOfBibitem
\bibitem[Luk \latin{et~al.}(2017)Luk, Feist, Toppari, and Groenhof]{Luk2017}
Luk,~H.-L.; Feist,~J.; Toppari,~J.~J.; Groenhof,~G. Multiscale Molecular
  Dynamics Simulations of Polaritonic Chemistry. \emph{J. Chem. Theory Comput.}
  \textbf{2017}, \emph{13}, 4324--4335\relax
\mciteBstWouldAddEndPuncttrue
\mciteSetBstMidEndSepPunct{\mcitedefaultmidpunct}
{\mcitedefaultendpunct}{\mcitedefaultseppunct}\relax
\EndOfBibitem
\bibitem[Duan \latin{et~al.}({2003})Duan, Wu, Chowdhury, Lee, Xiong, Zhang,
  Yang, Cieplak, Luo, Lee, Caldwell, Wang, and Kollman]{Duan2003}
Duan,~Y.; Wu,~C.; Chowdhury,~S.; Lee,~M.~C.; Xiong,~G.~M.; Zhang,~W.; Yang,~R.;
  Cieplak,~P.; Luo,~R.; Lee,~T.; Caldwell,~J.; Wang,~J.~M.; Kollman,~P. {A
  point-charge force field for molecular mechanics simulations of proteins
  based on condensed-phase quantum mechanical calculations}. \emph{J. Comput.
  Chem.} \textbf{{2003}}, \emph{{24}}, {1999--2012}\relax
\mciteBstWouldAddEndPuncttrue
\mciteSetBstMidEndSepPunct{\mcitedefaultmidpunct}
{\mcitedefaultendpunct}{\mcitedefaultseppunct}\relax
\EndOfBibitem
\bibitem[Jorgensen \latin{et~al.}(1983)Jorgensen, Chandrasekhar, Madura, Impey,
  and Klein]{Jorgensen1983}
Jorgensen,~W.~L.; Chandrasekhar,~J.; Madura,~J.~D.; Impey,~R.~W.; Klein,~M.~L.
  Comparison of simple potential functions for simulatin liquid water. \emph{J.
  Chem. Phys.} \textbf{1983}, \emph{79}, 926--935\relax
\mciteBstWouldAddEndPuncttrue
\mciteSetBstMidEndSepPunct{\mcitedefaultmidpunct}
{\mcitedefaultendpunct}{\mcitedefaultseppunct}\relax
\EndOfBibitem
\bibitem[Berendsen \latin{et~al.}(1984)Berendsen, Postma, van Gunsteren, la,
  and Haak]{Berendsen1984}
Berendsen,~H.; Postma,~J.; van Gunsteren,~W.; la,~A.~D.; Haak,~J. Molecular
  dynamics with coupling to an external bath. \emph{J. Chem. Phys.}
  \textbf{1984}, \emph{81}, 3684--3690\relax
\mciteBstWouldAddEndPuncttrue
\mciteSetBstMidEndSepPunct{\mcitedefaultmidpunct}
{\mcitedefaultendpunct}{\mcitedefaultseppunct}\relax
\EndOfBibitem
\bibitem[Hess \latin{et~al.}(1997)Hess, Bekker, Berendsen, and
  Fraaije]{Hess1997}
Hess,~B.; Bekker,~H.; Berendsen,~H. J.~C.; Fraaije,~J. G. E.~M. {LINCS: A
  linear constraint solver for molecular simulations}. \emph{J. Comput. Chem.}
  \textbf{1997}, \emph{18}, 1463--1472\relax
\mciteBstWouldAddEndPuncttrue
\mciteSetBstMidEndSepPunct{\mcitedefaultmidpunct}
{\mcitedefaultendpunct}{\mcitedefaultseppunct}\relax
\EndOfBibitem
\bibitem[Miyamoto and Kollman(1992)Miyamoto, and Kollman]{Miyamoto1992}
Miyamoto,~S.; Kollman,~P.~A. {SETTLE}: An analytical version of the {SHAKE} and
  {RATTLE} algorithms for rigid water molecules. \emph{J. Comp. Chem.}
  \textbf{1992}, \emph{18}, 1463--1472\relax
\mciteBstWouldAddEndPuncttrue
\mciteSetBstMidEndSepPunct{\mcitedefaultmidpunct}
{\mcitedefaultendpunct}{\mcitedefaultseppunct}\relax
\EndOfBibitem
\bibitem[Essmann \latin{et~al.}(1995)Essmann, Perera, Berkowitz, Darden, Lee,
  and Pedersen]{Essmann1995}
Essmann,~U.; Perera,~L.; Berkowitz,~M.~L.; Darden,~T.; Lee,~H.; Pedersen,~L.~G.
  A smooth particle mesh {E}wald potential. \emph{J. Chem. Phys} \textbf{1995},
  \emph{103}, 8577--8592\relax
\mciteBstWouldAddEndPuncttrue
\mciteSetBstMidEndSepPunct{\mcitedefaultmidpunct}
{\mcitedefaultendpunct}{\mcitedefaultseppunct}\relax
\EndOfBibitem
\bibitem[Groenhof \latin{et~al.}(2019)Groenhof, Climent, Feist, Morozov, and
  Toppari]{Groenhof2019}
Groenhof,~G.; Climent,~C.; Feist,~J.; Morozov,~D.; Toppari,~J.~J. Tracking
  Polariton Relaxation with Multiscale Molecular Dynamics Simulations. \emph{J.
  Chem. Phys. Lett.} \textbf{2019}, \emph{10}, 5476--5483\relax
\mciteBstWouldAddEndPuncttrue
\mciteSetBstMidEndSepPunct{\mcitedefaultmidpunct}
{\mcitedefaultendpunct}{\mcitedefaultseppunct}\relax
\EndOfBibitem
\bibitem[Hess \latin{et~al.}(2008)Hess, Kutzner, van~der Spoel, and
  Lindahl]{Hess2008}
Hess,~B.; Kutzner,~C.; van~der Spoel,~D.; Lindahl,~E. GROMACS 4: Algorithms for
  Highly Efficient, Load-Balanced, and Scalable Molecular Simulation. \emph{J.
  Chem. Theory Comput.} \textbf{2008}, \emph{4}, 435--447\relax
\mciteBstWouldAddEndPuncttrue
\mciteSetBstMidEndSepPunct{\mcitedefaultmidpunct}
{\mcitedefaultendpunct}{\mcitedefaultseppunct}\relax
\EndOfBibitem
\bibitem[Ufimtsev and Mart\'{i}nez(2009)Ufimtsev, and
  Mart\'{i}nez]{Ufimtsev2009}
Ufimtsev,~I.; Mart\'{i}nez,~T.~J. Quantum Chemistry on Graphical Processing
  Units. 3. Analytical Energy Gradients and First Principles Molecular
  Dynamics. \emph{J. Chem. Theory Comput.} \textbf{2009}, \emph{5},
  2619--2628\relax
\mciteBstWouldAddEndPuncttrue
\mciteSetBstMidEndSepPunct{\mcitedefaultmidpunct}
{\mcitedefaultendpunct}{\mcitedefaultseppunct}\relax
\EndOfBibitem
\bibitem[Titov \latin{et~al.}(2013)Titov, Ufimtsev, Luehr, and
  Mart\'{i}nez]{Titov2013}
Titov,~A.; Ufimtsev,~I.; Luehr,~N.; Mart\'{i}nez,~T.~J. Generating Efficient
  Quantum Chemistry Codes for Novel Architectures. \emph{J. Chem. Theory
  Comput.} \textbf{2013}, \emph{9}, 213--221\relax
\mciteBstWouldAddEndPuncttrue
\mciteSetBstMidEndSepPunct{\mcitedefaultmidpunct}
{\mcitedefaultendpunct}{\mcitedefaultseppunct}\relax
\EndOfBibitem
\bibitem[Runge and Gross(1984)Runge, and Gross]{Runge1984}
Runge,~E.; Gross,~E. K.~U. Density-Functional Theory for Time-Dependent
  Systems. \emph{Phys. Rev. Lett} \textbf{1984}, \emph{52}, 997--1000\relax
\mciteBstWouldAddEndPuncttrue
\mciteSetBstMidEndSepPunct{\mcitedefaultmidpunct}
{\mcitedefaultendpunct}{\mcitedefaultseppunct}\relax
\EndOfBibitem
\bibitem[Becke(1993)]{Becke1993}
Becke,~A.~D. {A new mixing of Hartree-Fock and local density-functional
  theories}. \emph{J. Chem. Phys.} \textbf{1993}, \emph{98}, 1372\relax
\mciteBstWouldAddEndPuncttrue
\mciteSetBstMidEndSepPunct{\mcitedefaultmidpunct}
{\mcitedefaultendpunct}{\mcitedefaultseppunct}\relax
\EndOfBibitem
\bibitem[Yanai \latin{et~al.}(2004)Yanai, Tew, and Handy]{Yanai2004}
Yanai,~T.; Tew,~D.~P.; Handy,~N.~C. A new hybrid exchange-correlation
  functional using the Coulomb-attenuating method (CAM-B3LYP. \emph{Chem. Phys.
  Lett.} \textbf{2004}, \emph{393}, 51--57\relax
\mciteBstWouldAddEndPuncttrue
\mciteSetBstMidEndSepPunct{\mcitedefaultmidpunct}
{\mcitedefaultendpunct}{\mcitedefaultseppunct}\relax
\EndOfBibitem
\bibitem[Dunning(1970)]{Dunning1970}
Dunning,~T.~H. Basis Functions for Use in Molecular Calculations. I.
  Contractions of (9s5p) Atomic Basis Sets for the First-Row Atoms. \emph{J.
  Chem. Phys.} \textbf{1970}, \emph{53}, 2823--2833\relax
\mciteBstWouldAddEndPuncttrue
\mciteSetBstMidEndSepPunct{\mcitedefaultmidpunct}
{\mcitedefaultendpunct}{\mcitedefaultseppunct}\relax
\EndOfBibitem
\bibitem[Roos(1999)]{Roos1999}
Roos,~B. O.~. Theoretical studies of electronically excited states of molecular
  systems using multiconfigurational perturbation theory. \emph{Acc. Chem.
  Res.} \textbf{1999}, \emph{32}, 137--144\relax
\mciteBstWouldAddEndPuncttrue
\mciteSetBstMidEndSepPunct{\mcitedefaultmidpunct}
{\mcitedefaultendpunct}{\mcitedefaultseppunct}\relax
\EndOfBibitem
\bibitem[Dunning(1989)]{Dunning1989}
Dunning,~T.~H. Gaussian basis sets for use in correlated molecular
  calculations. I. The atoms boron through neon and hydrogen. \emph{J. Chem.
  Phys.} \textbf{1989}, \emph{90}, 1007--1023\relax
\mciteBstWouldAddEndPuncttrue
\mciteSetBstMidEndSepPunct{\mcitedefaultmidpunct}
{\mcitedefaultendpunct}{\mcitedefaultseppunct}\relax
\EndOfBibitem
\bibitem[Granovsky(2011)]{Granovsky2011}
Granovsky,~A.~A. Extended multi-configuration quasi-degenerate perturbation
  theory: The new approach to multi-state multi-reference perturbation theory.
  \emph{J. Chem. Phys.} \textbf{2011}, \emph{134}, 214113\relax
\mciteBstWouldAddEndPuncttrue
\mciteSetBstMidEndSepPunct{\mcitedefaultmidpunct}
{\mcitedefaultendpunct}{\mcitedefaultseppunct}\relax
\EndOfBibitem
\bibitem[Frisch \latin{et~al.}()Frisch, Trucks, Schlegel, Scuseria, Robb,
  Cheeseman, Scalmani, Barone, Mennucci, Petersson, Nakatsuji, Caricato, Li,
  Hratchian, Izmaylov, Bloino, Zheng, Sonnenberg, Hada, Ehara, Toyota, Fukuda,
  Hasegawa, Ishida, Nakajima, Honda, Kitao, Nakai, Vreven, Montgomery, Peralta,
  Ogliaro, Bearpark, J.Heyd, Brothers, Kudin, Staroverov, Kobayashi, Normand,
  Raghavachari, Rendell, Burant, Iyengar, Tomasi, Cossi, Rega, Millam, Klene,
  Knox, Cross, Bakken, Adamo, Jaramillo, Gomperts, Stratmann, Yazyev, Austin,
  Cammi, Pomelli, Ochterski, Martin, Morokuma, Zakrzewski, Voth, Salvador,
  Dannenberg, Dapprich, Daniels, Farkas, Foresman, Ortiz, Cioslowski, and
  Fox]{Frisch2009}
Frisch,~M.~J.; Trucks,~G.~W.; Schlegel,~H.~B.; Scuseria,~G.~E.; Robb,~M.~A.;
  Cheeseman,~J.~R.; Scalmani,~G.; Barone,~V.; Mennucci,~B.; Petersson,~G.~A.;
  Nakatsuji,~H.; Caricato,~M.; Li,~X.; Hratchian,~H.~P.; Izmaylov,~A.~F.;
  Bloino,~J.; Zheng,~G.; Sonnenberg,~J.~L.; Hada,~M.; Ehara,~M.; Toyota,~K.;
  Fukuda,~R.; Hasegawa,~J.; Ishida,~M.; Nakajima,~T.; Honda,~Y.; Kitao,~O.;
  Nakai,~H.; Vreven,~T.; Montgomery,~J.~A.,~{Jr.}; Peralta,~J.~E.; Ogliaro,~F.;
  Bearpark,~M.; J.Heyd,~J.; Brothers,~E.; Kudin,~K.~N.; Staroverov,~V.~N.;
  Kobayashi,~R.; Normand,~J.; Raghavachari,~K.; Rendell,~A.; Burant,~J.~C.;
  Iyengar,~S.~S.; Tomasi,~J.; Cossi,~M.; Rega,~N.; Millam,~J.~M.; Klene,~M.;
  Knox,~J.~E.; Cross,~J.~B.; Bakken,~V.; Adamo,~C.; Jaramillo,~J.;
  Gomperts,~R.; Stratmann,~R.~E.; Yazyev,~O.; Austin,~A.~J.; Cammi,~R.;
  Pomelli,~C.; Ochterski,~J.~W.; Martin,~R.~L.; Morokuma,~K.;
  Zakrzewski,~V.~G.; Voth,~G.~A.; Salvador,~P.; Dannenberg,~J.~J.;
  Dapprich,~S.; Daniels,~A.~D.; Farkas,~{\"{O}}.; Foresman,~J.~B.;
  Ortiz,~J.~V.; Cioslowski,~J.; Fox,~D.~J. Gaussian~09 {R}evision {D}.1.
  Gaussian Inc. Wallingford CT 2009\relax
\mciteBstWouldAddEndPuncttrue
\mciteSetBstMidEndSepPunct{\mcitedefaultmidpunct}
{\mcitedefaultendpunct}{\mcitedefaultseppunct}\relax
\EndOfBibitem
\bibitem[Granovsky()]{Firefly}
Granovsky,~A.~A. Firefly version 8.2. Available at
  http://classic.chem.msu.su/gran/firefly/index.html\relax
\mciteBstWouldAddEndPuncttrue
\mciteSetBstMidEndSepPunct{\mcitedefaultmidpunct}
{\mcitedefaultendpunct}{\mcitedefaultseppunct}\relax
\EndOfBibitem
\bibitem[Baieva \latin{et~al.}(2017)Baieva, Hakamaa, Groenhof, Heikkil{\"{a}},
  and Toppari]{Baieva2017}
Baieva,~S.; Hakamaa,~O.; Groenhof,~G.; Heikkil{\"{a}},~T.~T.; Toppari,~J.~J.
  Dynamics of strongly coupled modes between surface plasmon polaritons and
  photoactive molecules: the effect of the Stokes shift. \emph{ACS Photonics}
  \textbf{2017}, \emph{4}, 28--37\relax
\mciteBstWouldAddEndPuncttrue
\mciteSetBstMidEndSepPunct{\mcitedefaultmidpunct}
{\mcitedefaultendpunct}{\mcitedefaultseppunct}\relax
\EndOfBibitem
\bibitem[Huang \latin{et~al.}(2011)Huang, Lin, and van Gunsteren]{Gromos96}
Huang,~W.; Lin,~Z.; van Gunsteren,~W.~F. Validation of the GROMOS 54A7 Force
  Field with Respect to $\beta$-Peptide Folding. \emph{Journal of Chemical
  Theory and Computation} \textbf{2011}, \emph{7}, 1237--1243, PMID:
  26610119\relax
\mciteBstWouldAddEndPuncttrue
\mciteSetBstMidEndSepPunct{\mcitedefaultmidpunct}
{\mcitedefaultendpunct}{\mcitedefaultseppunct}\relax
\EndOfBibitem
\bibitem[Berghuis \latin{et~al.}(2022)Berghuis, Tichauer, de~Jong, Sokolovskii,
  Bai, Ramezani, Murai, Groenhof, and Rivas]{Berghuis2022}
Berghuis,~M.~A.; Tichauer,~R.~H.; de~Jong,~L.; Sokolovskii,~I.; Bai,~P.;
  Ramezani,~M.; Murai,~S.; Groenhof,~G.; Rivas,~J.~G. Controlling Exciton
  Propagation in Organic Crystals through Strong Coupling to Plasmonic
  Nanoparticle Arrays. \emph{ACS Photonics} \textbf{2022}, \emph{9}, 123\relax
\mciteBstWouldAddEndPuncttrue
\mciteSetBstMidEndSepPunct{\mcitedefaultmidpunct}
{\mcitedefaultendpunct}{\mcitedefaultseppunct}\relax
\EndOfBibitem
\bibitem[Bussi \latin{et~al.}(2007)Bussi, Donadio, and Parrinello]{Bussi2007}
Bussi,~G.; Donadio,~D.; Parrinello,~M. Canonical sampling through velocity
  rescaling. \emph{J. Chem. Phys.} \textbf{2007}, \emph{126}, 014101\relax
\mciteBstWouldAddEndPuncttrue
\mciteSetBstMidEndSepPunct{\mcitedefaultmidpunct}
{\mcitedefaultendpunct}{\mcitedefaultseppunct}\relax
\EndOfBibitem
\bibitem[Becke(1997)]{Becke97}
Becke,~A.~D. Density-functional thermochemistry. V. Systematic optimization of
  exchange-correlation functionals. \emph{J. Chem. Phys.} \textbf{1997},
  \emph{107}, 8554--8560\relax
\mciteBstWouldAddEndPuncttrue
\mciteSetBstMidEndSepPunct{\mcitedefaultmidpunct}
{\mcitedefaultendpunct}{\mcitedefaultseppunct}\relax
\EndOfBibitem
\bibitem[Schwartz \latin{et~al.}(2013)Schwartz, Hutchison, Leonard, Genet,
  Haacke, and Ebbesen]{Schwartz2013}
Schwartz,~T.; Hutchison,~J.~A.; Leonard,~J.; Genet,~C.; Haacke,~S.;
  Ebbesen,~T.~W. Polariton Dynamics under Strong Light-Molecule Coupling.
  \emph{ChemPhysChem} \textbf{2013}, \emph{14}, 125--131\relax
\mciteBstWouldAddEndPuncttrue
\mciteSetBstMidEndSepPunct{\mcitedefaultmidpunct}
{\mcitedefaultendpunct}{\mcitedefaultseppunct}\relax
\EndOfBibitem
\bibitem[George \latin{et~al.}(2015)George, Wang, Chervy, Canaguier-Durand,
  Schaeffer, Lehn, Hutchison, Genet, and Ebbesen]{George2015}
George,~J.; Wang,~S.; Chervy,~T.; Canaguier-Durand,~A.; Schaeffer,~G.;
  Lehn,~J.-M.; Hutchison,~J.~A.; Genet,~C.; Ebbesen,~T.~W. Ultra-strong
  coupling of molecular materials: spectroscopy and dynamics. \emph{Faraday
  Discuss.} \textbf{2015}, \emph{178}, 281--294\relax
\mciteBstWouldAddEndPuncttrue
\mciteSetBstMidEndSepPunct{\mcitedefaultmidpunct}
{\mcitedefaultendpunct}{\mcitedefaultseppunct}\relax
\EndOfBibitem
\bibitem[Rozenman \latin{et~al.}(2018)Rozenman, Akulov, Golombek, and
  Schwartz]{Rozenman2018}
Rozenman,~G.~G.; Akulov,~K.; Golombek,~A.; Schwartz,~T. Long-Range Transport of
  Organic Exciton-Polaritons Revealed by Ultrafast Microscopy. \emph{ACS
  Photonics} \textbf{2018}, \emph{5}, 105--110\relax
\mciteBstWouldAddEndPuncttrue
\mciteSetBstMidEndSepPunct{\mcitedefaultmidpunct}
{\mcitedefaultendpunct}{\mcitedefaultseppunct}\relax
\EndOfBibitem
\bibitem[Frisch \latin{et~al.}(2016)Frisch, Trucks, Schlegel, Scuseria, Robb,
  Cheeseman, Scalmani, Barone, Petersson, Nakatsuji, Li, Caricato, Marenich,
  Bloino, Janesko, Gomperts, Mennucci, Hratchian, Ortiz, Izmaylov, Sonnenberg,
  Williams-Young, Ding, Lipparini, Egidi, Goings, Peng, Petrone, Henderson,
  Ranasinghe, Zakrzewski, Gao, Rega, Zheng, Liang, Hada, Ehara, Toyota, Fukuda,
  Hasegawa, Ishida, Nakajima, Honda, Kitao, Nakai, Vreven, Throssell,
  Montgomery, Peralta, Ogliaro, Bearpark, Heyd, Brothers, Kudin, Staroverov,
  Keith, Kobayashi, Normand, Raghavachari, Rendell, Burant, Iyengar, Tomasi,
  Cossi, Millam, Klene, Adamo, Cammi, Ochterski, Martin, Morokuma, Farkas,
  Foresman, and Fox]{g16}
Frisch,~M.~J.; Trucks,~G.~W.; Schlegel,~H.~B.; Scuseria,~G.~E.; Robb,~M.~A.;
  Cheeseman,~J.~R.; Scalmani,~G.; Barone,~V.; Petersson,~G.~A.; Nakatsuji,~H.;
  Li,~X.; Caricato,~M.; Marenich,~A.~V.; Bloino,~J.; Janesko,~B.~G.;
  Gomperts,~R.; Mennucci,~B.; Hratchian,~H.~P.; Ortiz,~J.~V.; Izmaylov,~A.~F.;
  Sonnenberg,~J.~L.; Williams-Young,~D.; Ding,~F.; Lipparini,~F.; Egidi,~F.;
  Goings,~J.; Peng,~B.; Petrone,~A.; Henderson,~T.; Ranasinghe,~D.;
  Zakrzewski,~V.~G.; Gao,~J.; Rega,~N.; Zheng,~G.; Liang,~W.; Hada,~M.;
  Ehara,~M.; Toyota,~K.; Fukuda,~R.; Hasegawa,~J.; Ishida,~M.; Nakajima,~T.;
  Honda,~Y.; Kitao,~O.; Nakai,~H.; Vreven,~T.; Throssell,~K.;
  Montgomery,~J.~A.,~{Jr.}; Peralta,~J.~E.; Ogliaro,~F.; Bearpark,~M.~J.;
  Heyd,~J.~J.; Brothers,~E.~N.; Kudin,~K.~N.; Staroverov,~V.~N.; Keith,~T.~A.;
  Kobayashi,~R.; Normand,~J.; Raghavachari,~K.; Rendell,~A.~P.; Burant,~J.~C.;
  Iyengar,~S.~S.; Tomasi,~J.; Cossi,~M.; Millam,~J.~M.; Klene,~M.; Adamo,~C.;
  Cammi,~R.; Ochterski,~J.~W.; Martin,~R.~L.; Morokuma,~K.; Farkas,~O.;
  Foresman,~J.~B.; Fox,~D.~J. Gaussian˜16 {R}evision {C}.01. 2016; Gaussian
  Inc. Wallingford CT\relax
\mciteBstWouldAddEndPuncttrue
\mciteSetBstMidEndSepPunct{\mcitedefaultmidpunct}
{\mcitedefaultendpunct}{\mcitedefaultseppunct}\relax
\EndOfBibitem
\bibitem[Lidzey \latin{et~al.}(2000)Lidzey, Bradley, Armitage, Walker, and
  Skolnick]{Lidzey2000}
Lidzey,~D.; Bradley,~D.; Armitage,~A.; Walker,~S.; Skolnick,~M. Photon-Mediated
  Hybridization of Frenkel Excitons in Organic Semiconductor Microcavities.
  \emph{Science} \textbf{2000}, \emph{288}, 1620--1623\relax
\mciteBstWouldAddEndPuncttrue
\mciteSetBstMidEndSepPunct{\mcitedefaultmidpunct}
{\mcitedefaultendpunct}{\mcitedefaultseppunct}\relax
\EndOfBibitem
\bibitem[Tichauer \latin{et~al.}(2022)Tichauer, Morozov, Sokolovskii, Toppari,
  and Groenhof]{Tichauer2022}
Tichauer,~R.~H.; Morozov,~D.; Sokolovskii,~I.; Toppari,~J.~J.; Groenhof,~G.
  Identifying Vibrations that Control Non-Adiabatic Relaxation of Polaritons in
  Strongly Coupled Molecule-Cavity Systems. \emph{J. Phys. Chem. Lett.}
  \textbf{2022}, \emph{13}, 6259--6267\relax
\mciteBstWouldAddEndPuncttrue
\mciteSetBstMidEndSepPunct{\mcitedefaultmidpunct}
{\mcitedefaultendpunct}{\mcitedefaultseppunct}\relax
\EndOfBibitem
\bibitem[Pandya \latin{et~al.}(2021)Pandya, Chen, Gu, Sung, Schnedermann,
  Ojambati, Chikkaraddy, Gorman, Jacucci, Onelli, Willhammar, Johnstone,
  Collins, Midgley, Auras, Baikie, Jayaprakash, Mathevet, Soucek, Du, Alvertis,
  Ashoka, Vignolini, Lidzey, Baumberg, Friend, Barisien, Legrand, Chin,
  Yuen-Zhou, Saikin, Kukura, Musser, and Rao]{Pandya2021}
Pandya,~R.; Chen,~R. Y.~S.; Gu,~Q.; Sung,~J.; Schnedermann,~C.;
  Ojambati,~O.~S.; Chikkaraddy,~R.; Gorman,~J.; Jacucci,~G.; Onelli,~O.~D.;
  Willhammar,~T.; Johnstone,~D.~N.; Collins,~S.~M.; Midgley,~P.~A.; Auras,~F.;
  Baikie,~T.; Jayaprakash,~R.; Mathevet,~F.; Soucek,~R.; Du,~M.;
  Alvertis,~A.~M.; Ashoka,~A.; Vignolini,~S.; Lidzey,~D.~G.; Baumberg,~J.~J.;
  Friend,~R.~H.; Barisien,~T.; Legrand,~L.; Chin,~A.~W.; Yuen-Zhou,~J.;
  Saikin,~S.~K.; Kukura,~P.; Musser,~A.~J.; Rao,~A. Microcavity-like
  exciton-polaritons can be the primary photoexcitation in bare organic
  semiconductors. \emph{Nat. Commun.} \textbf{2021}, \emph{12}, 6519\relax
\mciteBstWouldAddEndPuncttrue
\mciteSetBstMidEndSepPunct{\mcitedefaultmidpunct}
{\mcitedefaultendpunct}{\mcitedefaultseppunct}\relax
\EndOfBibitem
\bibitem[Freixanet \latin{et~al.}(2000)Freixanet, Sermage, Tiberj, and
  Planel]{Freixanet2000}
Freixanet,~T.; Sermage,~B.; Tiberj,~A.; Planel,~R. In-plane Propagation of
  Excitonic Cavity Polaritons. \emph{Phys. Rev. B} \textbf{2000}, \emph{61},
  7233\relax
\mciteBstWouldAddEndPuncttrue
\mciteSetBstMidEndSepPunct{\mcitedefaultmidpunct}
{\mcitedefaultendpunct}{\mcitedefaultseppunct}\relax
\EndOfBibitem
\bibitem[Flick \latin{et~al.}(2017)Flick, Ruggenthaler, Appel, and
  Rubio]{Flick2017b}
Flick,~J.; Ruggenthaler,~M.; Appel,~H.; Rubio,~A. Atoms and Molecules in
  Cavities: From Weak to Strong Coupling in QED Chemistry. \emph{Proc. Natl.
  Acad. Sci. USA} \textbf{2017}, \emph{114}, 3026--3034\relax
\mciteBstWouldAddEndPuncttrue
\mciteSetBstMidEndSepPunct{\mcitedefaultmidpunct}
{\mcitedefaultendpunct}{\mcitedefaultseppunct}\relax
\EndOfBibitem
\end{mcitethebibliography}


\end{document}


\setlength{\marginparwidth}{3cm} 
\newpage
\tableofcontents

\newpage

\setcounter{page}{1}

\section{Molecular Dynamics Simulation Model}

\subsection{Multiscale Tavis-Cummings Hamiltonian}

\begin{figure*}[!htb]
\centerline{
  \epsfig{figure=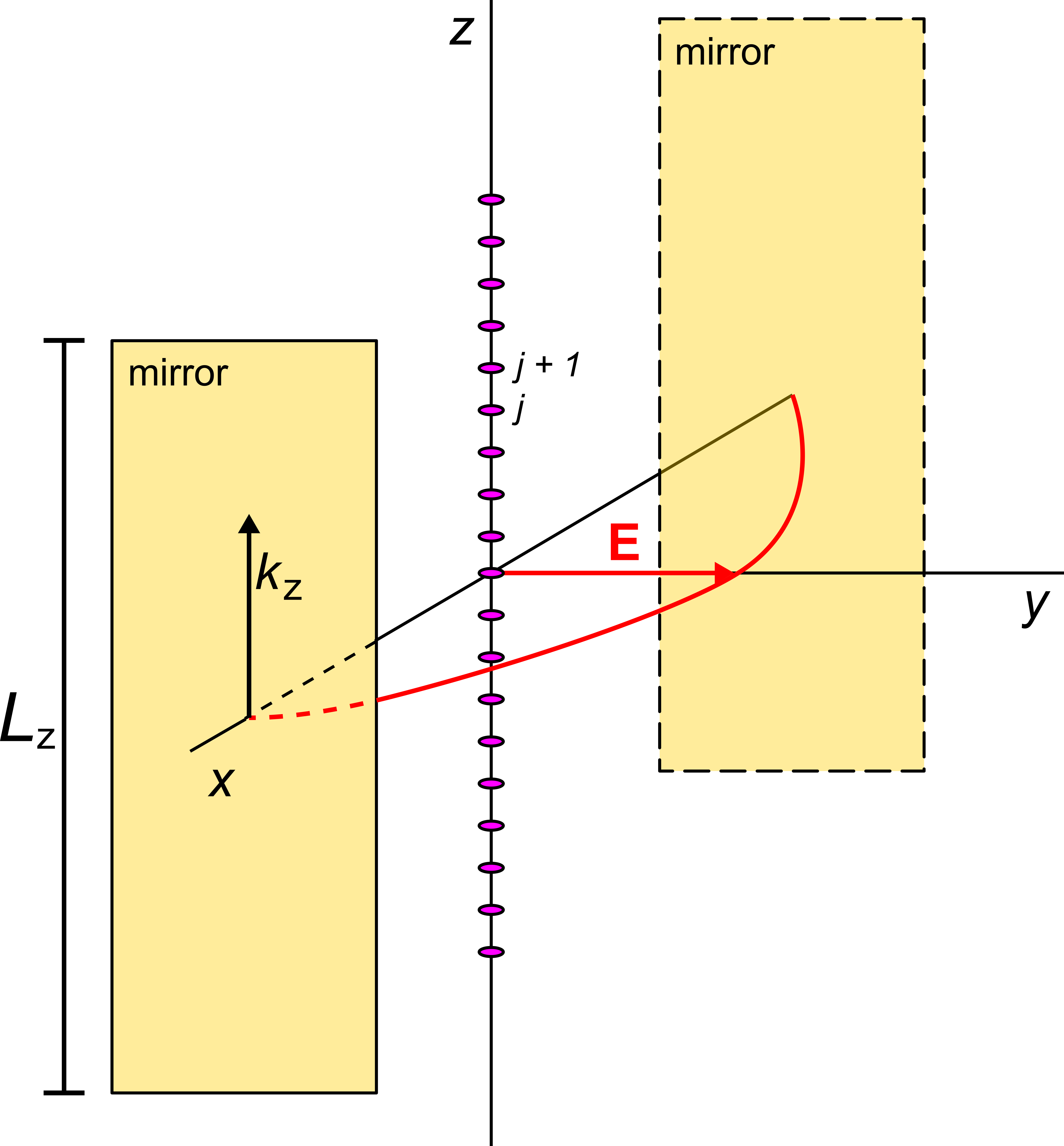,width=7cm,angle=0}}
  \caption{One-dimensional (1D) Fabry-P\'{e}rot micro-cavity model. Two reflecting mirrors located at $-\frac{1}{2}x$ and $\frac{1}{2}x$, confine light modes along this direction, while free propagation along the $z$ direction is possible for plane waves with in-plane momentum $k_z$ and energy $\hslash\omega_{\text{cav}}(k_z)$. The vacuum field vector (red) points along the $y$-axis, reaching a maximum amplitude at $x=0$ where the $N$ molecules (magenta ellipses) are placed, distributed along the $z$-axis at positions $z_j$ with $1\le j \le N$. 
  }
  \label{fig:1D_cavity}
\end{figure*}

Within the single-excitation subspace, probed experimentally under weak driving conditions, and employing the rotating wave approximation (RWA), valid for light-matter coupling strengths below 10\% of the material excitation energy,\cite{Forn-Diaz2019} the interaction between $N$ molecules and $n_\text{mode}$ confined light modes in a one-dimensional (1D)  Fabry-P\'{e}rot cavity (Figure~\ref{fig:1D_cavity}) is modelled with the multi-scale Molecular Dynamics (MD) extension of the traditional Tavis-Cummings model of quantum optics:\cite{Jaynes1963,Tavis1969,Michetti2005,Tichauer2021}
\begin{equation}
\begin{array}{ccl}
    {\hat{H}}^{\text{TC}} &=& \sum_j^N h\nu_j(\mathbf R_j)\hat{\sigma}^+_j\hat{\sigma}^-_j+\sum_{k_z}^{n_\text{mode}}\hslash\omega_{\text{cav}}(k_z)\hat{a}_{k_z}^\dagger\hat{a}_{k_z}+\\
    \\
    &&\sum_j^N\sum_{k_z}^{n_\text{mode}}\hslash g_j(k_z) \left(\hat{\sigma}^+_j\hat{a}_{k_z}e^{ik_zz_j}+\hat{\sigma}^-_j\hat{a}_{k_z}^\dagger e^{-ik_zz_j}\right)+\\
    \\
    &&\sum_{i}^N V^\text{mol}_{\text{S}_0}({\bf{R}}_i)
    \end{array}
    \label{eq:dTCH}
\end{equation}
Here, $\hat{\sigma}_j^+$ ($\hat{\sigma}_j^-$) is the operator that excites (de-excites) molecule $j$ from the electronic ground (excited) state $|\text{S}_0^j({\bf{R}}_j)\rangle$ ($|\text{S}_1^{j}({\bf{R}}_j)\rangle$) to the electronic excited (ground) state $|\text{S}_1^{j}({\bf{R}}_j)\rangle$ ($|\text{S}_0^j({\bf{R}}_j)\rangle$); ${\bf{R}}_j$ is the vector of the Cartesian coordinates of all atoms in molecule $j$, centered at $z_j$;
$\hat{a}_{k_z}$ ($\hat{a}_{k_z}^\dagger$) is the annihilation (creation) operator of an excitation of a cavity mode with wave-vector $k_z$; $h\nu_j(\mathbf R_j)$ is the excitation energy of molecule $j$, defined as: 
\begin{equation}
  h\nu_j(\mathbf R_j)=V_{\text{S}_1}^{\text{mol}}({\bf{R}}_j)-V_{\text{S}_0}^{\text{mol}}({\bf{R}}_j)
  \label{eq:Hnu}
\end{equation}
with $V_{\text{S}_0}^{\text{mol}}({\bf{R}}_j)$ and $V_{\text{S}_1}^{\text{mol}}({\bf{R}}_j)$ the adiabatic potential energy surfaces (PESs) of molecule $j$ in the electronic ground (S$_0$) and excited (S$_1$) state, respectively. 
The last term in Equation~\ref{eq:dTCH} is the total potential energy of the system in the absolute ground state ({\it i.e.}, with no excitations in neither the molecules nor the cavity modes), defined as the sum of the ground-state potential energies of all molecules in the cavity. The $V_{\text{S}_0}^{\text{mol}}({\bf{R}}_j)$ and $V_{\text{S}_1}^{\text{mol}}({\bf{R}}_j)$ adiabatic PESs are modelled at the hybrid quantum mechanics / molecular mechanics (QM/MM) level of theory.\cite{Warshel1976b,Boggio-Pasqua2012}

The third term in Equation~\ref{eq:dTCH} describes the light-matter interaction within the dipolar approximation through $g_j(k_z)$:
\begin{equation}
    g_j(k_z) = -{\boldsymbol{\upmu}}^\text{TDM}_j({\bf{R}}_j) \cdot {\bf{u}}_{\text{cav}} \sqrt{\frac{\hslash\omega_{\text{cav}}(k_z)}{2\epsilon_0 V_{\text{cav}}}} 
    \label{eq:dipole_coupling}
\end{equation}
where ${\boldsymbol{\upmu}}_j^\text{TDM}({\bf{R}}_j)$ is the transition dipole moment of molecule $j$ that depends on the molecular geometry (${\bf{R}}_j)$; ${\bf{u}}_\text{cav}$ the unit vector in the direction of the electric component of cavity vacuum field ({\it{i.e.}}, $|{\bf{E}}|=\sqrt{\hslash\omega_\text{cav}(k_z)/2\epsilon_0V_\text{cav}}$), here along the $y$-direction (Figure~\ref{fig:1D_cavity}); $\epsilon_0$ the vacuum permittivity; and $V_{\text{cav}}$ the cavity mode volume.

\subsection{One-dimensional Periodic Cavity}

Following Michetti and La Rocca,~\cite{Michetti2005} we impose periodic boundary conditions in the $z$-direction of our cavity, and thus restrict the wave vectors, $k_z$, to discrete values: $k_{z,p}=2\pi p/L_z$ with $p \in \mathbb Z$ and $L_{z}$ the length of the 1D cavity. With this approximation the molecular Tavis-Cummings Hamiltonian in Equation~\ref{eq:dTCH} can be represented as an $(N+n_{\text{mode}})$ by $(N+n_{\text{mode}})$ matrix with four blocks:\cite{Tichauer2021}
\begin{equation}
  {\bf{H}}^{\text{TC}} = \left(\begin{array}{cc} 
  {\bf{H}}^{\text{mol}} & {\bf{H}}^{\text{int}}\\
  {\bf{H}}^{\text{int}\dagger} & {\bf{H}}^{\text{cav}}\\
  \end{array}\right)\label{eq:TavisCummings}
\end{equation}
We compute the elements of this matrix in the product basis of adiabatic molecular states times cavity mode excitations:
\begin{equation}
\begin{array}{ccl}
|\phi_j\rangle &=& \hat{\sigma}_j^+|\text{S}_0^1\text{S}_0^2..\text{S}_0^{N-1}\text{S}_0^N\rangle\otimes|00..0\rangle\\
\\
&=&\hat{\sigma}_j^+|\Pi_i^N\text{S}_0^i\rangle\otimes|\Pi_k^{n_\text{mode}}0_k\rangle \\
\\
&=&\hat{\sigma}_j^+|\phi_0\rangle
\end{array}
\end{equation}
for $1\le j\le N$, and 
\begin{equation}
\begin{array}{ccl}
|\phi_{j >N}\rangle &=& \hat{a}_{j-N}^\dagger|\text{S}_0^1\text{S}_0^2..\text{S}_0^{N-1}\text{S}_0^N\rangle\otimes|00..0\rangle\\
\\
&=&\hat{a}_{j-N}^\dagger|\Pi_i^N\text{S}_0^i\rangle\otimes|\Pi_k^{n_\text{mode}}0_k\rangle \\
\\
&=&\hat{a}_{j-N}^\dagger|\phi_0\rangle
\end{array}
\end{equation}
for $N < j\le N+n_\text{mode}$. In these expressions $|00..0\rangle$ indicates that the Fock states for all $n_\text{mode}$ cavity modes are empty. The basis state $|\phi_0\rangle$ is the ground state of the molecule-cavity system with no excitations of neither the molecules nor cavity modes:
\begin{equation}
|\phi_0\rangle = |\text{S}_0^1\text{S}_0^2..\text{S}_0^{N-1}\text{S}_0^N\rangle\otimes|00..0\rangle
=|\Pi_i^N\text{S}_0^i\rangle\otimes|\Pi_k^{n_\text{mode}}0_k\rangle\label{eq:phi0}
\end{equation}

The upper left block, ${\bf{H}}^{\text{mol}}$, is an $N\times N$ matrix containing the single-photon excitations of the molecules. Because we neglect direct excitonic interactions between molecules, this block is diagonal, with elements labeled by the molecule indices $j$:
\begin{equation}
    H_{j,j}^{\text{mol}}=\langle \phi_0|\hat{\sigma}_j\hat{H}^{\text{TC}}\hat{\sigma}_j^+|\phi_0\rangle\label{eq:molecular_diagonal}
\end{equation}
for $1 \le j \le N$. 
Each matrix element of ${\bf{H}}^\text{mol}$ thus represents the potential energy of a molecule, $j$, in the electronic excited state $|\text{S}_1^{j}({\bf{R}}_j)\rangle$ while all other molecules, $i\neq j$, are in the electronic ground state $|\text{S}_0^i({\bf{R}}_i)\rangle$:
\begin{equation}  H_{j,j}^{\text{mol}}=V_{\text{S}_1}^{\text{mol}}({\bf{R}}_j)+\sum^N_{i\neq j}V_{\text{S}_0}^{\text{mol}}({\bf{R}}_i)
  \label{eq:Hj}
\end{equation}

The lower right block,  ${\bf{H}}^{\text{cav}}$, is an $n_\text{mode}\times n_\text{mode}$ matrix (with $n_{\text{mode}}=n_{\text{max}}-n_{\text{min}}+1$) containing the single-photon excitations of the cavity modes, and is also diagonal:
\begin{equation}
    H_{p,p}^{\text{cav}}= \langle \phi_0|\hat{a}_p\hat{H}^{\text{TC}}\hat{a}^\dagger_p|\phi_0\rangle\label{eq:photonic_diagonal}
\end{equation}
for $n_{\text{min}}\le p \le n_{\text{max}}$. 
Here, $\hat{a}_p^\dagger$ excites cavity mode $p$ 
with wave-vector $k_{z,p}= 2\pi p/L_z$. In these matrix elements, all molecules are in the electronic ground state (S$_0$). The energy is therefore the sum of the cavity energy at $k_{z,p}$, and the molecular ground state energies:
\begin{equation}
    H_{p,p}^{\text{cav}}= \hslash\omega_{\text{cav}}(2\pi p/L_z)+\sum^N_{j}V_{\text{S}_0}^{\text{mol}}({\bf{R}}_j)
    \label{eq:Hn}
\end{equation}
where, $\omega_{\text{cav}}(k_{z,p})$ is the cavity dispersion (dashed curve in Figure 1b, main text): 
\begin{equation}
  \omega_\text{cav}(k_{z,p})=\sqrt{\omega^2_0+ c^2k_{z,p}^2/n^2}
  \label{eq:dispersion}
\end{equation}
with $\hslash\omega_0$ the energy at $k_0=0$, $n$ the refractive index of the medium and $c$ the speed of light in vacuum.

The two $N\times n_\text{mode}$ off-diagonal blocks ${\bf{H}}^{\text{int}}$ and ${\bf{H}}^{\text{int}\dagger}$ in the multi-mode Tavis-Cummings Hamiltonian (Equation~\ref{eq:TavisCummings}) model the light-matter interactions between the molecules and the cavity modes. These matrix elements are approximated as the overlap between the transition dipole moment of molecule $j$ and the transverse electric field of the cavity mode $p$ at the geometric center $z_j$ of that molecule:
\begin{equation}
\begin{array}{ccl}
  H_{j,p}^{\text{int}} &=& 
  -{\boldsymbol{\upmu}}_j^\text{TDM}({\bf{R}}_j)\cdot {\bf{u}}_{\text{cav}} \sqrt{\frac{\hslash\omega_{\text{cav}}(2\pi p/L_z)}{2\epsilon_0 V_{\text{cav}}}} \langle\phi_0|\hat{\sigma}_j\hat{\sigma}^+_j\hat{a}_p e^{i 2\pi p z_j/L_z}\hat{a}_p^\dagger|\phi_0\rangle \\
  \\
  &=&
  -{\boldsymbol{\upmu}}_j^\text{TDM}({\bf{R}}_j) \cdot {\bf{u}}_{\text{cav}} \sqrt{\frac{\hslash\omega_{\text{cav}}(2\pi p/L_z)}{2\epsilon_0 V_{\text{cav}}}} e^{i 2\pi p z_j/L_z}     
  \end{array}
  \label{eq:QMMMdipole_coupling}
\end{equation}
for $1 \le j \le N$ and $n_{\text{min}}\le p \le n_{\text{max}}$. 

Diagonalization of the multi-scale Tavis-Cummings Hamiltonian in Equation~\ref{eq:TavisCummings} yields the $N+n_\text{mode}$ hybrid light-matter states $\lvert\psi^m\rangle$:\cite{Michetti2005,Agranovich2007}
\begin{equation}
\vert\psi^{m}\rangle=\left(\sum_{j}^N\beta^m_j\hat{\sigma}^+_j + \sum_{p}^{n_{\text{mode}}}\alpha^m_p\hat{a}_p^\dagger\right)\vert\phi_0\rangle
\label{eq:Npolariton}
\end{equation}
with eigenenergies $E_m$.
The $\beta_j^m$ and $\alpha_p^m$ expansion coefficients reflect the contribution of the molecular excitons ($\vert\text{S}_1^j(\mathbf{R}_j)\rangle$) and the cavity mode excitations ($\vert 1_p\rangle$) to polariton $\vert \psi^m\rangle$. 

\subsection{Mean-Field Non-Adiabatic Molecular Dynamics Simulations}

Semi-classical Molecular Dynamics (MD) trajectories of all molecules are computed by numerically integrating Newton's equations of motion. The multi-mode Tavis-Cummings Hamiltonian (Equation~\ref{eq:TavisCummings}) is diagonalized at each time-step to obtain the $N+n_\text{mode}$ (adiabatic) polaritonic eigenstates $|\psi^m\rangle$ and energies $E^m$.
The \emph{total} polaritonic wavefunction $|\Psi(t)\rangle$ is coherently propagated along with the classical degrees of freedom of the $N$ molecules as a time-dependent superposition of the $N+n_\text{mode}$ time-independent adiabatic polaritonic states:
\begin{equation}
|\Psi(t)\rangle=\sum_m^{N+n_{\text{mode}}}c_m(t)|\psi ^m\rangle\label{eq:totalwf}
\end{equation}
where $c_m(t)$ are the time-dependent expansion coefficients of the time-independent polaritonic basis functions $|\psi^m\rangle$ defined in Equation~\ref{eq:Npolariton}. A unitary propagator in the \emph{local} diabatic basis is used to integrate these coefficients,\cite{Granucci2001} while the nuclear degrees of freedom of the $N$ molecules evolve on the mean-field PES:
\begin{equation}
V({\bf{R}})=\langle\Psi(t)|\hat{H}^\text{TC}|\Psi(t)\rangle
\end{equation}
Population transfers between polaritonic states $|\psi^m\rangle$ and $|\psi^l\rangle$ are driven by the non-adiabatic coupling, or vibronic coupling $V_{m,l}^\text{NAC}$.~\cite{Yarkony2012,Worth2004,Azumi1977} $V_{m,l}^\text{NAC}$ depends on the overlap between the non-adiabatic coupling vector ${\bf{d}}_{m.l}$ and the velocities of the atoms $\dot{\bf{R}}$:\cite{Crespo2018}
\begin{equation}
    V_{m,l}^\text{NAC} = {\bf{d}}_{m,l}\cdot \dot{\bf{R}}
    \label{eq:nac}
\end{equation}
where $\dot{\bf{R}}$ is a vector containing the velocities of all atoms in the system, and the non-adiabatic coupling vector ${\bf{d}}_{m,l}$ is given by:
\begin{equation}
 {\bf{d}}_{m,l}=\frac{\langle\psi^m|\nabla_{\bf{R}}\hat{H}^\text{TC}|\psi^l\rangle}{E_l-E_m}
    \label{eq:nacv}
\end{equation}
where $E_m$ and $E_l$ are the eigenvalues associated with $|\psi^m\rangle$ and $|\psi^l\rangle$, respectively. The direction of ${\bf{d}}_{m,l}$ determines the direction of population transfer, while its magnitude determines the amount of population that is transferred.

\subsection{Cavity Decay}

Radiative loss through the cavity mirrors is modeled as a first-order decay process into the overall ground state of the system ({\it{i.e.}}, no excitation in neither the molecules nor the cavity modes).\cite{Luk2017} Assuming that the intrinsic decay rates $\gamma_{\text{cav}}$ are the same for all modes, the total loss rate is calculated as the product of $\gamma_{\text{cav}}$ and the total photonic weight, $\sum_{p}^{n_\text{mode}}|\alpha^m_p|^2$, of state $|\psi^m\rangle$. Thus, after an MD step $\Delta t$, $\rho_m(t)= |c_m(t)|^2$ becomes:
\begin{equation}
  \rho_m(t+\Delta t) = \rho_m(t) \exp\left[-\gamma_{\text{cav}}\sum_p^{n_\text{mode}}|\alpha^m_p(t)|^2\Delta t\right]\label{eq:decay_tot}
\end{equation}
Since $\rho_m = (\Re[c_m])^2+(\Im[c_m])^2$, changes in the real and imaginary parts of the  (complex) expansion coefficients $c_m(t)$, due to spontaneous photonic loss in a low-Q cavity, are:
\begin{equation*}
    \Re[c_m (t+\Delta t)] = \Re[c_m
                              (t)] \exp\left[-\frac{1}{2}\gamma_{\text{cav}}\sum_p^{n_\text{mode}}|\alpha^m_p(t)|^2\Delta
                              t\right]
\end{equation*}
\begin{equation*}
 \Im[c_m (t+\Delta t)] = \Im[c_m
                              (t)] \exp\left[-\frac{1}{2}\gamma_{\text{cav}}\sum_p^{n_\text{mode}}|\alpha_p^m(t)|^2\Delta
                              t\right]
\end{equation*}
Simultaneously, the population of the zero-excitation subspace, or ground state, $\rho_0(t+\Delta t)$, increases as:
\begin{equation}
\rho_0(t+\Delta t)=\rho_0(t)+\sum_m\rho_m(t)\left(1-\exp\left[-\gamma_{\text{cav}}\sum_p^{n_\text{mode}}|\alpha_p^m(t)|^2\Delta t\right]\right)
\end{equation}

\newpage

\section{Simulation details} 
\label{subsec:sim_details}

\subsection{Rhodamine Model}

\begin{figure*}[!htb]
\centerline{ \epsfig{figure=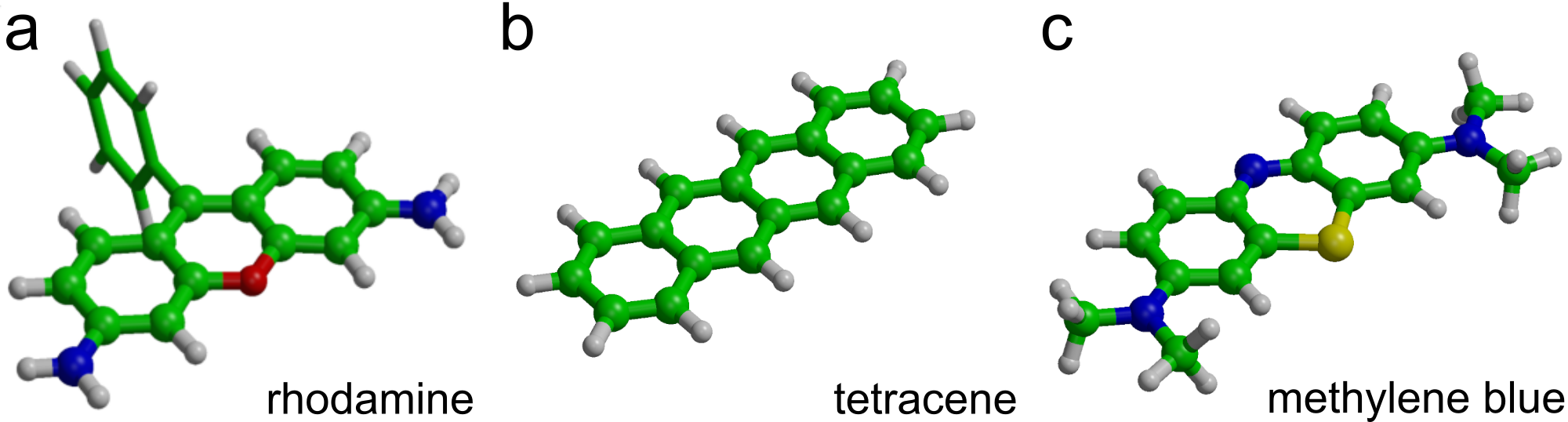,width=\textwidth,angle=0}}
  \caption{Panel \textbf{a}: Rhodamine model used in our simulations. The QM subsystem, shown in ball-and-stick representation, is described at the HF/3-21G level of theory in the electronic ground state (S$_0$), and at the CIS/3-21G level of theory in the first singlet excited state (S$_1$). The MM subsystem, consisting of the atoms shown in stick representation, and the water molecules (not shown), are modelled with the Amber03 force field. Panel \textbf{b}: Tetracene model. All atoms are included in the QM subsystem, modelled at the RHF/3-21G level of the theory in S$_0$ and at the CIS/3-21G level of theory in S$_1$. The cyclohexane solvent (not shown) is included in the MM subsystem and described with the Gromos96-54a7 force field. Panel \textbf{c}: Methylene Blue model. All atoms are included in the QM subsystem, modelled at the B97/3-21G level of the theory in S$_0$ and at the TDA-B97/3-21G level of theory in S$_1$. The water solvent (not shown) is included in the MM subsystem and described with the Amber03 force field.
  }
  \label{fig:rhodamine}
\end{figure*}

Rhodamine molecules, one of which is shown schematically in Figure~\ref{fig:rhodamine}a, were modelled with the Amber03 force field,\cite{Duan2003} using the parameters provided by Luk {\it et al.}\cite{Luk2017} After a geometry optimization at the force field level, the molecule was placed at the center of a rectangular box and 3,684 TIP3P water molecules,\cite{Jorgensen1983} were added. The simulation box, which contained 11,089 atoms, was equilibrated for 2~ns with harmonic restraints on the heavy atoms of Rhodamine (force constant 1000~kJmol$^{-1}$nm$^{-1}$). Subsequently, a 200~ns classical MD trajectory was computed at constant temperature (300~K) using a stochastic dynamics integrator with a friction coefficient of 0.1~ps$^{-1}$. The pressure was kept constant at 1 bar using the Berendsen isotropic pressure coupling algorithm\cite{Berendsen1984} with a time constant of 1~ps. The LINCS algorithm was used to constrain bond lengths,\cite{Hess1997} while SETTLE was applied to constrain the internal degrees of freedom of water molecules,\cite{Miyamoto1992} enabling a time step of 2~fs in the classical MD simulations. A 1.0~nm cut-off was used for Van der Waals' interactions, which were modeled with Lennard-Jones potentials. Coulomb interactions were computed with the smooth particle mesh Ewald method,\cite{Essmann1995} using a 1.0~nm real space cut-off and a grid spacing of 0.12~nm. The relative tolerance at the real space cut-off was set to 10$^{-5}$.


From the second half of the 200~ns MD trajectory, snapshots were extracted and subjected to further equilibration for 10~ps at the QM/MM level. The time step was reduced to 1~fs. As in previous work,\cite{Luk2017,Groenhof2019,Tichauer2021} the fused ring system was included in the QM region and described at the RHF/3-21G level, while the rest of the molecule as well as the water solvent were modelled with the Amber03 force field\cite{Duan2003} and TIP3P water model,\cite{Jorgensen1983} respectively (Figure~\ref{fig:rhodamine}). The bond connecting the QM and MM subsystems was replaced by a constraint and the QM part was capped with a hydrogen atom. The force on the cap atom was distributed over the two atoms of the bond. The QM system experienced the Coulomb field of all MM atoms within a 1.6~nm cut-off sphere and Lennard-Jones interactions between MM and QM atoms were added. The singlet electronic excited state (S$_1$) was modeled with the Configuration Interaction method, truncated at single electron excitations, for the QM region ({\it {i.e.,}} CIS/3-21G//Amber03). At this level of QM/MM theory, the excitation energy is 4.18~eV~\cite{Luk2017}. The QM/MM simulations were performed with GROMACS version 4.5.3,\cite{Hess2008} interfaced to TeraChem version 1.93.\cite{Ufimtsev2009,Titov2013}


\subsubsection{Comparison of CIS/3-21G to higher levels of theory}\label{subsubsec:QMvalidation}

The level of theory used in this work is a necessary compromise between efficiency and accuracy. Because no experimental data is available for the Rhodamine model used here, we instead verified the validity of our model by recomputing the electronic energies along an excited-state molecular dynamics trajectory of a single Rhodamine molecule in water at higher levels of theory:
\begin{itemize}
    \item Time-dependent density functional theory (TD-DFT)\cite{Runge1984} with the CAM-B3LYP functional,\cite{Becke1993,Yanai2004} in combination with the 6-31G(d) basis set;\cite{Dunning1970} \item Complete Active Space Self Consistent Field (CASSCF) with 12 electrons in 12 $\pi$ orbitals,\cite{Roos1999} averaged over the first two states (SA2) and expanded in a cc-pVDZ basis set;\cite{Dunning1989}  
    \item Extended Multi-Configuration Quasi-Degenerate Perturbation Theory to second order (xMCQDPT2),\cite{Granovsky2011} based on the SA2-CASSCF(12,12)/cc-pVDZ reference.
\end{itemize}
The TD-DFT calculations were carried out with the Gaussian09 program,\cite{Frisch2009} while the CASSCF and xMCQDPT2 calculations were performed with Firefly.\cite{Firefly}

\begin{figure*}[!htb]
\centerline{
    \epsfig{figure=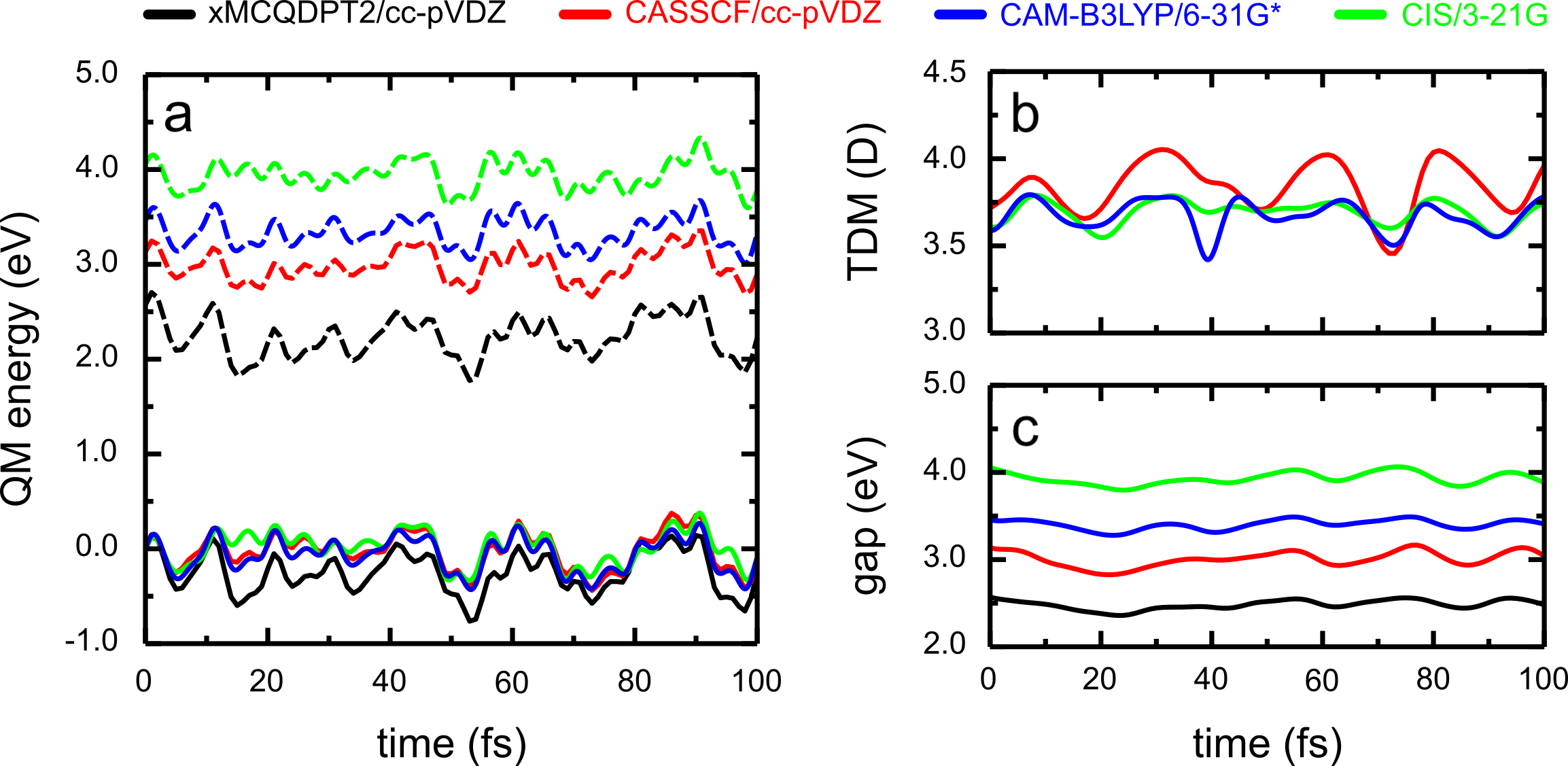,width=14cm,angle=0}}\caption{Panel a: potential energies in the electronic ground (S$_0$, continuous lines) and excited state (S$_1$, dashed lines) as a function of time in an excited-state MD trajectory of a single Rhodamine at the CIS/3-21G level or theory (green), re-evaluated at the xMCQDPT2//SA2-CASSCF(12,12)/cc-pVDZ (black), SA2-CASSCF(12,12)/cc-pVDZ (red), TD-CAM-B3LYP/6-31G(d) (blue) levels of theory. Panel \textbf{b}: the transition dipole moment at the various levels of theory. Note that because the xMCQDPT2 method only corrects the energy, but not the wave function, the transition dipole moment is the same as that of the underlying CASSCF reference. Panel \textbf{c}: the energy gaps between the excited and ground state.}\label{fig:validate_CIS}
  \end{figure*}

As shown in Figure~\ref{fig:validate_CIS}, the absolute excitation energy is overestimated at the CIS/3-21G level of theory with respect to the more accurate xMCQDPT2//SA2-CASSCF(12,12)/cc-pVDZ approach. However, the positions of the minima and maxima coincide at all levels of theory. Therefore, we conclude that the topology of the PES is not very sensitive to the level of theory and that the nuclear dynamics will be sufficiently similar at all levels of theory. 

Furthermore, the Stokes shift, which was shown to have an important effect on polariton relaxation,\cite{Baieva2017} was evaluated by first optimizing the molecule in the ground and excited state and then taking the difference between the vertical excitation energies at these minima. At the CIS/3-21G level of theory the Stokes shift of 0.12 meV is in somewhat better agreement with the xMCQDPT2//SA2-CASSCF(12,12)/cc-pVDZ value of 0.200~meV than SA2-CASSCF(12,12)/cc-pVDZ, which yields 338~meV, or TD-CAMB3LYP/6-31G(d), which yields 88~meV.

Because the polaritonic states are determined by the coupling between the cavity field and the molecular transition dipole moments, we also compared the transition dipole moment between the different levels of theory (Figure~\ref{fig:validate_CIS}b). As with the electronic energies, there are deviations, but the average transition dipole moments are within 0.1~D.

Because we can compensate the overestimated excitation energy by adding an offset to the cavity dispersion in our simulations, we conclude that for the purposes of our study, the CIS/3-21G method provides an acceptable compromise between desired accuracy and computational efficiency for calculating trajectories of ensembles of Rhodamine molecules strongly coupled to the confined light modes of a Fabry-P\'{e}rot micro-cavity.

\subsubsection{Minimum Energy Conical Intersection}\label{subsubsec:MECI}

There are three important points on the S$_1$ PES of the Rhodamine molecule: (i) the Frank-Condon (FC) vertical excitation point, (ii) the (local) minimum on the S$_1$ state and (iii) the minimum energy conical intersection (MECI) point between S$_1$ and S$_0$ states. We have optimized these points at the unified SA2-CASSCF(12,12)/cc-pVDZ level of theory and compared their energies to understand whether on the timescales of our simulations  radiationless deactivation of the Rhodamine molecule via the conical intersection could play a role in the transport process. 

First, we optimized the ground state geometry of the QM subsystem at the PBE0/cc-pVDZ level of theory. With this geometry, we evaluated the vertical excitation energy into the FC point at the SA2-CASSCF(12,12)/cc-pVDZ level of theory. Next, we optimized the geometry on the S$_1$ excited state PES at the same CASSCF level of theory. Finally, we optimized the MECI point with the Firefly \cite{Firefly} package at the same SA2-CASSCF(12,12)/cc-pVDZ level of theory. The convergence criterion for all optimizations was that the maximum Cartesian gradient component is below 0.0001~a.u.

Results of the optimizations are summarized in  Table~\ref{tab:rho_energies}, in which the first colomn contains the absolute S$_1$ energies in Hartrees, while the second colomn lists the relative energies with respect to the FC energy in kJmol$^{-1}$. Because the MECI point is more than 120 kJmol$^{-1}$ higher in energy than FC point, it is extremely unlikely that the system would sample the MECI region and undergo a radiationless decay into the S$_0$ ground state on the time scale of our simulations. Instead, the most probably deactivation process for bare Rhodamine is fluorescence, in line with experiment.

\begin{table}[]
\caption{Computed S$_1$ state energies for the model rhodamine molecule in a gas-phase at the SA2-CASSCF(12,12)/cc-pVDZ level of theory}
\label{tab:rho_energies}
\begin{tabular}{c|c|c}
\hline
Geometry  & E(S$_1$), a.u. & $\Delta$E(S$_1$), kJmol$^{-1}$ \\ \hline
FC        & -682.337367 & 0             \\
S$_1$ minimum & -682.342427 & -13.29        \\
MECI      & -682.289575 & 125.48        \\ 
\end{tabular}
\end{table}

\subsection{Tetracene Model}

Tetracene (Figure~\ref{fig:rhodamine}b) was placed at the center of a rectangular box that was filled with 265 cyclohexane molecules. Intermolecular interactions were modelled with the Gromos96-54a7 united-atom force field.\cite{Gromos96} The simulation box, which contained 2,785 atoms, was equilibrated for 100~ns. During equilibration, the coordinates of the Tetracene atoms were kept fixed. The LINCS algorithm was used to constrain bond lengths in cyclohexane,\cite{Hess1997} enabling a time step of 2 fs. Temperature and pressure were maintained at 300~K and 1~atmosphere by means of weak-coupling to an external bath ($\tau_T = $~0.1~ps, $\tau_p = $ 1~ps).\cite{Berendsen1984} Because cyclohexane atoms are uncharged, the interactions between Tetracene and cyclohexane were modeled with Lennard-Jones potentials only. The Lennard-Jones potential was truncated at 1.4~nm. 

Snapshots were extracted from the equilibration trajectory and further equilibrated at the QM/MM level with a time step of 1~fs. In all QM/MM simulations, Tetracene was modelled at the restricted Hartree-Fock level of theory in combination with a 3-21G basis set and mechanically embedded into the MM region, which consisted of the cyclohexane solvent described with the Gromos96-54A7 united-atom force field.\cite{Gromos96}. The Lennard-Jones interactions between the QM and MM subsystems were truncated at 1.4~nm. As in previous work,\cite{Berghuis2022}, we used the Configuration Interaction truncated at the single excitations (CIS), in combination with the 3-21G basis set,\cite{Dunning1970} to model the singlet excited electronic state of Tetracene. At this level of theory the absorption maximum is 3.87~eV, which is overestimated with respect to experiments due to the lack of electron-electron correlation in the CIS method and the limited flexibility of the 3-21G basis set. 
The QM/MM simulations were performed with GROMACS version 4.5.3,\cite{Hess2008} interfaced to TeraChem version 1.93.\cite{Ufimtsev2009,Titov2013}

\subsection{Methylene Blue Model}

Methylene Blue (Figure~\ref{fig:rhodamine}c) was placed at the center of a rectangular box that was filled with 2,031 TIP3P water molecules.\cite{Jorgensen1983} Non-covalent interactions between Methylene Blue and water were modelled with the Amber03 force field.\cite{Duan2003} A 1.0~nm cut-off was used for the Van der Waals' interactions, which were modeled with Lennard-Jones potentials, while the Coulomb interactions were computed with the smooth particle mesh Ewald method,\cite{Essmann1995} using a 1.0~nm real space cut-off and a grid spacing of 0.12~nm and a relative tolerance at the real space cut-off of 10$^{-5}$. The simulation box, containing 6,131 atoms, was equilibrated for 1~ns at the force field level of theory. During this equilibration, the coordinates of the Methylene atoms were kept fixed. The temperature was kept at 300~K with the v-rescale thermostat,\cite{Bussi2007} while 
the pressure was kept constant at 1~atmosphere using the Berendsen isotropic pressure coupling algorithm\cite{Berendsen1984} with a time constant of 1~ps. The SETTLE algorithm was applied to constrain the internal degrees of freedom of water molecules,\cite{Miyamoto1992} enabling a time step of 2~fs in the classical MD simulations. 

After the equilibration at the force field (MM) level, the system was further equilibrated for 10~ps at the QM/MM level. The time step was reduced to 1~fs. The Methylene Blue molecule was modelled with density functional theory, using the B97 functional,\cite{Becke97} in combination with the 3-21G basis set.\cite{Dunning1970} The water solvent was modelled with the TIP3P force field.\cite{Jorgensen1983} The QM system experienced the Coulomb field of all MM atoms within a 1.0~nm cut-off sphere and Lennard-Jones interactions between MM and QM atoms were added. The singlet electronic excited state (S$_1$) was modeled with time-dependent density functional theory,\cite{Runge1984} using the B97 functional in combination with the 3-21G basis set for the QM region ({\it {i.e.,}} TD-B97/3-21G),\cite{Dunning1970}. At this level of QM/MM theory, the excitation energy is 2.5~eV. The overestimation of the vertical excitation energy is due to the limited accuracy of the employed level of theory, but was compensated by adding an off-set to the cavity resonance energy. The QM/MM simulations were performed with GROMACS version 4.5.3,\cite{Hess2008} interfaced to TeraChem version 1.93.\cite{Ufimtsev2009,Titov2013}

\subsection{Molecular Dynamics of Rhodamine-Cavity Systems}


After equilibration at the QM/MM level, the Rhodamine molecules in solution were placed with equal inter-molecular distances on the $z$-axis of a 1D,~\cite{Michetti2005} 50~$\upmu$m long, optical Fabry-P\'erot micro-cavity (Figure~\ref{fig:1D_cavity}). The dispersion of this cavity was modelled  with 160~discrete modes ({\it i.e.}, $k_{z,p}=2\pi p/L_z$ with $0\leq p\leq 159$ and $L_z=$~50 $\upmu$m). The microcavity was red-detuned with respect to the Rhodamine absorption maximum, which is 4.18~eV at the CIS/3-21G//Amber03 level of theory used in this work, such that the energy of the cavity photon at zero incidence angle ($k_0=0$) was $\hslash\omega_0=3.81$ eV, corresponding to a distance of $L_x=$~0.163 $\upmu$m between the mirrors (Figure~\ref{fig:1D_cavity}). To maximize the collective light-matter coupling strength, the transition dipole moments of the Rhodamine molecules were aligned to the vacuum field at the start of the simulation. Both lossless ($\hslash\gamma_{\text{cav}}=$ 0~eV or $\gamma_{\text{cav}}=$ 0~ps$^{-1}$, Equation \ref{eq:decay_tot}) and lossy ( $\hslash\gamma_{\text{cav}}=$ 0.04~eV or $\gamma_{\text{cav}}=$ 66.7~ps$^{-1}$) mirrors were considered such that the lifetime, $\tau_\text{cav}$, of the lossy cavity is comparable to the 2-14~fs lifetimes of metallic Fabry-P\'{e}rot cavities used in experiments of strong coupling with organic molecules.\cite{Schwartz2013,George2015,Rozenman2018} 
With a lifetime of 15~fs and a resonance at 3.81~eV, the Quality-factor, defined as $\text{Q}=\omega_\text{cav}\tau_\text{cav}$, for the lossy cavity would be 87. We note, however, that this Q-factor is artificially high, because to compensate for the overestimated excitation energies at the CIS/3-21G level of {\it{ab initio}} theory (section~\ref{subsubsec:QMvalidation}), we added an offset to the cavity dispersion. Under the rotation wave approximation (RWA) and within the single-excitation subspace, valid under the weak driving conditions that are typically employed in experiments on exciton transport in organic micro-cavities, the absolute energy scale is irrelevant and therefore adding such offset has no direct impact on the dynamics of the molecule-cavity system.

We assessed the robustness of our approach with respect to the ensemble size by investigating the effect of varying the number of molecules $N$. In a first set of simulations the same starting coordinates were used for all Rhodamines, but different initial velocities were selected randomly from a Maxwell-Boltzmann distribution at 300~K. To vary $N$ while keeping the Rabi splitting the same ($\sim$ 325~meV), we scaled the cavity volume by the number of Rhodamine molecules (Equation~\ref{eq:dipole_coupling}), yielding cavity vacuum fields of ${\bf{E}}_y =$ 0.0001~a.u. (0.514~MVcm$^{-1}$) for 256~molecules, ${\bf{E}}_y =$ 0.0000707107~a.u. (0.364~MVcm$^{-1}$) for 512~molecules and ${\bf{E}}_y =$ 0.00005~a.u (0.257 MVcm$^{-1}$) for 1024~molecules. With identical Rabi splitting energies, the polariton dispersion, and hence also the polariton group velocities, defined as $v_\text{gr}^\text{LP}(k_{z,p})=\frac{\partial \omega(k_{z,p})}{\partial k_{z,p}}$, are the same.


To also explore the effect of initial energy disorder, the starting coordinates of Rhodamines were varied in a second set of simulations. For these simulations,  the initial conformations and velocities were taken from the QM/MM equilibrium trajectories. 

To investigate the effect of the Rabi splitting on polariton transport, we performed a third set of simulations with 1024~molecules at multiple cavity vacuum field strengths: ${\bf{E}}_y = $~0.000025, 0.000037 and 0.00005~a.u., yielding Rabi splittings of $\hbar\Omega^\text{Rabi} = $~159, 240, and 325~meV. In these simulations, the same starting coordinates were used for all Rhodamines, but with different initial velocities, selected randomly from a Maxwell-Boltzmann distribution at 300~K. For each Rabi splitting, the simulations were started with the same random velocities.

For systems without disorder, bright states and dark states are easily distinguishable because dark states lack contributions from cavity mode excitations. In contrast, in the presence of disorder due to molecules adopting different conformations in the simulations (and in reality), the cavity modes are smeared out over many states. Nevertheless, because the cavity modes are not distributed evenly over all states, the upper (UP) and lower (LP) polaritonic branches remain visible, even if the number of eigenstates that make up these branches exceeds the number of cavity modes. Therefore, to have sufficient dark states, loosely defined as states with a negligible cavity mode contribution, the ratio between the number of dark ($N_\text{dark}$) and polaritonic states ($N_\text{pol}$) must be larger than zero. While this is the case when there are 1024 ($N_\text{dark}/N_\text{pol} = $~2.7 ), 512 ($N_\text{dark}/N_\text{pol} = $~1.1), or 256 molecules ($N_\text{dark}/N_\text{pol} = $~0.3) in the cavity, the smallest cavity systems with 160 molecules ($N_\text{dark}/N_\text{bright} = $ 0), has no real dark state manifold, which, as we show below, affects the propagation mechanism.

We used a leap-frog Verlet algortihm with a 0.1~fs timestep (0.5~fs, for simulations at diffent Rabi splittings) to perform the integration of the nuclear dynamics, while the unitary propagator in the local diabatic basis was used to propagate the polaritonic degrees of freedom.\cite{Granucci2001} To avoid self-interference of the wavepacket, the simulations were terminated far before a wavepacket could reach the periodic boundary of the 1D cavity at $z_{N+1}=z_1$. All results were obtained as averages of at least two simulations. For all cavity simulations we used Gromacs~4.5.3,\cite{Hess2008} in which the multi-mode Tavis-Cummings QM/MM model was implemented,\cite{Tichauer2021} in combination with Gaussian16.~\citep{g16} 

\subsection{Molecular Dynamics of Tetracene-Cavity Systems}


After equilibration at the QM/MM level 1024~Tetracene molecules in cyclohexane were placed with equal inter-molecular distances on the $z$-axis of a 1D,~\cite{Michetti2005} 50~$\upmu$m long, optical Fabry-P\'erot micro-cavity (Figure~\ref{fig:1D_cavity}). The dispersion of this cavity was modelled  with 160~discrete modes ({\it i.e.}, $k_{z,p}=2\pi p/L_z$ with $0\leq p\leq 159$ and $L_z=$~50 $\upmu$m). The microcavity was red-detuned with respect to the Tetracene absorption maximum, which is 3.87~eV at the CIS/3-21G//Gromos96-54a7 level of theory used in this work, such that the energy of the cavity photon at zero incidence angle ($k_0=0$) was $\hslash\omega_0 =$~3.55~eV. To maximize the collective light-matter coupling strength, the transition dipole moments of Tetracene molecules were aligned to the vacuum field at the start of the simulation. With a vacuum field strength of ${\bf{E}}_y =$ 0.0001~a.u. (0.51~MVcm$^{-1}$) this configuration yields a Rabi splitting of $\hbar\Omega^\text{Rabi}= $~301~meV. Simulations were performed for both lossless ($\hslash\gamma_{\text{cav}}=$ 0~ps$^{-1}$) and a lossy cavities ($\hslash\gamma_{\text{cav}}=$ 66.7~ps$^{-1}$) such that the lifetime, $\tau_\text{cav}$, of the lossy cavity is comparable to the 2-14~fs lifetimes of metallic Fabry-P\'{e}rot cavities used in experiments of strong coupling with organic molecules.\cite{Schwartz2013,George2015,Rozenman2018} .

A leap-frog Verlet algortihm with a 0.1~fs timestep was used to integrate the nuclear dynamics, while the unitary propagator in the local diabatic basis was used to propagate the polaritonic degrees of freedom.\cite{Granucci2001} To avoid self-interference of the wavepacket, the simulation was stopped before a wavepacket could reach the periodic boundary of the 1D cavity at $z_{N+1}=z_1$. Results were obtained as averages of three independent simulations that were performed with Gromacs~4.5.3,\cite{Hess2008} in which the multi-mode Tavis-Cummings QM/MM model was implemented,\cite{Tichauer2021} in combination with Gaussian16.~\citep{g16} 

\subsection{Molecular Dynamics of Methylene-Blue Cavity System}

After equilibration at the QM/MM level 1024~Methylene Blue molecules in solution were placed with equal inter-molecular distances on the $z$-axis of a 1D,~\cite{Michetti2005} 50~$\upmu$m long, optical Fabry-P\'erot micro-cavity (Figure~\ref{fig:1D_cavity}). The dispersion of this cavity was modelled  with 160~discrete modes ({\it i.e.}, $k_{z,p}=2\pi p/L_z$ with $0\leq p\leq 159$ and $L_z=$~50 $\upmu$m). The microcavity was red-detuned with respect to the Methylene Blue absorption maximum, which is 2.5~eV at the TD-B97/3-21G//Amber03 level of theory used in this work, such that the energy of the cavity photon at zero incidence angle ($k_0=0$) was $\hslash\omega_0 =$~1.9~eV. To maximize the collective light-matter coupling strength, the transition dipole moments of the Methylene Blue molecules were aligned to the vacuum field at the start of the simulation. With a vacuum field strength of ${\bf{E}}_y =$ 0.000025 a.u. (0.128~MVcm$^{-1}$) this configuration yields a Rabi splitting of $\hbar\Omega^\text{Rabi}= $~175~meV. Because the purpose of this additional simulation was to confirm the general transport mechanism, rather than explore the effect of molecular or cavity parameters, we only considered an ideal lossless cavity ($\hslash\gamma_{\text{cav}}=$ 0~eV or $\gamma_{\text{cav}}=$ 0~ps$^{-1}$, Equation \ref{eq:decay_tot}).

A leap-frog Verlet algortihm with a 0.5~fs timestep was used to integrate the nuclear dynamics, while the unitary propagator in the local diabatic basis was used to propagate the polaritonic degrees of freedom.\cite{Granucci2001} To avoid self-interference of the wave packet, the simulation was ended before a wavepacket could reach the periodic boundary of the 1D cavity at $z_{N+1}=z_1$. The simulation was performed with Gromacs~4.5.3,\cite{Hess2008} in which the multi-mode Tavis-Cummings QM/MM model was implemented,\cite{Tichauer2021} in combination with Gaussian16.~\citep{g16} 

\subsection{Initial Conditions}

\subsubsection{On-resonant Excitation}

Following Agranovich and Garstein,~\cite{Agranovich2007} on-resonant excitation of a wavepacket of LP states was achieved by assigning to the expansion coefficients $c_m(t=0)$ values of a Gaussian distribution centered at $k$-vector $k_c$~= $55\frac{2\pi}{L_z}$ = 6.91 $\upmu$m$^{-1}$ corresponding to the maximum group velocity of the LP:
\begin{equation}
    c_m(0) = \left(\frac{\beta}{2\pi^3}\right)^{\frac{1}{4}}\exp[-\beta(k_{z}^m-k_c)^2]\label{eq:wavepacket_excitation}
\end{equation}
where $\beta = 10^{-12}$ $\text{m}^2$ is a coefficient characterising the shape of the wavepacket and $k_{z,m}$ is the in-plane momentum of polariton $|\psi_m\rangle$, given by: 
\begin{equation}
    \langle k_{z}^m \rangle = \frac{\sum_p^{n_\text{mode}} |\alpha_p^m|^2 k_{z,p} }{ \sum_p^{n_\text{mode}}|\alpha_p^m|^2}
\end{equation}
with $k_{z,p} = 2\pi p/L_z$ the discrete wavevector in a periodic 1D cavity of length $L_z$.

\subsubsection{Off-resonant Excitation}

In experiments off-resonant excitation is often realized by pumping a higher-energy electronic excited state of a single molecule. According to Kasha's rule, this molecule then rapidly relaxes into the lowest energy electronic excited state (S$_1$), which is resonant with the cavity. To avoid the computationally highly demanding description of this ultra-fast initial relaxation process, we started our simulations in the S$_1$ electronic state of one of the molecules. Because we performed our simulations in the adiabatic representation, in which the polaritonic wave functions are delocalized over all molecules, we took a linear combination of adiabatic states that localizes the excitation onto that molecule as initial conditions. To find the expansion coefficients for such localisation, we used the completeness relation to transform the adiabatic eigenstates of the Tavis-Cummings Hamiltonian ($|\psi^\alpha\rangle$, Equation~\ref{eq:Npolariton}) back into the basis of molecular excitations and cavity mode Fock states ({\it{i.e.}}, $|\phi_j\rangle=\hat{\sigma}^+_j
|\phi_0\rangle$ for $j\le N$ and $|\phi_j\rangle=\hat{a}^\dagger_j|\phi_0\rangle$ for $i>N$):
\begin{equation}
\begin{array}{ccl}
   \langle \psi^\alpha | \hat{H}^{\text{TC}}|\psi^\beta\rangle &=& \sum_i\sum_j\langle \psi^\alpha|\phi_i\rangle\langle \phi_i|\hat{H}^{\text{TC}}|\phi_j\rangle \langle \phi_j|\psi^\beta\rangle\\
   \\
   &=& \sum_i\sum_j U_{\beta i}^{\dagger}\langle \phi_i|\hat{H}^{\text{TC}}|\phi_j\rangle U_{j\alpha}
   \end{array}
   \label{eq:unitary_transformation}
\end{equation}
where ${\bf{H}}^\text{TC}$ is the matrix representation of the Tavis-Cummings Hamiltonian in Equation~\ref{eq:dTCH} with elements in Equations~\ref{eq:molecular_diagonal}, \ref{eq:photonic_diagonal} and \ref{eq:QMMMdipole_coupling}, and
{\bf{U}} is the unitary matrix that diagonalizes ${\bf{H}}^\text{TC}$ at the start of the simulation. Thus, the expansion coefficients for the linear combination of adiabatic states (Equation~\ref{eq:totalwf}) that initially localizes the single-photon excitation onto molecule $j$ are $c_\alpha(0) = U_{\alpha j}^\dagger$, {\it{i.e.}}:
\begin{equation}
   |\Psi(0)\rangle^{(\text{loc. on }j\text{)}} =  \sum_\alpha^{N+n_\text{mode}}c_\alpha(0)|\psi^\alpha\rangle =\sum_\alpha^{N+n_\text{mode}}U_{\alpha j}^{\dagger}|\psi^\alpha\rangle
   \label{eq:WF_in_diabatic_basis}
\end{equation}

\newpage

\section{Simulation Analysis}

\subsection{Contributions of Molecular and Cavity Mode Excitations to the Wavepacket}
\label{sec:mol-pho_parts}

The total contribution of the molecular excitations ($|\Psi_\text{exc}|^2$) to the wavepacket (black curve in the fourth column of Figures~\ref{fig:wps_off_decay0}, \ref{fig:wps_off}, \ref{fig:wps_on} and \ref{fig:wps_on_decay0} and panels {\bf d} and {\bf h} of Figures 2 and 4 of the main article), was obtained by summing the projections of the $N$ excitonic basis states ({\it i.e.}, $|\phi_j\rangle=\hat{\sigma}_j^+|\phi_0\rangle$
for $1\le j\le N$, Equation~\ref{eq:phi0}) onto |$\Psi(t)\rangle$ (Equation~\ref{eq:totalwf}) at each time-step of the simulation:
\begin{equation}
\begin{array}{ccl}
    |\Psi_\text{exc}(t)|^2 &=&\sum_{j=1}^N\left|\langle\phi_0|\hat{\sigma}_j|\Psi(t)\rangle\right|^2\\
    \\
    &=&\sum_{j=1}^N\left|\sum_m^{N+n_\text{mode}} c_m(t)\beta^m_j\right|^2 
\end{array}
\end{equation}
where $N+n_\text{mode}$ is the number of polaritonic states. 

Likewise, the total contribution of the cavity mode excitations ($|\Psi_\text{pho}|^2$) to the wavepacket (red curve in the fourth column of Figures~\ref{fig:wps_off_decay0}, \ref{fig:wps_off}, \ref{fig:wps_on} and \ref{fig:wps_on_decay0} and panels {\bf d} and {\bf h} of Figures 2 and 4 of the main article) was obtained by summing the projections of the $n_\text{mode}$ cavity mode excitation basis states ({\it i.e.}, $|\phi_p\rangle=\hat{a}_p^\dagger|\phi_0\rangle$
for $0\le p <  n_\text{mode}$) onto $|\Psi(t)\rangle$ (Equation~\ref{eq:totalwf}) at each time step of the simulation:
\begin{equation}
\begin{array}{ccl}
     |\Psi_\text{pho}(t)|^2 &=&\sum_{p=0}^{n_\text{mode}-1}\left|\langle\phi_0|\hat{a}_p|\Psi(t)\rangle\right|^2\\
     \\
     &=&\sum_{p=0}^{n_{\text{mode}}-1}\left|\sum_m^{N+n_\text{mode}} c_m(t)\alpha^m_p\right|^2
\end{array}
\end{equation}

\subsection{Monitoring Wavepacket Dynamics}
\label{sec:analysis-wps}

To monitor the propagation of the wavepacket, we plotted the probability density of the total time-dependent wave function $|\Psi(t)|^2$ 
at the positions of the molecules,  $z_j$, as a function of time (first column of Figures~\ref{fig:wps_off_decay0}, \ref{fig:wps_off}, \ref{fig:wps_on} and \ref{fig:wps_on_decay0} and panels {\bf a} and {\bf e} of Figures 2 and 4 in the main text). We thus represent the density as a \emph{discrete} distribution at grid points that correspond to the molecular positions, rather than a continuous distribution. In addition to the total probability density, $|\Psi(t)|^2$ , we also plotted the probability densities of the molecular $|\Psi_{\text{mol}}(t)|^2$ and photonic $|\Psi_{\text{pho}}(t)|^2$ contributions separately (second and third columns of Figures~\ref{fig:wps_off_decay0}, \ref{fig:wps_off}, \ref{fig:wps_on} and \ref{fig:wps_on_decay0} and panels {\bf b}, {\bf f} and {\bf c}, {\bf g}, respectively, of Figures 2 and 4 in the main text). 

The amplitude of $|\Psi_{\text{exc}}(t)\rangle$ at position $z_j$ in the 1D cavity (with $z_j=(j-1)L_z/N$ for $1\le j \le N$) is obtained by projecting the excitonic basis state in which molecule $j$ at position $z_j$ is excited ($|\phi_j\rangle=\hat{\sigma}_j^+|\phi_0\rangle$) to the total wave function (Equation~\ref{eq:totalwf}):
\begin{equation}
\begin{array}{ccl}
    |\Psi^{\text{exc}}(z_j,t)\rangle&=&\hat{\sigma}_j^+|\phi_0\rangle
        \langle\phi_0|\hat{\sigma}_j|\Psi(t)\rangle\\
    \\
    &=& \sum_m^{N+n_\text{mode}}c_m(t)\beta_j^m\hat{\sigma}_j^+|\phi_0\rangle
   \end{array}\label{eq:molecularpart}
\end{equation}
where the $\beta_j^m$ are the expansion coefficients of the excitonic basis states in polaritonic state $|\psi^m\rangle$ (Equation~\ref{eq:Npolariton}) and $c_m(t)$ the time-dependent expansion coefficients of the total wavefunction $|\Psi(t)\rangle$ (Equation~\ref{eq:totalwf}).

The cavity mode excitations are described as plane waves that are delocalized in real space. We therefore obtain the amplitude of the cavity modes in polaritonic eigenstate $|\psi^m\rangle$ at position $z_j$ by Fourier transforming the projection of the cavity mode excitation basis states, in which cavity mode $p$ is excited ($|\phi_p\rangle=\hat{a}_{p}^\dagger|\phi_0\rangle$):
\begin{equation}
\begin{array}{ccl}
    |\psi^m_\text{pho}(z_j)\rangle&=&\mathcal{FT}^{-1}\left[\sum_p^{n_\text{mode}}\hat{a}_p^\dagger|\phi_0\rangle\langle\phi_0|\hat{a}_p|
    \psi^m\rangle\right]\\
    \\
    &=&\frac{1}{\sqrt N}\sum_p^{n_{\text{mode}}} \alpha_p^me^{i2\pi z_j p} \hat{a}_p^\dagger|\phi_0\rangle
    \end{array}\label{eq:photonicpart1}
\end{equation}
where the $\alpha_p^m$ are the expansion coefficients of the cavity mode excitations in polaritonic state $|\psi^m\rangle$ (Equation~\ref{eq:Npolariton}) and we normalize by $1/\sqrt{N}$ rather than $1/\sqrt{L_z}$, as we represent the density on the grid of molecular positions. The total contribution of the cavity mode excitations to the wavepacket at position $z_j$ at time $t$ is then obtained as the weighted sum over the Fourier transforms:
\begin{equation}
\begin{array}{ccl}
|\Psi^\text{pho}(z_j,t)\rangle &=& \sum_m^{N+n_\text{mode}}c_m(t)\times\mathcal{FT}^{-1}\left[\sum_p^{n_\text{mode}}\hat{a}_p^\dagger|\phi_0\rangle\langle\phi_0|\hat{a}_p|\psi^m\rangle\right]\\
\\
&=& \sum_m^{N+n_{\text{mode}}}c_m(t)\frac{1}{\sqrt N}\sum_p^{n_{\text{mode}}} \alpha_p^me^{i2\pi z_jp}\hat{a}_p^\dagger|\phi_0\rangle 
\end{array}\label{eq:photonicpart2}
\end{equation}
with $c_m(t)$ the time-dependent expansion coefficients of the adiabatic polaritonic states $|\psi^m\rangle$ in the total wavefunction $|\Psi(t)\rangle$ (Equation~\ref{eq:totalwf}).

\subsection{Average Velocity}
\label{sec:ave}

To determine the average velocity, $v_{\text{av}}$, of the total wave packet, we first computed the expectation value of the position as a function of time:
\begin{equation}
    \langle z(t)\rangle = \frac{\langle\Psi(t)|\hat{z}|\Psi(t)\rangle}{\langle\Psi(t)|\Psi(t)\rangle}.
\end{equation}
Then, $v_{\text{av}}$ was obtained from the slope of a linear fit to $\langle z(t)\rangle$.

\subsection{Photo-Absorption Spectra}
\label{sec:absorption}

Following Lidzey and coworkers,\cite{Lidzey2000} we define the "visibility", $I^m$, of  polaritonic state $\psi^m$ as the total photonic contribution to that state ({\it{i.e.}}, $I^m\propto \sum_p^{n_\text{max}}|\alpha_p^m|^2$). Thus, the angle-resolved, or wave vector-dependent, one-photon absorption spectra of the Rhodamine cavity systems were computed from the QM/MM trajectory of the uncoupled Rhodamine as follows: For each frame of this trajectory, the polaritonic states were computed and the energy gaps of these states with respect to the overall ground state ({\it{i.e.}}, $E^0$, with all molecules in S$_0$, no photon in the cavity, Equation~\ref{eq:phi0}), were extracted for all cavity modes, $p$, and summed up into a superposition of Gaussian functions, weighted by $|\alpha_p^m|^2$:
\begin{equation}
  I^{\text{abs}}(E,k_z) \propto \sum_t^s\left[ \sum_m^{N+n_\text{max}+1}|\alpha_{p,t}^m|^2\exp[-\frac{(E-\Delta
    E_t^m)^2}{2\sigma^2}]\right]\label{eq:abs_mol}
\end{equation}
Here, $I^{\text{abs}}(E,k_z)$ is the absorption intensity as a function of excitation energy $E$ and in-plane momentum $k_z = 2\pi p/L_z$, $s$ the number of trajectory frames included in the analysis, $\Delta E_t^m$ the excitation energy of polaritonic state $m$ in frame $t$ ({\it{i.e.}}, $\Delta E^m_t = E^m_t-E^0_t$) and $\alpha_{p,t}^m$ the expansion coefficient of cavity mode $p$ in polaritonic state $m$ in that frame (equation~\ref{eq:Npolariton}). A width of $\sigma = 0.05$~eV was chosen for the convolution.

\subsection{Photo-Luminescence Spectra}
\label{sec:mask}

As in Berghuis {\it{et al}}.,\cite{Berghuis2022} we analysed the photo-luminescence of the propagating wavepacket after non-resonant excitation through a (virtual) mask placed at various distances from the molecule that was excited initially. This analysis thus mimics the Fourier Transform microscopy technique that is used experimentally to determine how different polaritonic states contribute to the propagation.\cite{Berghuis2022} In such imaging experiments, the photoluminescence from a specific area of the sample is collected through a pinhole and converted into momentum space. In the analysis of our simulations we modeled this pinhole, or mask, as a rectangular function of width $\Delta z = $~10~$\upmu$m centered at $z = $~10 and 20~$\upmu$m. At each time-step $t$, the cavity photoluminescence at the position of the pinhole was collected as follows:

First, the excitonic (Equation~\ref{eq:molecularpart}) and cavity mode contributions (Equation~\ref{eq:photonicpart2}) to the wave function (Equation~\ref{eq:totalwf}) were convoluted with the rectangular mask function:
\begin{align}
|\Psi_\text{hole}^{\text{exc}}(z,t)\rangle&=\sum_{z_j \ge z-\Delta z/2}^{z_j\le z+\Delta z/2} |\Psi^{\text{exc}}(z_j,t) \rangle \\
\\
|\Psi_\text{hole}^{\text{pho}}(z,t)\rangle&=\sum_{z_j \ge z-\Delta z/2}^{z_j\le z+\Delta z/2}|\Psi^{\text{pho}}(z_j,t) \rangle
\end{align}
Second, $|\Psi_\text{hole}^{\text{pho}}(z,t)\rangle$ was  Fourier transformed from real space into $k_{z}$ space:
\begin{equation}
    |\Psi_\text{hole}^{\text{pho}}(k_{z},t)\rangle=\mathcal{FT}\left[\Psi^{\text{pho}}_{\text{hole}}(z,t)\right]
\end{equation}
such that the total wave function under the mask can be written as  $|\Psi_\text{hole}(t)\rangle=|\Psi_\text{hole}^{\text{exc}}(z,t)\rangle\otimes|\Psi_\text{hole}^{\text{pho}}(k_{z},t)\rangle$. Since the polaritonic states $\{|\psi^m\rangle\}$ form a complete basis (Equation~\ref{eq:totalwf}), the part of the wavepacket under the mask at time $t$ can be expanded as a linear combination of these eigenstates:
\begin{equation}
    |\Psi_\text{hole}(t)\rangle=\sum_m^{N+n_\text{mode}} d_m(t)|\psi^m\rangle
\end{equation}
To obtain the $N+n_\text{mode}$ expansion coefficients, $d_m(t)$, we computed the overlap between the polaritonic eigenstates and the fraction of the wavepacket visible through the pinhhole:
\begin{equation}
    d_m(t) = \langle \psi^m|\Psi_\text{hole}(z,t)\rangle
\end{equation}
Following Lidzey and co-workers,\cite{Lidzey2000} we approximated the photoluminescence detected through the pinhole at time $t$ as the ``visibility'' of the polaritonic states weighted by their expansion coefficients under the mask, $d_m(t)$:\cite{Tichauer2021}
\begin{equation}
    I_{\text{hole}}(E,p,z,t)\propto\sum_m^{N+n_\text{mode}}|d_m(t)|^2|\alpha^m_p|^2\exp\left[-\frac{(E-\Delta E^m)^2}{2\sigma^2}\right]
\end{equation}
with $I_{\text{hole}}(E,p,z,t)$ the photoluminescence intensity at time $t$ as a function of excitation energy $E$, in-plane momentum $p$ ($k_{z,p}=2\pi p/L_z$) and position of the mask $z$; $\Delta E^m$ the excitation energy of polaritonic state $|\psi^m\rangle$ at that time frame ($\Delta E^m = E^m-E^0$, with $E^0$ the energy of the ground state, $|\phi_0\rangle$, Equation~\ref{eq:phi0}); and $\sigma = 0.05$~eV the convolution width. The visibility through the pinhole, $I_\text{hole}(E,p,z)$, was accumulated over 100fs:
\begin{equation}
    I_\text{hole}(E,p,z)=\sum_t^sI_{\text{hole}}(E,p,z,t)
\end{equation}
where $s$ is the number of time steps 
(1000 used for the pinhole analysis of simulations under off-resonant excitation conditions with no photon losses in this work).

\clearpage

\section{Additional Results}

\subsection{Polariton Transport for Different Numbers of Molecules $N$}

Because the number of molecules we can include in our simulations is necessarily much smaller than in experiments, we investigated how that number affects energy transport by performing simulations with 160, 256, 368, 512, 768 and 1024~molecules. To keep the Rabi splitting ($\sim$~325~meV) constant, and hence polariton dispersion the same, we scaled the cavity mode volume by the number of molecules $N$ (see Section~\ref{subsec:sim_details} for details).

\subsubsection{On-resonant Excitation}

\begin{figure}[!htb]
\centerline{
\epsfig{figure=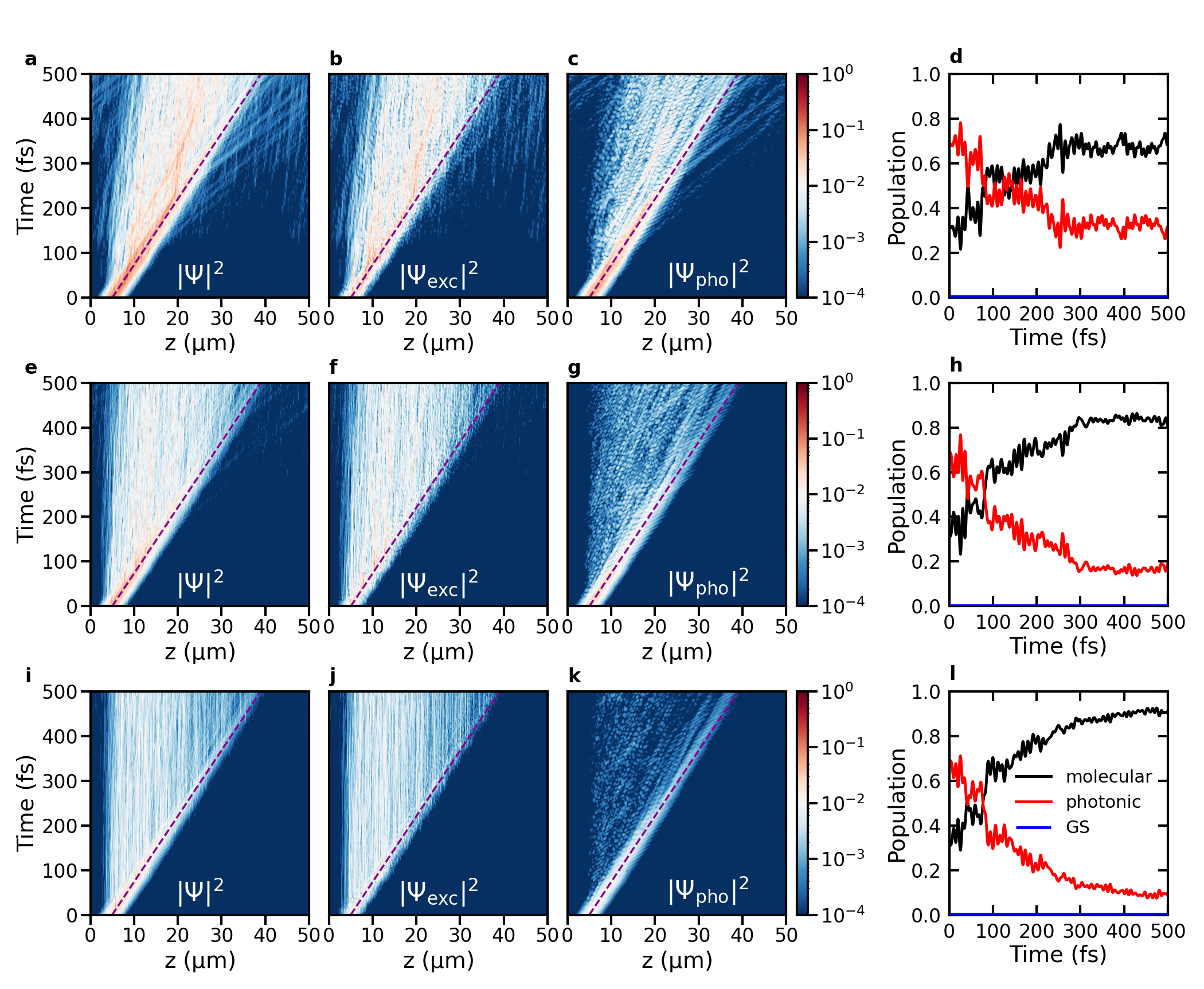,width=15cm,angle=0}
}
   \caption{Polariton propagation in ensembles of 256, 512 and 1024~molecules (first, second and third row, respectively) strongly coupled to a {\bf{lossless cavity}} ({\it i.e.}, $\gamma_{\text{cav}} = $~0~ps$^{-1}$), after {\bf on-resonant} excitation of a wavepacket of LP states centered at the $k_z$-vector corresponding to the maximum LP group velocity. First, second and third columns: total probability density, $|\Psi(t)|^2$, probability density of the molecular excitons, $|\Psi_{\text{exc}}(t)|^2$, and of the cavity mode excitations, $|\Psi_{\text{pho}}(t)|^2$, respectively, as a function of distance (horizonal axis) and time (vertical axis). The magenta dashed line indicates propagation at the maximum group velocity of the LP (68~$\mu$mps$^{-1}$). Fourth column: Contributions of the  molecular excitons (black) and cavity mode excitations (red) to $|\Psi(t)|^2$ as a function of time. Without cavity decay, the population of the ground state (GS, blue) remains zero.}
  \label{fig:wps_on_decay0}
\end{figure}

\begin{figure}[!htb]
\centerline{
\epsfig{figure=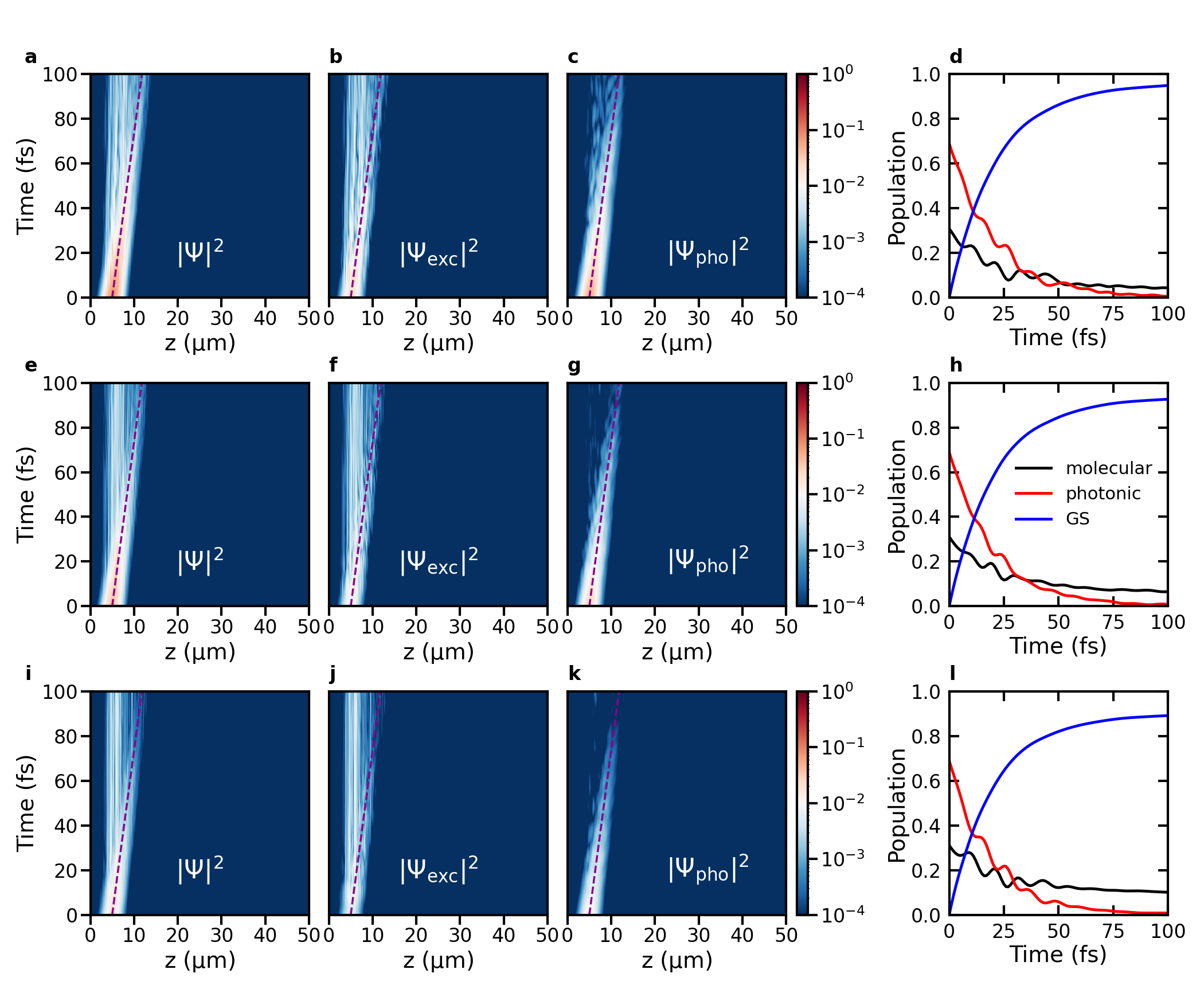,width=15cm,angle=0}
}
  \caption{Polariton propagation in an ensemble of 256, 512 and 1024~molecules (first, second and third row, respectively) strongly coupled to a {\bf{lossy cavity}} ({\it i.e.}, $\gamma_{\text{cav}} = $~66.7~ps$^{-1}$), after {\bf on-resonant} excitation of a wavepacket of LP states centered at the $k_z$-vector corresponding to the maximum LP group velocity. First, second and third columns: total probability density, $|\Psi(t)|^2$, probability density of the molecular excitons, $|\Psi_{\text{exc}}(t)|^2$, and of the cavity mode excitations, $|\Psi_{\text{pho}}(t)|^2$, respectively, as a function of distance (horizontal axis) and time (vertical axis). The magenta dashed line indicates propagation at the maximum group velocity of the LP (68~$\mu$mps$^{-1}$). Fourth column: Contributions of molecular excitons (black) and cavity mode excitations (red) to $|\Psi(t)|^2$ and population of the ground state (GS, blue) as a function of time.}
  \label{fig:wps_on}
\end{figure}

\begin{SCfigure}
 \includegraphics[width=0.65\textwidth]{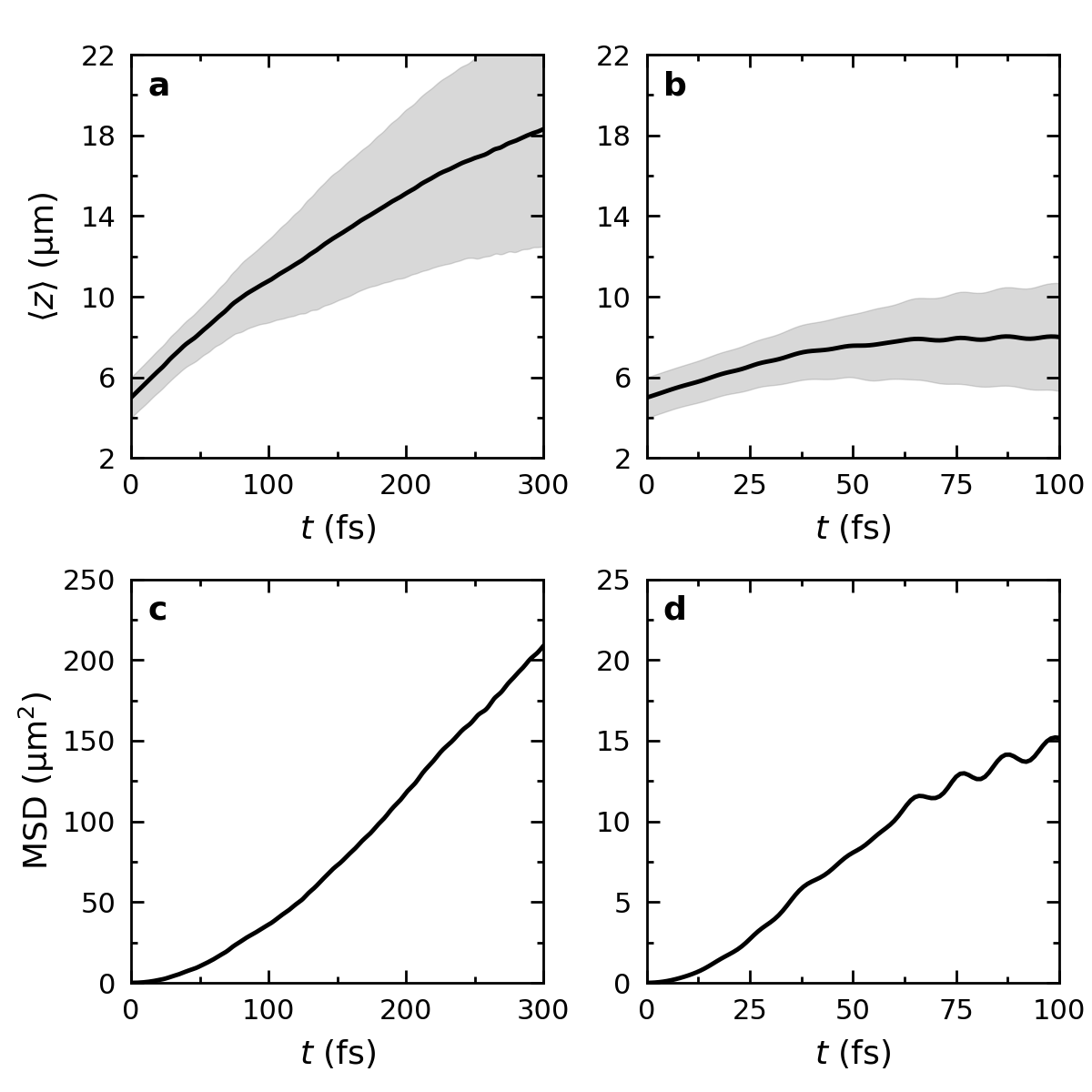}
  \caption{Expectation value of the position ($\langle z\rangle$) (top panels) and mean square displacement (MSD) (bottom panels) of the total time-dependent wavefunction $|\Psi(t)\rangle$ of {\bf{256 Rhodamine chromophores}} strongly coupled to an ideal cavity ({\it i.e.}, $\gamma_\text{cav}=0$~ps$^{-1}$, panels {\bf a} and {\bf c}) and a lossy cavity with $\gamma_\text{cav}=66.7$~ps$^{-1}$ (panels {\bf b} and {\bf d}), under {\bf on-resonant} excitation conditions. In the top panels, the black lines represents  $\langle z\rangle =  \langle\Psi(t)| \hat{z}|\Psi(t)\rangle/\langle\Psi(t)| \Psi(t)\rangle$ while the shaded area indicates the root mean squared deviation $\sqrt{\langle (z(t)-\langle z(t)\rangle)^2\rangle}$.} 
  \label{fig:z_256_on}
\end{SCfigure}

\begin{SCfigure}
\includegraphics[width=0.65\textwidth]{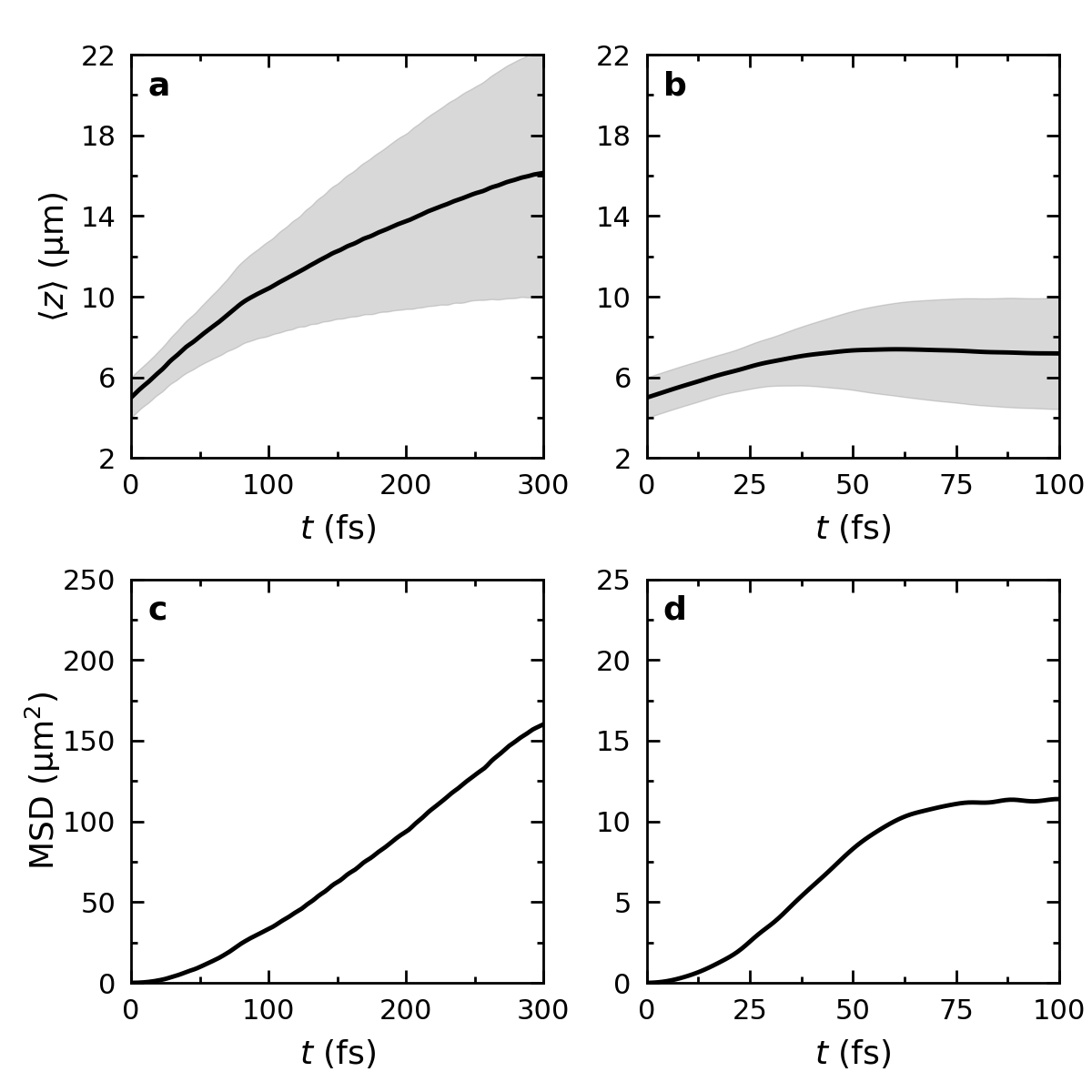}
  \caption{Expectation value of the position ($\langle z\rangle$) (top panels) and mean square displacement (MSD) (bottom panels) of the total time-dependent wavefunction $|\Psi(t)\rangle$ of {\bf{512 Rhodamine chromophores}} strongly coupled to an ideal cavity ({\it i.e.}, $\gamma_\text{cav}=0$~ps$^{-1}$, panels {\bf a} and {\bf c}) and a lossy cavity with $\gamma_\text{cav}=66.7$~ps$^{-1}$ (panels {\bf b} and {\bf d}), under {\bf on-resonant} excitation conditions. In the top panels, the black lines represents  $\langle z\rangle =  \langle\Psi(t)| \hat{z}|\Psi(t)\rangle/\langle\Psi(t)| \Psi(t)\rangle$ while the shaded area indicates the root mean squared deviation $\sqrt{\langle (z(t)-\langle z(t)\rangle)^2\rangle}$.}
    \label{fig:z_512_on}
\end{SCfigure}

In Figure~\ref{fig:wps_on_decay0} we show the time-evolution of the probability density of the polaritonic wave function, $|\Psi(t)|^2$, after resonant excitation of a Gaussian wavepacket of LP states with a broad-band laser pulse in a perfect lossless cavity ($\gamma=$~0~ps$^{-1}$, Equation~\ref{eq:decay_tot}) containing 256, 512 and 1024~Rhodamine molecules. In Figure~\ref{fig:wps_on} we show the time evolution of the probability density of $|\Psi(t)|^2$ after resonant excitation in a lossy cavity ($\gamma=$~66.7~ps$^{-1}$ for all cavity modes) containing 256, 512 and 1024~Rhodamine molecules. In Figures~\ref{fig:z_256_on} and \ref{fig:z_512_on} we plot the expectation value of the position, $\langle z\rangle$, and mean square displacement (MSD) of the total time-dependent wavefunction, $|\Psi(t)\rangle$, for $N=$~256 and 512 molecules under on-resonant excitation conditions in the lossless and lossy cavities, respectively. Plots of $\langle z\rangle$ and MSD for 1024 molecules are discussed in the main text (Figure 3).

The underlying polariton-mediated transport mechanism is the same for all $N$: Initially the propagation is ballistic and the wavepacket moves with the central group velocity until the contribution of the molecular excitons to the total wavefunction (Equation~\ref{eq:totalwf}) exceeds the contributions of the cavity mode excitations. Then, the propagation continues as a diffusion process, as indicated by the change in the slope of $\langle z\rangle$ and the change from a quadratic to a linear time-dependence in the Mean Square Displacement (Figures~\ref{fig:z_256_on} and \ref{fig:z_512_on}).

When  photon loss is included (Figure~\ref{fig:wps_on}), population builds up in the ground state (blue curves) at the expense of states with cavity mode excitations (red). Because the number of dark states increases with $N$, and the dark states "protect" against decay, the total decay decreases with $N$, albeit that for the ensemble sizes simulated here, the effect is rather small.

\clearpage

\subsubsection{Off-resonant Excitation}

\begin{figure}[!htb]
\centerline{
\epsfig{figure=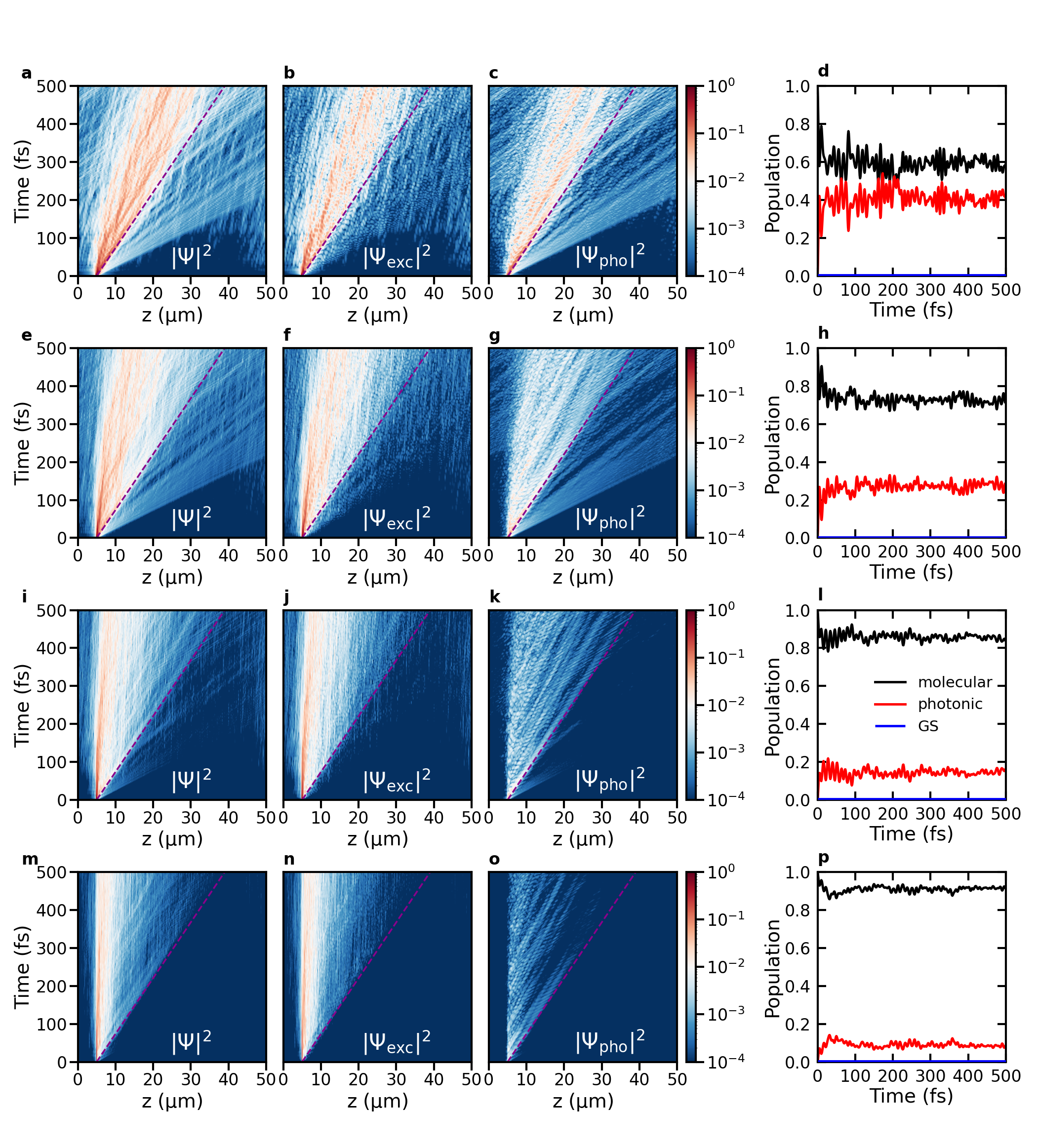,width=15cm,angle=0}
}
   \caption{Polariton propagation after {\bf off-resonant} excitation into the S$_1$ state of a single molecule located at 5~$\upmu$m in an ensemble of 160, 256, 512 and 1024~molecules (first, second, third and fourth row, respectively) strongly coupled to 160~confined light modes of a 1D Fabry-P\'{e}rot optical cavity with {\bf perfect mirrors} ({\it i.e.}, $\gamma_{\text{cav}} =$~0 ps$^{-1}$). First, second and third column: total probability density, $|\Psi(t)|^2$, probability density of the molecular excitons, $|\Psi_{\text{exc}}(t)|^2$, and of the cavity mode excitations, $|\Psi_{\text{pho}}(t)|^2$, respectively, as a function of distance (horizontal axis) and time (vertical axis). The magenta dashed line indicates propagation at the maximum group velocity of the LP (68~$\mu$mps$^{-1}$). Fourth column: Contributions of the  molecular excitons (black) and cavity mode excitations (red) to $|\Psi(t)|^2$ as a function of time. Without cavity decay, the population of the ground state (GS, blue) remains zero.}
  \label{fig:wps_off_decay0}
\end{figure}

\begin{figure}[!htb]
\centerline{
\epsfig{figure=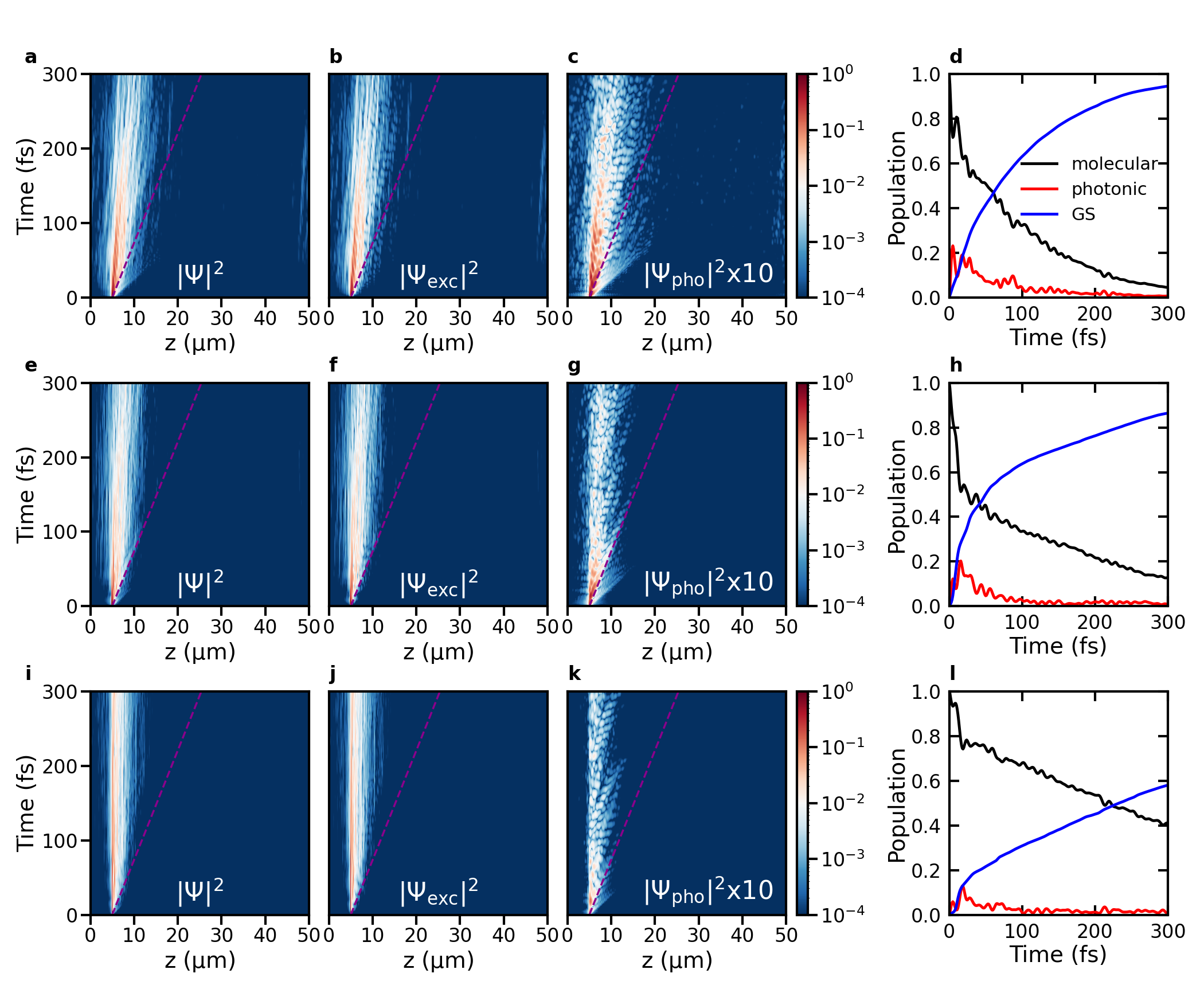,width=15cm,angle=0}
}
   \caption{Polariton propagation after {\bf off-resonant} excitation into the S$_1$ state of a single molecule located at 5~$\upmu$m in an ensemble of 256, 512 and 1024~molecules (first, second and third row, respectively) strongly coupled to 160~confined light modes of a 1D Fabry-P\'{e}rot optical cavity with {\bf lossy mirrors} ({\it i.e.}, $\gamma_{\text{cav}} =$~66.7 ps$^{-1}$). First, second and third column: total probability density, $|\Psi(t)|^2$, probability density of the molecular excitons, $|\Psi_{\text{exc}}(t)|^2$, and of the cavity mode excitations, $|\Psi_{\text{pho}}(t)|^2$, respectively, as a function of distance (horizontal axis) and time (vertical axis). The magenta dashed line indicates propagation at the maximum group velocity of the LP (68~$\mu$mps$^{-1}$). Fourth column: Contributions of molecular excitons (black) and cavity mode excitations (red) to $|\Psi(t)|^2$ and population of the ground state (GS, blue) as a function of time.}
  \label{fig:wps_off}
\end{figure}

In Figure~\ref{fig:wps_off_decay0} we show the propagation of the wavepacket after off-resonant excitation in a loss-less cavity ($\gamma=$~0~ps$^{-1}$, Equation~\ref{eq:decay_tot}) with 160, 256, 512 and 1024~Rhodamine molecules. In Figure~\ref{fig:wps_off} we show the propagation of the wavepacket after off-resonant excitation in a lossy cavity ($\gamma=$~66.7~ps$^{-1}$ for all cavity modes) with 256, 512 and 1024~Rhodamine molecules.

\begin{figure*}[!htb]
\centerline{
\epsfig{figure=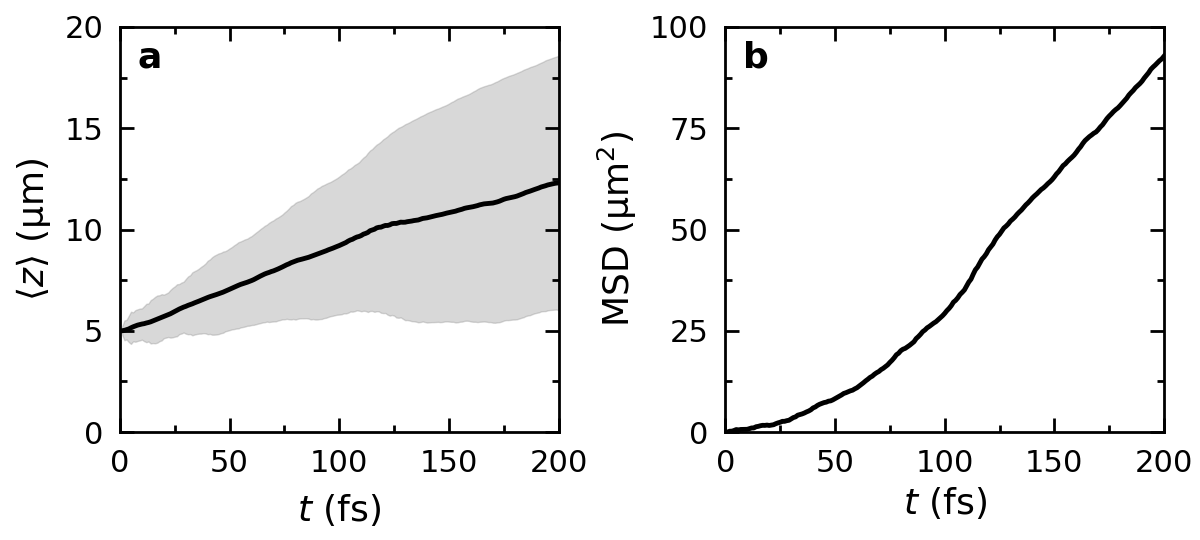,width=12cm,angle=0}
 }
  \caption{Expectation value of the position ($\langle z\rangle$) ({\bf a}) and mean square displacement (MSD) ({\bf b}) of the total time-dependent wavefunction $|\Psi(t)\rangle$ of {\bf{160~molecules}} strongly coupled to a {\bf{perfect}} Fabry-P\'{e}rot cavity ({\it i.e.}, $\gamma_{\text{cav}} =$~0~ps$^{-1}$), after {\bf off-resonant} excitation into the S$_1$ state of a single molecule located at 5~$\upmu$m. In panel {\bf a}, the black lines represents  $\langle z\rangle =  \langle\Psi(t)| \hat{z}|\Psi(t)\rangle/\langle\Psi(t)| \Psi(t)\rangle$ while the shaded area shows the root mean squared deviation $\sqrt{\langle (z(t)-\langle z(t)\rangle)^2\rangle}$.}
  \label{fig:z_160_off}
\end{figure*}

\begin{SCfigure}
 \includegraphics[width=0.67\textwidth]{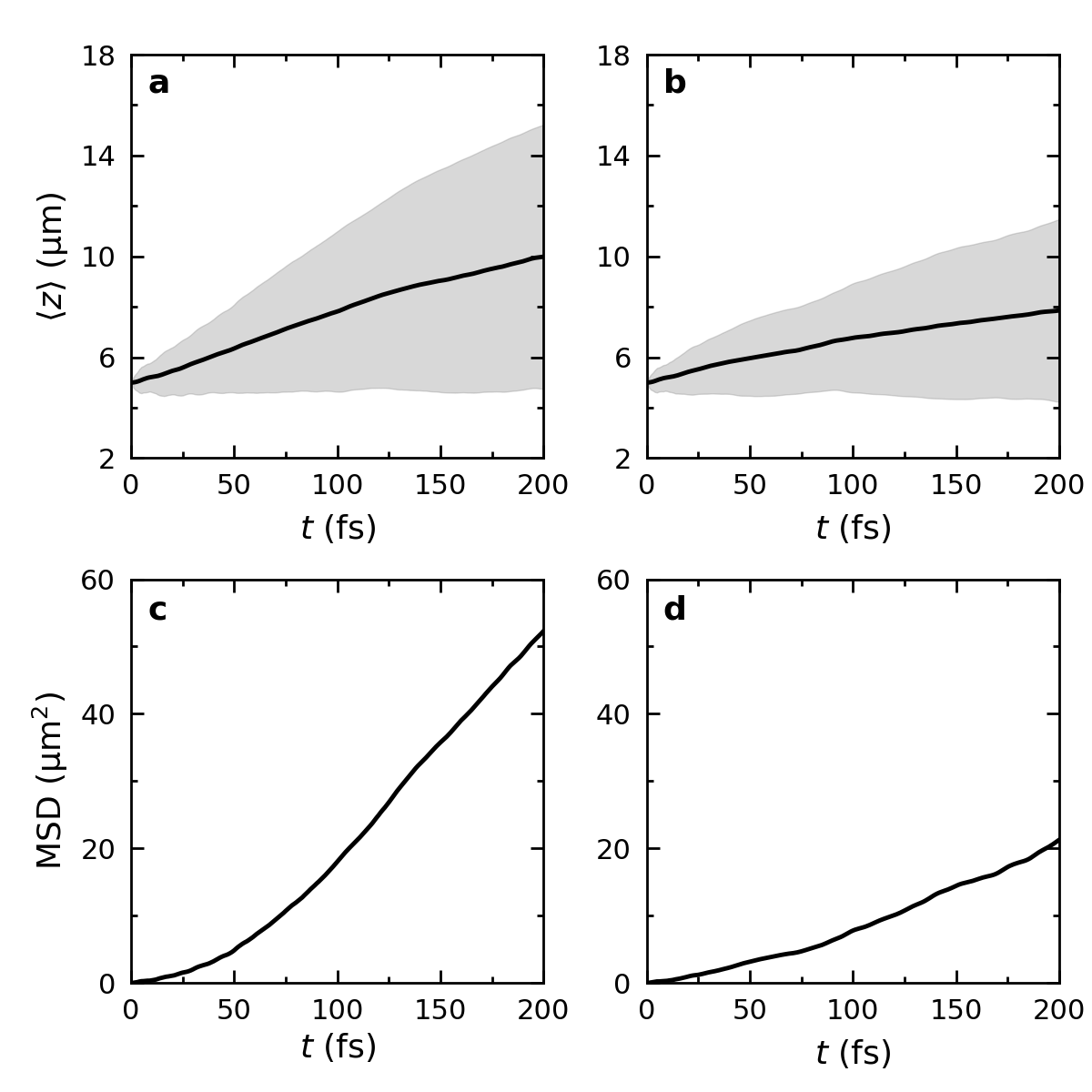}
  \caption{Expectation value of the position ($\langle z\rangle$) (top panels) and mean square displacement (MSD) (bottom panels) of the total time-dependent wavefunction $|\Psi(t)\rangle$ of {\bf{256~Rhodamine molecules}} strongly coupled to an {\bf{ideal cavity}} {\it i.e.}, $\gamma_\text{cav}=0$~ps$^{-1}$ (panels {\bf a} and {\bf c}) and a {\bf{lossy cavity}} with $\gamma_\text{cav}=66.7$~ps$^{-1}$ (panels {\bf b} and {\bf d}), after {\bf off-resonant} excitation of a single molecule. In the top panels, the black lines represents  $\langle z\rangle =  \langle\Psi(t)| \hat{z}|\Psi(t)\rangle/\langle\Psi(t)| \Psi(t)\rangle$ while the shaded area shows the root mean squared deviation $\sqrt{\langle (z(t)-\langle z(t)\rangle)^2\rangle}$.}
  \label{fig:z_256_off}
\end{SCfigure}

\begin{SCfigure}  \includegraphics[width=0.67\textwidth]{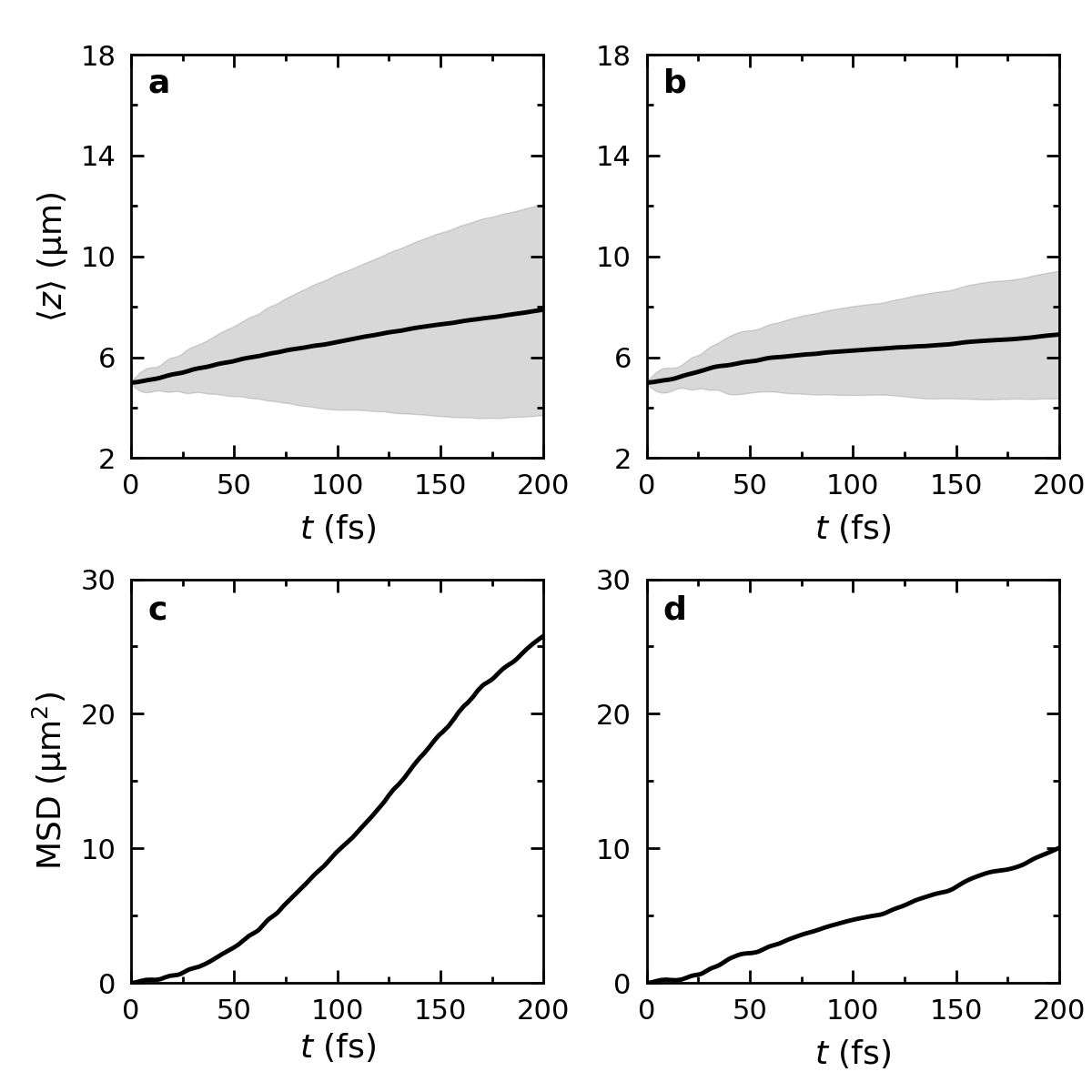}
  \caption{Expectation value of the position ($\langle z\rangle$) (top panels) and mean square displacement (MSD) (bottom panels) of the total time-dependent wavefunction $|\Psi(t)\rangle$ of {\bf{512~Rhodamine molecules}} strongly coupled to an {\bf{ideal cavity}} {\it i.e.}, $\gamma_\text{cav}=0$~ps$^{-1}$ (panels {\bf a} and {\bf c}) and a {\bf{lossy cavity}} with $\gamma_\text{cav}=66.7$~ps$^{-1}$ (panels {\bf b} and {\bf d}), after {\bf off-resonant} excitation of a single molecule. In the top panels, the black lines represents  $\langle z\rangle =  \langle\Psi(t)| \hat{z}|\Psi(t)\rangle/\langle\Psi(t)| \Psi(t)\rangle$ while the shaded area the root mean squared deviation $\sqrt{\langle (z(t)-\langle z(t)\rangle)^2\rangle}$.} 
  \label{fig:z_512_off}
\end{SCfigure}

Because the initial wavefunction that localizes the excitation onto a single molecule at the start of the simulation, is a superposition of all states, including the highest-energy, highest-momentum UP states, the fastest propagation in the smaller ensembles ({\it{i.e.}}, $N<512)$ is dominated by wavepackets of these states. Since UP states posses much higher group velocities compared to LP states, these wavepackets cross the periodic boundary around 200~fs and start self-interfering in these ensembles. We therefore only plot $\langle z\rangle$ and MSD until such crossings occur in Figures~\ref{fig:z_160_off}, \ref{fig:z_256_off} and \ref{fig:z_512_off}.


In the smallest ensemble, with 160 molecules coupled to the 160 modes of the 1D Fabry-P\'{e}rot cavity, there are no dark states and all states are hybrids of molecular and cavity mode excitations. Therefore, in contrast to the larger ensembles with dark states, all states have group velocity due to their cavity mode excitations, and all of the population propagates ballistically, as evidenced by the quadratic dependence of the mean-square displacement (MSD) on time (Figure~\ref{fig:z_160_off}). Because in the largest ensembles with many dark states, the propagation appears diffusive, the 160~molecule simulation \emph{without} dark states provides additional support for the conclusion that
transient trapping of the population in stationary dark states is the origin for such diffusion.

Comparing the relative contributions to the wavepacket of the molecular excitons (black) and cavity mode excitations (red) in the population plots for 160, 256, 512 and 1024 Rhodamine molecules in the right column of Figure~\ref{fig:wps_off_decay0} suggests that as the number of molecules, and hence the number of dark states, increases, the total population that resides in bright light-matter states decreases. As explained in Tichauer {\it{et al.}}~\cite{Tichauer2022}, such decrease in bright state population is due to $1/N$ scaling of the rate at which population transfers between dark and bright states.\cite{Tichauer2022} Because the number of dark states is proportional to $N$, while (for large $N$) the number of bright states is constant, this dependency affects the ratio between the population in the dark and bright states, with the latter rapidly decreasing with increasing $N$. 

\begin{figure*}[!htb]
\centerline{
 \epsfig{figure=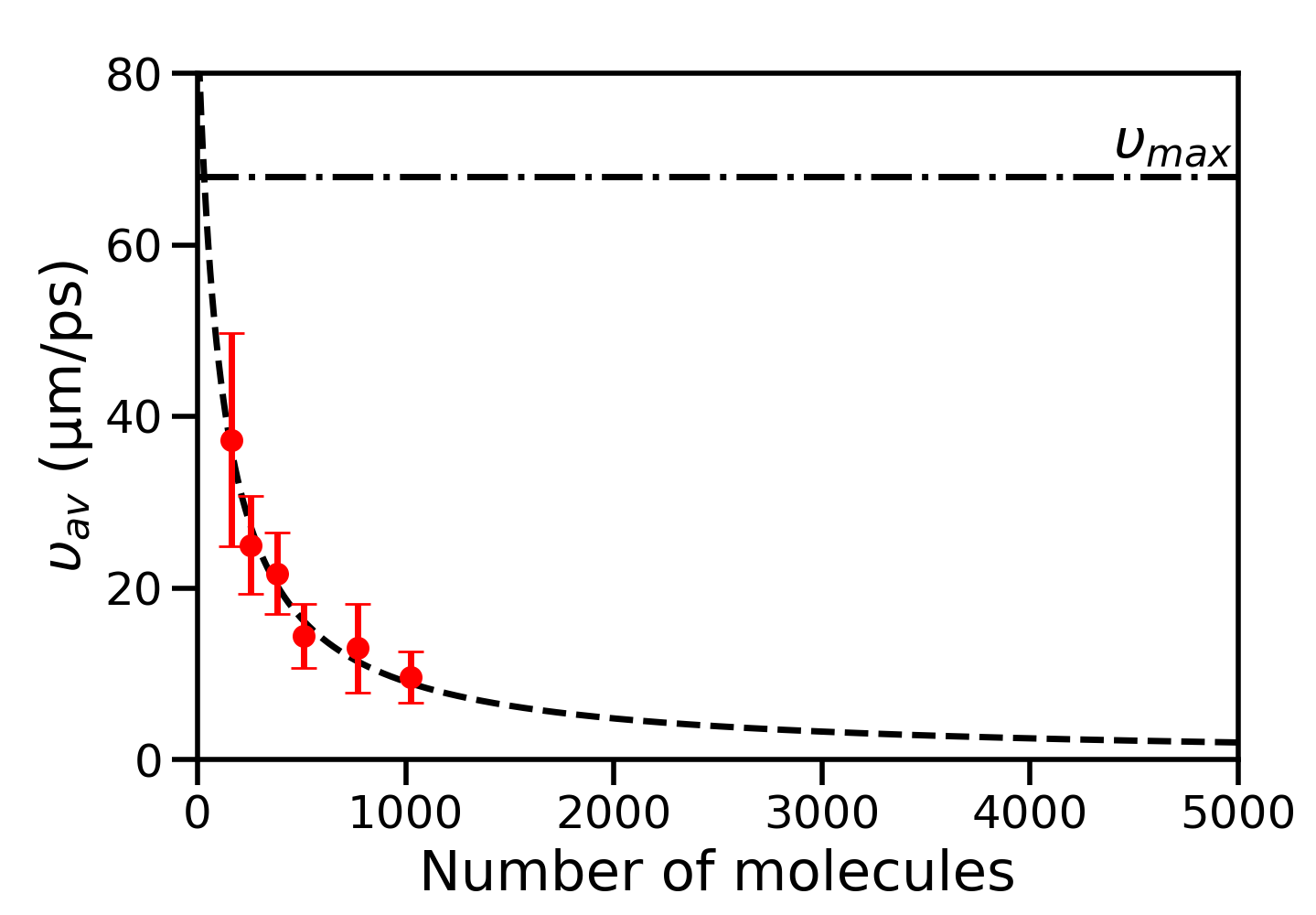,width=12cm,angle=0}
 }
  \caption{Average propagation velocity $\upsilon_{\text{av}}$ as a function of $N$ in molecule-cavity systems after {\bf{off-resonant}} excitation of a single molecule. Red dots are the values of $\upsilon_{\text{av}}$ extracted from the fits to $\langle z\rangle$ for different ensemble sizes. Red bars are the associated standard deviation. The dashed black line shows the $a/\left(N+b\right)$ fit to $\upsilon_{\text{av}}(N)$ with $a = $ 10130 $\upmu$m ps$^{-1}$ and $b=$ 117.24.
  The dash-dotted line shows the value of the maximum group velocity $\upsilon_{\text{LP}}^{\text{max}}$ of the LP.}
  \label{fig:av_velocity}
\end{figure*}

With a larger fraction of the population residing in the stationary dark states, the wavepacket propagates more slowly, as evidenced by comparing the expectation value of $\langle z\rangle$ in Figure~\ref{fig:z_160_off} for 160~molecules (without dark states),  Figure~\ref{fig:z_256_off} for 256 molecules, Figure~\ref{fig:z_512_off} for 512~molecules and Figure~5 (main text) for 1024~molecules. Average propagation velocities ($\upsilon_\text{av}$) obtained by fitting a linear function to $\langle z\rangle$, are plotted as a function of the number of molecules in Figure~\ref{fig:av_velocity}. In this plot we also include the average propagation velocity obtained for additional simulations with 384 and 768~Rhodamines strongly coupled to the cavity with the same Rabi splitting. Because the overall propagation velocity is determined by the population of the bright states, and that population is inversely proportional to $N$ due to the inverse scaling of the non-adiabatic population transfer rate,\cite{Tichauer2022} also the velocity is inversely proportional to $N$. Indeed, fitting
the average velocities to $\upsilon_\text{av}(N) = a/\left(N+b\right)$
yields a satisfactory description of the data (dashed line in Figure~\ref{fig:av_velocity}).

Because of the $1/N$ scaling, the value of $\upsilon_\text{av}$ approaches the "thermodynamic limit"  already around 1000 molecules. We therefore consider the results of the simulations with 1024 Rhodamines sufficiently representative for experiment and for providing qualitative insights into polariton propagation. Indeed, the $\upsilon_{\text{av}}$ of 9.6~${\upmu}$mps$^{-1}$ in the cavity containing 1024~molecules is about an order of magnitude below the maximum group velocity of the LP (68 ${\upmu}$mps$^{-1}$) in line with experiments on organic microcavities, \cite{Rozenman2018} and cavity-free polaritons.\cite{Pandya2021}

Furthermore, with a larger population in bright states, the motion of the wavepacket in the smaller ensembles is also more ballistic, as shown in the Mean-Square Displacement (MSD) plots in Figures~\ref{fig:z_160_off}, \ref{fig:z_256_off}, \ref{fig:z_512_off}, and Figure 3 of the main text for 160, 256, 512, and 1024~molecules, respectively. For the smaller ensembles the MSD increases more quadratically with time, in contrast to the largest ensemble with 1024~Rhodamine molecules, for which a more linear dependency is observed (Figure~3{\bf{c}}, main text). Because a linear dependency of the MSD on time is associated with diffusion, these trends further confirm that stationary dark states are responsible for diffusive propagation.

When cavity losses are added, population transfer into the ground state competes with polariton propagation (Figure~\ref{fig:wps_off}). Because only states with a significant contribution from cavity mode excitations can emit, the overall population decays faster for the smaller ensembles, suggesting that dark states can "protect" the excitation from decay under off-resonance excitation conditions and hence increase the lifetime of the excitation.\cite{Groenhof2019}

\clearpage

\subsection{Comparison between Uni- and Bi-directional Cavity Models}

\begin{figure}[!htb]
\centerline{
\epsfig{figure=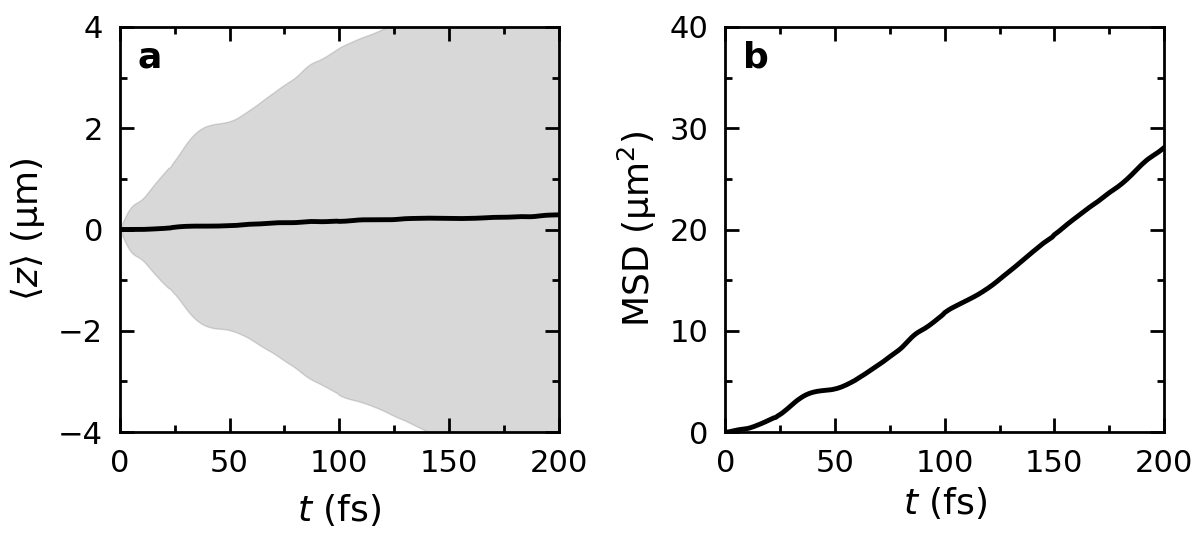,width=14cm,angle=0}}
    \caption{({\bf{a}}) Expectation value of the position ($\langle z\rangle$)  and ({\bf{b}}) mean square displacement (MSD)  of the total wavepacket $|\Psi(t)|^2$ of {\bf{1024 Rhodamine molecules}} strongly coupled to a symmetric, {\it i.e., bi-directional}  cavity {\bf{without losses}} ($\gamma_\text{cav}=0$~ps$^{-1}$) after {\bf off-resonant} excitation of a molecule located at 0 $\mu$m at 300~K.}
  \label{fig:z_MSD_symcav}
\end{figure}

In MD simulations of 1024~Rhodamine molecules strongly coupled to a symmetric cavity ({\it{i.e.}}, $n_{\text{min}}=-n_{\text{max}}=-80$) we observe that under  off-resonant excitation conditions, the 
expectation value of the position operator, $\langle\Psi(t)|\hat{z}|\Psi(t)\rangle/\langle\Psi(t)| \Psi(t)\rangle$, remains zero in average, while the MSD increases approximately linearly (Figure~\ref{fig:z_MSD_symcav}). The lack of net motion is due to cancellation between wavepackets propagating in opposite directions. Nevertheless, the excitation diffuses  trough the cavity as evidenced by the linear increase of the Mean Square Displacement (MSD), which is very similar to the MSD in the uni-directional cavity (Figure~5{\bf{c}}) that only included positive wavevectors.

\clearpage

\subsection{Simulations at 0K}

We could previously show that 
population transfer between bright and dark states is mediated by nuclear dynamics.\cite{Tichauer2022} Therefore, in \emph{classical} MD simulations, a finite temperature is required for such non-adiabatic transitions. To understand the effect of the thermally activated nuclear dynamics on the transport, we repeated the simulations of 1024~Rhodamine molecules strongly coupled to the confined light modes of a cavity ($\omega_0 =$ 3.81~eV, 160 modes) with perfect mirrors ($\gamma_\text{cav}=$~0~ps$^{-1}$) at 0~K. This hypothetical temperature was modelled by freezing all nuclear degrees of freedom. 

\subsubsection{On-resonant Excitation}

\begin{figure}[!htb]
\centerline{
\epsfig{figure=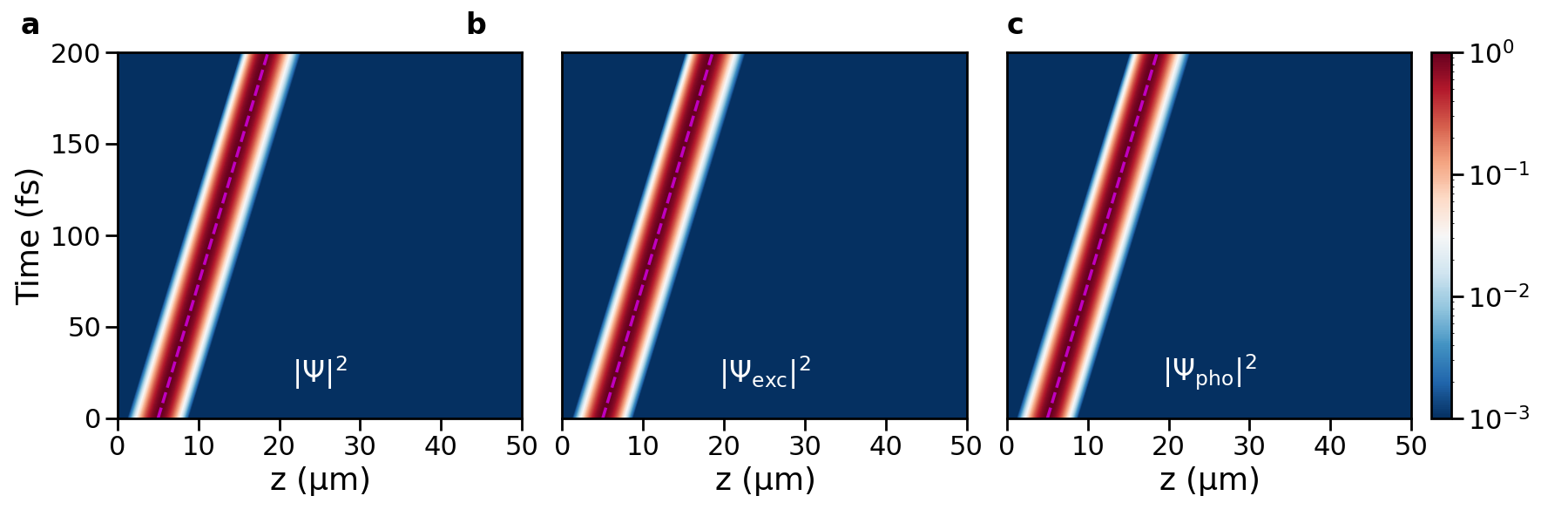,width=17cm,angle=0}}
\caption{Polariton propagation after {\bf on-resonant} excitation of a wavepacket of LP states centered at the $k_z$-vector corresponding to the maximum LP
group velocity in a {\bf{lossless}} cavity ({\it i.e.}, $\gamma_{\text{cav}} = 0~\text{ps}^{-1}$) containing {\bf{1024~Rhodamine molecules}} at {\bf{0~K}}. Panels {\bf{a}}, {\bf{b}} and {\bf{c}}: total probability density $\vert\Psi(t)\vert^2$, probability density of the molecular excitons $\vert\Psi_{\text{exc}}(t)\vert^2$, and of the cavity mode excitations $\vert\Psi_{\text{pho}}(t)\vert^2$, respectively, as a function of distance (horizontal axis) and time (vertical axis). The purple dashed line indicates propagation at the maximum group velocity of the LP (68~$\mu$m ps$^{-1}$).}
  \label{fig:wps_on_frozen}
\end{figure}

\begin{figure}[!htb]
\centerline{
\epsfig{figure=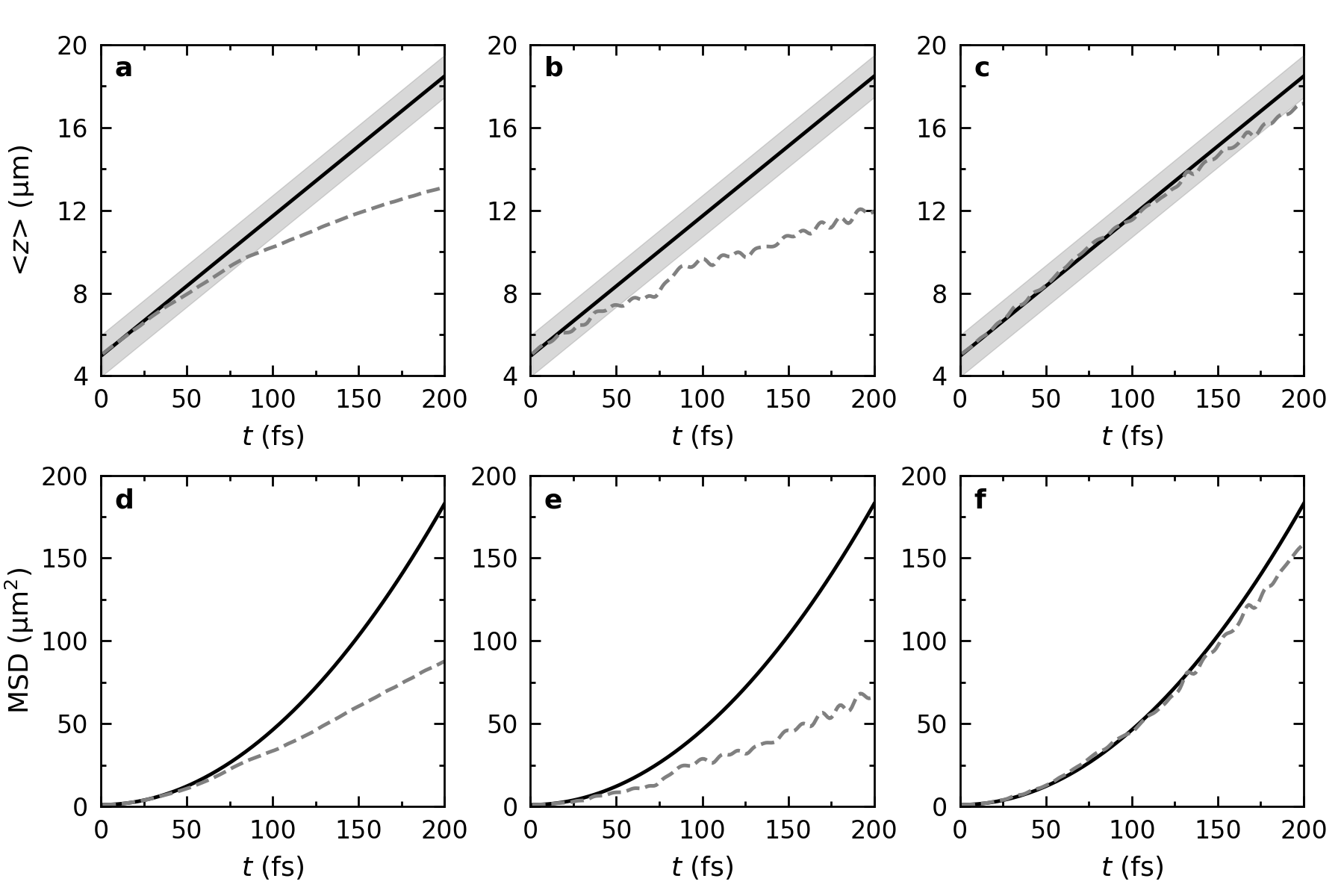,width=14cm,angle=0}}
    \caption{Expectation value of the position ($\langle z\rangle$, top panels) and mean square displacement (MSD, bottom panels) of the total $|\Psi(t)\rangle$ (left panels), excitonic $|\Psi_{\text{exc}}(t)\rangle$ (middle panels), and photonic $|\Psi_{\text{pho}}(t)\rangle$ (right panels) wave functions of {\bf{1024 Rhodamine molecules}} strongly coupled to a {\bf{lossless}} cavity ($\gamma_\text{cav}=0$~ps$^{-1}$) after {\bf on-resonant} excitation at {\bf{0~K}}. As a reference, the curves corresponding to the same system at {\bf{300~K}} are plotted as grey dashed lines.}
  \label{fig:z_MSD_on_frozen}
\end{figure}

In Figure~\ref{fig:wps_on_frozen} we plot the probability density of the wavepacket as a function of time and position after on-resonant excitation at 0K. The wavepacket  propagates ballistically at the central group velocity and retains its smooth Gaussian shape, in contrast to simulations at 300~K, in which peaks appear at the locations of molecules due to the fluctuations in their geometries and hence energies {\it i.e.} diagonal energy disorder.\cite{Agranovich2007} Importantly, at 0K, no transition from ballistic to diffusive motion is observed (Figure~\ref{fig:z_MSD_on_frozen}{\bf{d}}), in line with experiments on inorganic materials at low temperatures.\cite{Freixanet2000}

\clearpage

\subsubsection{Off-resonant Excitation}

\begin{figure}[!htb]
\centerline{
\epsfig{figure=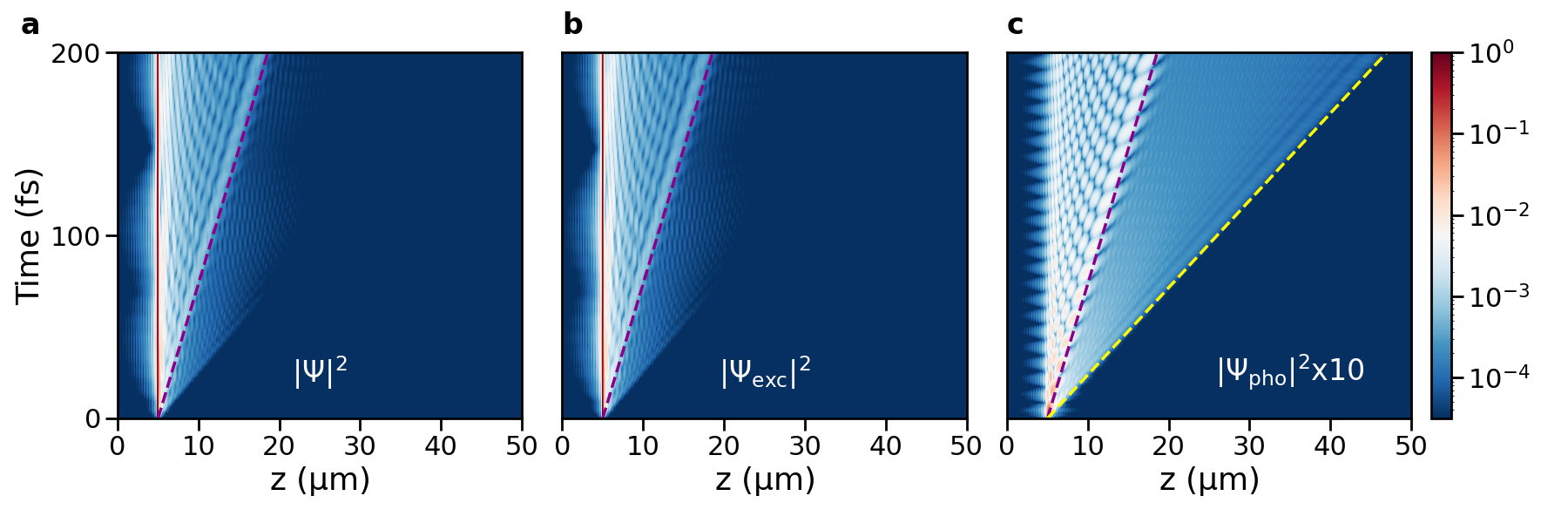,width=17cm,angle=0}}
\caption{Polariton propagation after {\bf off-resonant} excitation of a single molecule located at $z = $~5~$\mu$m into S$_1$ at {\bf{0~K}}. Panels {\bf{a}}, {\bf{b}} and {\bf{c}}: total probability density $\vert\Psi(t)\vert^2$, probability density of the molecular excitons $\vert\Psi_{\text{exc}}(t)\vert^2$, and of the cavity mode excitations $\vert\Psi_{\text{pho}}(t)\vert^2$, respectively, as a function of distance (horizontal axis) and time (vertical axis), in a cavity with perfect mirrors ({\it i.e.}, $\gamma_{\text{cav}} =$~0~ps$^{-1}$). The purple and yellow dashed lines indicate propagation at the maximum group velocity of the LP (68 $\mu$m ps$^{-1}$) and UP (212 $\mu$mps$^{-1}$), respectively.}
  \label{fig:wps_off_frozen}
\end{figure}

\begin{figure}[!htb]
\centerline{
\epsfig{figure=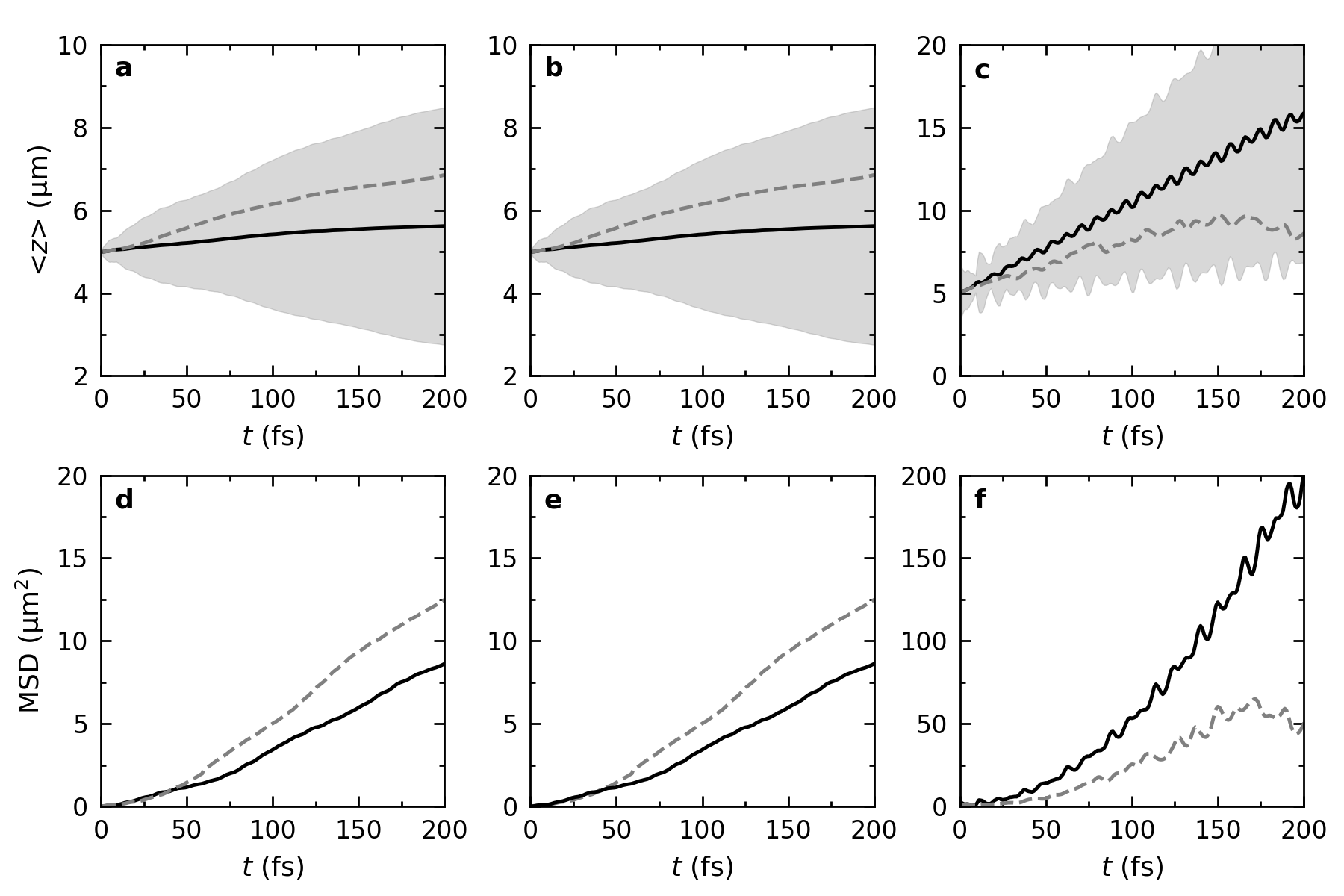,width=14cm,angle=0}}
\caption{Expectation value of the position ($\langle z\rangle$, top panels) and mean square displacement (MSD, bottom panels) of the total, $|\Psi(t)\rangle$ (left panels), excitonic, $|\Psi_{\text{exc}}(t)\rangle$ (middle panels), and photonic, $|\Psi_{\text{pho}}(t)\rangle$ (right panels), wave functions of {\bf{1024 Rhodamine molecules}} strongly coupled to an {\bf{lossless}} cavity ($\gamma_\text{cav}=$~0~ps$^{-1}$) after {\bf off-resonant} excitation at {\bf{0~K}}. As a reference, the curves corresponding to the same system at 300~K are plotted as grey dashed lines. Note that we plot both $\langle z\rangle$ and MSD only until $t =$~200~fs, since at later times the part of the wave packet composed of upper polariton states has crossed the periodic boundary at $z = $~50~$\mu$m.}
  \label{fig:z_MSD_off_frozen}
\end{figure}

In Figure~\ref{fig:wps_off_frozen} we plot the probability density of the wavepacket  as a function of time and position after off-resonant excitation at 0K. Compared to the time-evolution of the wavepacket at 300~K, the propagation is reduced at 0K, and most of the excitation remains localized at the molecule that was initially excited (Figure~\ref{fig:wps_off_frozen}{\bf{b}}). The reduced mobility of the \emph{total} wavepacket is visible also in the evolution of the expectation value of $\langle z\rangle$ in Figure~\ref{fig:z_MSD_off_frozen}{\bf{a}}. In contrast to the total wavepacket and the contribution of the molecular excitons, the part of the wavepacket that consists of the cavity mode excitations propagates ballistically (Figure~\ref{fig:z_MSD_off_frozen}{\bf{c}} and {\bf{f}}). Thus, while most of the wavepacket remains localized on the molecule that was excited initially, a small fraction of bright states propagate ballistically. Because there are no nuclear dynamics at 0K, this propagation is solely driven by constructive and destructive interferences between bright states that were occupied in the initial superposition, and that are due to differences in their phases ({\it{i.e.}}, Rabi oscillations: $c_m(t) = c_m(0)e^{-i E_m t/\hslash}$).  

\clearpage

\subsection{Non-Adiabatic Transitions}

\begin{figure*}[!htb]
\centerline{
\epsfig{figure=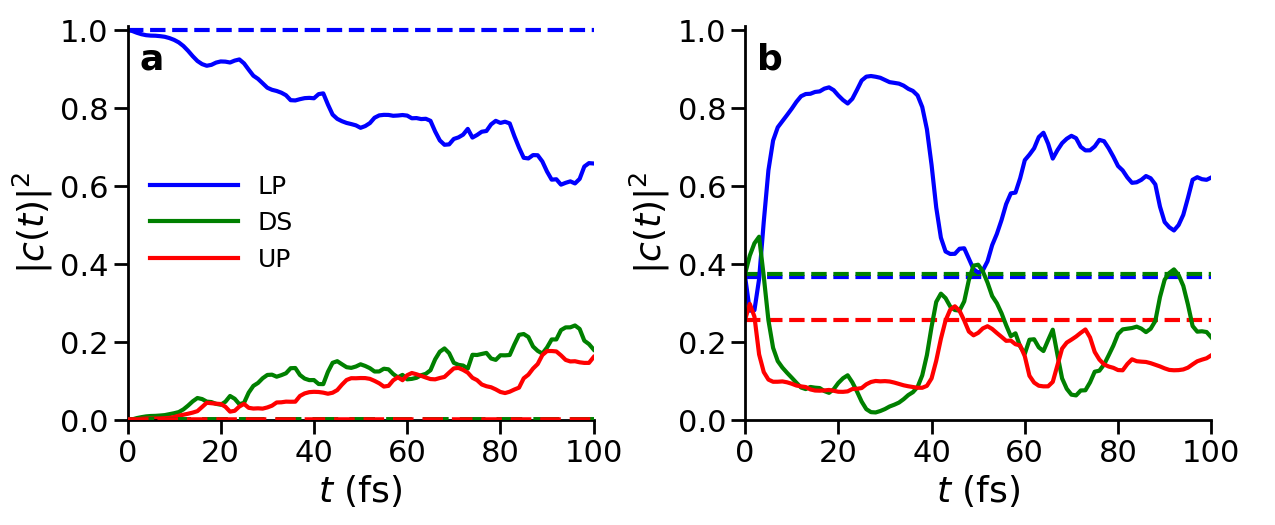,width=16cm,angle=0}
  }
  \caption{Time-evolution of the sum of expansion coefficients $\sum_m|c_m(t)|^2$ of states belonging to the LP (blue line), the UP (red line) and the dark states manifold (green line) at 300~K (solid lines) and at 0K (dashed lines) for the system with 256 Rhodamine molecules strongly coupled to an ideal cavity after off-resonant (panel {\bf a}) and on-resonant (panel {\bf b}) excitation.}
  \label{fig:expan_coeffs}
\end{figure*}

As shown in the previous section, polariton propagation at 300~K differs significantly from motion at 0K. To understand if these differences solely arise from dynamically induced disorder or also involve non-adiabatic transitions, we plot the time-evolution of the expansion coefficients of the adiabatic LP, UP and dark states ({\it{i.e.}}, $\sum_m|c_m(t)|^2$ with $m \in$~LP (160 low-energy states), UP (160 higher-energy states), and DS (the remaining states), respectively) in Figure~\ref{fig:expan_coeffs} for 256~molecules in an ideal lossless cavity after on-resonant and off-resonant excitation.
The variation in the magnitude of these coefficients with time suggest that there is significant non-adiabatic population transfer between bright and dark states in our simulations.

\clearpage
\subsection{Effect of Positional Disorder on Polariton Propagation}

\begin{figure}[!htb]
\centerline{
\epsfig{figure=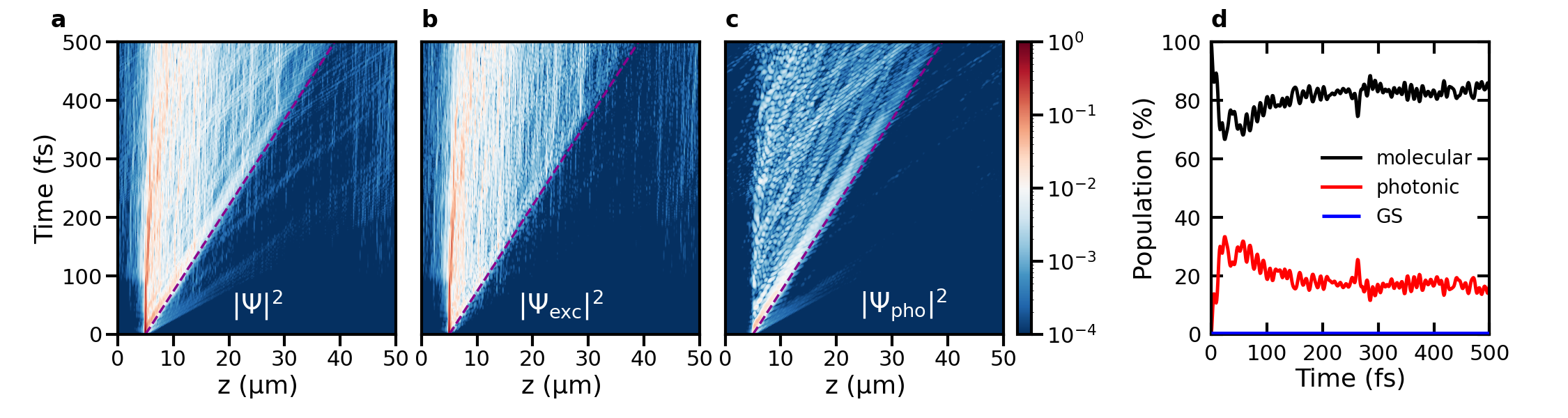,width=17cm,angle=0}
}
  \caption{Polariton propagation after an {\bf off-resonant} excitation into the S$_1$ electronic state of a single molecule located at $z=5~\upmu$m, in an ensemble of {\bf{512~Rhodamine molecules}} {\bf{randomly distributed}} along the $z$-axis of an {\bf{ideal}} cavity ({\it i.e.}, $\gamma_{\text{cav}} = 0~\text{ps}^{-1}$). Panels {\bf a}, {\bf b} and {\bf c}: total probability density, $|\Psi(t)|^2$, probability density of the molecular excitons, $|\Psi_{\text{exc}}(t)|^2$, and of the cavity mode excitations, $|\Psi_{\text{pho}}(t)|^2$, respectively, as a function of distance (horizontal axis) and time (vertical axis). Panel {\bf d}: Contributions of molecular excitons (black) and cavity mode excitations (red) to $|\Psi(t)|^2$ and population of the ground state (GS, blue) as a function of time.}
  \label{fig:wps_512_off_disord}
\end{figure}

In our simulations the solvated Rhodamine molecules were located equidistantly along the $z$-axis of the 1D cavity. To verify if such ordering affects the propagation process, we repeated a simulation under off-resonant excitation conditions with 512 Rhodamines \emph{randomly} distributed along the $z$-axis of the cavity. This was achieved by adding a random displacement to the initial equidistant positions. We only considered a perfect cavity with no losses ($\gamma_\text{cav} =$ 0~ps$^{-1}$) and with a vacuum field strength of 0.0000707107 au (0.364~MVcm$^{-1}$), yielding a Rabi splitting of $\sim$~325~meV. Direct visual comparison between the wavepacket dynamics in this system, shown in Figure~\ref{fig:wps_512_off_disord}, to that in the system with an equal spacing between the molecules (third row in Figure~\ref{fig:wps_off_decay0}), suggest that positional disorder does not influence polariton propagation, nor the time evolution of the wave function composition (Figure~\ref{fig:wps_512_off_disord}{\bf{d}} and Figure~\ref{fig:wps_off_decay0}{\bf{h}}).

\clearpage

\subsection{Effect of Initial Energy Disorder on Ballistic Transport}\label{sec:static_disorder}

Starting simulations with the same molecular geometries, and thus identical excitation energies and transition dipole moments, facilitates the identification of bright and dark states. However, in experiments the molecular geometries span a distribution weighted by their Boltzmann factors. To assess the effect of such initial (thermal) disorder on polariton propagation after resonant excitation, we repeated a simulation of 256 Rhodamines inside a lossless cavity ($\gamma_\text{cav} =$ 0~ps$^{-1}$) with a vacuum field strength of ${\bf{E}}_y =$ 0.0001~a.u. (0.514~MVcm$^{-1}$), with different initial geometries and velocities. These configurations were selected from the equilibrium QM/MM trajectory described in section~\ref{subsec:sim_details}.

\begin{figure*}[!htb]
\centerline{
\epsfig{figure=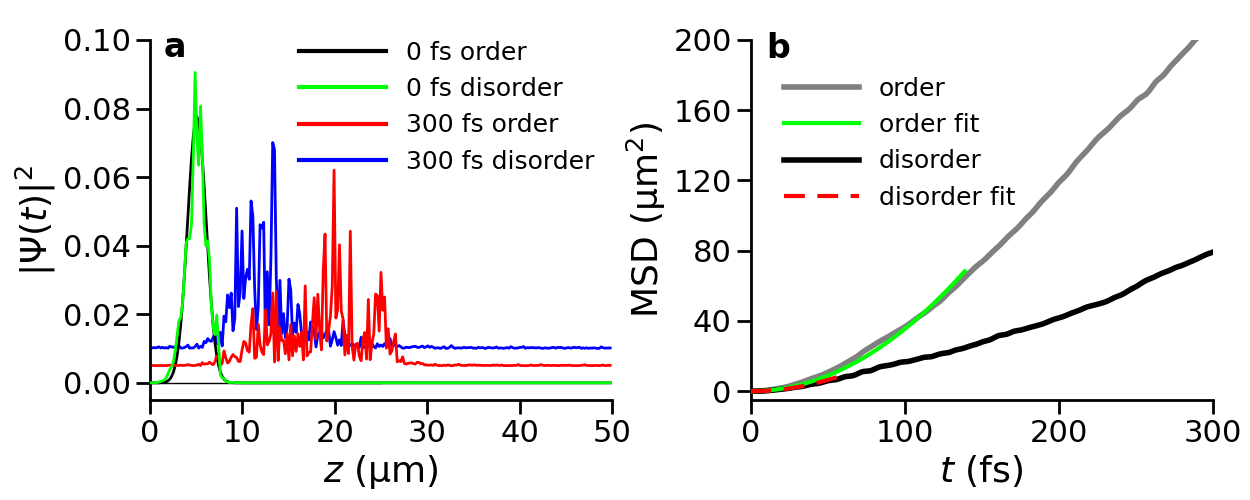,width=15cm,angle=0}
  }
  \caption{Panel {\bf{a}}: Total probability density, $\left|\Psi\right|^2$, of {\bf{256 Rhodamine molecules}} strongly coupled to an {\bf{ideal cavity}} with (green line at $t=0$~fs and blue line at $t=100$~fs) and without (black line at $t=0$~fs and red line at $t=100$~fs) initial energy disorder after {\bf on-resonant} excitation. Panel {\bf{b}}:  Mean square displacement (MSD) of the total wave function, $|\Psi(t)\rangle$, with (black) and without (gray) initial geometric disorder among the molecules. The green and red dashed lines are quadratic fits to the MSD in the ballistic phase.}
  \label{fig:disorder}
\end{figure*}

As shown in panel {\bf a} of Figure~\ref{fig:disorder}, with energetic disorder, the initial wavepacket is no longer a smooth Gaussian curve (green line), but its propagation is very similar to that of a wavepacket starting without such disorder (black line):
Initially, there is ballistic motion at a velocity close to the maximum group velocity of the LP, followed by diffusion (see Figure~\ref{fig:disorder}{\bf b}). However, because with the initial energy disorder non-adiabatic population transfers into lower energy dark states occur more readily from the start of the simulation, the transition between ballistic and diffusive transport happens earlier.

\clearpage
\subsection{Effect of Rabi Splitting on Polariton Propagation}

The results of our MD simulations suggest that reversible population transfers between bright polaritonic states with group velocity and  dark states without group velocity, play an important role in the propagation of a polaritonic wavepacket. Because the non-adiabatic coupling vectors that control these population transfers are inversely proportional to the energy gap between the states,\cite{Tichauer2022} altering the overlap between dark and bright states by changing the Rabi splitting could impact polariton propagation. To investigate how the Rabi splitting affects the propagation of the polariton wavepacket, we repeated simulations for 1024~molecules strongly coupled to the confined light modes of a lossy cavity ($\gamma_\text{cav} =$~66.7~ps$^{-1}$) after on-resonant and off-resonant excitation at three different Rabi splittings. In these simulations, the timestep of the MD simulations was increased to 0.5~fs. To alter the Rabi splitting without changing the number of states, we varied the vacuum field strengths: $|{\bf{E}}_y|=$~0.000025, 0.000037 and 0.00005~a.u., yielding Rabi splittings of $\sim$~159, 240 and 325~meV. We carried out two simulations per Rabi splitting value with different initial velocities. To facilitate direct comparison, the same random velocities were used for each Rabi splitting. The latter was realized by using the index of the molecule as a random seed for selecting the velocities in the first series of simulations, and adding 1024 to that index in the second series.



\subsubsection{On-resonant Excitation}

\begin{figure*}[!htb]
\centerline{  \epsfig{figure=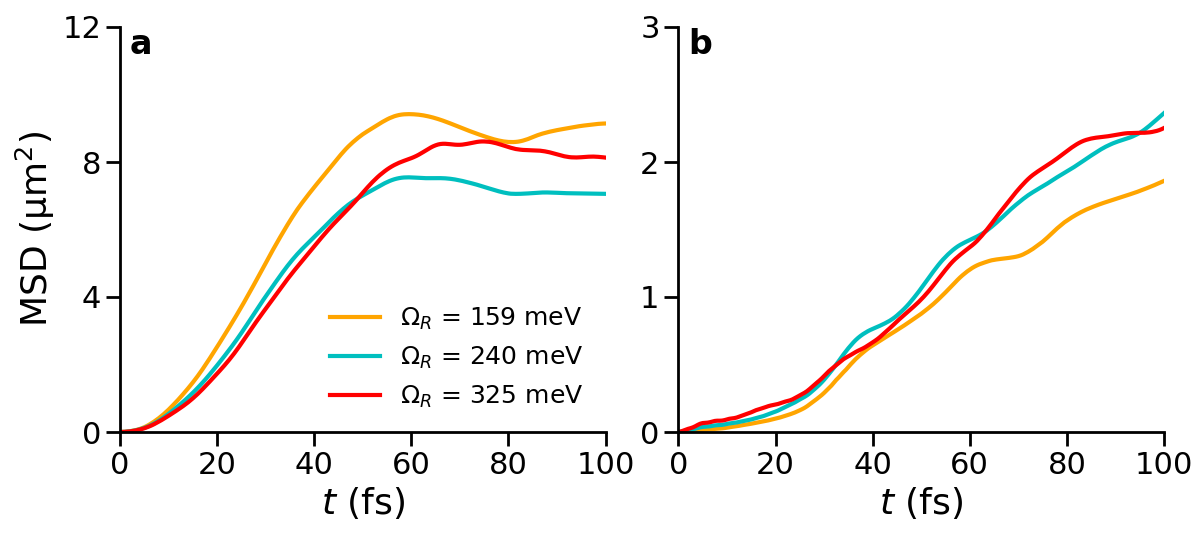,width=14cm,angle=0}
  }
  \caption{Effect of Rabi splitting on the Mean Square Displacement of the total polariton wave function ($\Psi(t)$) after {\bf{on-resonant }} (panel \textbf{a}) and \textbf{off-resonant} (panel \textbf{b}) excitation into the LP branch of a system with 1024 molecules strongly coupled to a \textit{lossy} cavity ($\gamma_\text{cav}=66.7$~ps$^{-1}$). 
  }
  \label{fig:diff_Rabi}
\end{figure*}

In Figure~\ref{fig:diff_Rabi}a we show the Mean Square Displacement (MSD) of the total wave function for various Rabi splitting energies after on-resonant excitation of a wavepacket of LP states. While the initial velocity in the ballistic phase reflects the central group velocity of the wavepacket, which is highest for the smallest Rabi splitting (81~$\mu$m/ps \textit{vs.} 68~$\mu$m/ps for the largest Rabi splittings), there is no clear trend at longer timescales, suggesting that more simulations are needed.

Based on previous results, which suggest that increasing the vacuum field strength decreases the overlap between the bright polaritonic branches and dark state manifold,\cite{Groenhof2019} we speculate that the decrease in group velocity upon increasing the Rabi splitting, may be compensated by a reduction of the population transfer into the dark states. To minimize that population transfer and extend the ballistic phase, the Rabi splitting should exceed the absorption line width of the molecules,\cite{Groenhof2019} which for typical organic materials would require entering the ultra-strong coupling regime, where the Rabi splitting is more than 10\% of the excitation energy.\cite{Forn-Diaz2019} However, in this regime, the Tavis-Cummings models is no longer strictly valid as the RWA breaks down, and the counter-rotating terms ($\hat{\sigma}_j^+\hat{a}_{k_z}^\dagger$ and $\hat{\sigma}_j\hat{a}_{k_z}$), as well as the dipole self energy ($\hat{H}^\text{DSE} = \frac{1}{2\epsilon_0V_\text{cav}}\left[{\bf{u}}_\text{y}\cdot\left(\sum_i^N\hat{{\boldsymbol{\mu}}}_i\right)\right]^2$) would need to be included,\cite{Flick2017b} which is computationally intractable for the system sizes used in this work.


Nevertheless, even if our results do not reveal a clear trend, there may be a trade-off between the absorption line width of the material and the Rabi splitting, for optimal exciton transport. However, because the focus of the current work is on the \emph{mechanism} of polariton propagation, we consider finding that trade-off beyond the scope of our work, and will address the effect of the Rabi splitting on polariton transport in future work.

\subsubsection{Off-resonant Excitation}

In Figure~\ref{fig:diff_Rabi}b we show the Mean Square Displacement of the total wavepacket for the various Rabi splittings after off-resonant excitation of a single molecule located at $z= $ 5 $\upmu$m. As for on-resonance excitation, we do not observe a clear correlation between the Rabi splitting on the one hand and the Mean Square Displacement on the other hand. Very many additional simulations are probably needed to extract a trend. However, because the focus of the current work is on the \emph{mechanism} of polariton propagation, we consider a thorough investigation of the effect of the Rabi splitting beyond the scope of the current work. Yet, since we anticipate that the Rabi splitting may have an effect that could possibly be exploited and can be directly controlled via the molecular concentration, we will address the effect of the concentration on polariton transport in future work.

\clearpage
\subsection{Contribution of High-Energy Polariton States to the Wavepacket Propagation
}

\begin{figure*}[!htb]
\centerline{
\epsfig{figure=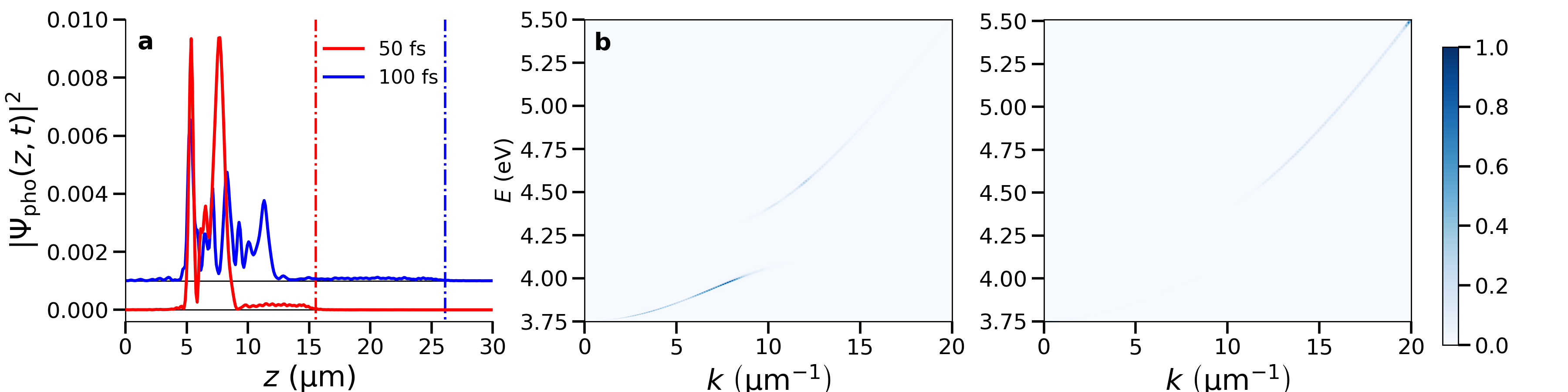,width=17cm,angle=0}
  }
  \caption{Panel {\bf{a}}: Probability densities of $\left|\Psi_{\text{pho}}\right|^2$ at $t=$ 50 and 100~fs after {\bf off-resonant} excitation of a cavity including {\bf{512 Rhodamine molecules}}. The initial excitation is localised on molecule placed at $z=$~5~$\upmu$m. The dashed-dotted lines indicate the product between the maximum group velocity $\upsilon_\text{UP}^\text{max}$ = 212 ${\upmu}$mps$^{-1}$ and the simulation time, and thus corresponds to where something moving at that group velocity would be located. Panels {\bf{b}} and {\bf{c}}: Photoluminescence spectra collected through pinholes located at  $z=$ 10~$\upmu$m and $z =$ 20~$\upmu$m, respectively.
  }
  \label{fig:fast_tail}
\end{figure*}

Following \emph{off-resonant} excitation into the S$_1$ electronic state of a single Rhodamine molecule, we observe that the photonic part of the wavepacket has a small front that propagates faster than the rest of the wavepacket. In Figure~\ref{fig:fast_tail} we show the probability density of the cavity mode excitations in the 512~molecule cavity system at 50 and 100~fs. The wavepacket is composed of a slower and faster component, with the front of the faster component reaching distances corresponding to the maximum group velocity $\upsilon_{\text{UP}}^{\text{max}}$ = 212 ${\upmu}$mps$^{-1}$ of the UP branch. 

To confirm that the fastest components in the wavepacket are due to propagation of population in the UP states that get rapidly populated due to Rabi oscillations (Equation~\ref{eq:WF_in_diabatic_basis}), we analyzed the photoluminesce through 10~$\upmu$m wide pinholes located at $z$ = 10~$\upmu$m and at $z$ = 20~$\upmu$m (see section~\ref{sec:mask} for details of these calculations).\cite{Berghuis2022} The photoluminescence spectra collected through these pinholes are plotted as a function of the in-plane momentum ($k_z$) in Figures~\ref{fig:fast_tail}{\bf{b}} and {\bf{c}}. While at $z$ = 10~$\upmu$m the photoluminescence is due to emission from  both LP and UP states, emission through the pinhole at $z$ = 20~$\upmu$m originates exclusively from UP states. These observations thus confirm that the fastest moving part of the wave packet is composed of UP states.

\clearpage

\subsection{Polariton Propagation in Tetracene and Methylene Blue cavities}

With an absorption linewidth of 100~meV (FWHM), a Stokes shift of 120~meV, and a minimum energy conical intersection sufficiently above the Franck Condon region to prevent internal conversion (1.3 eV), our Rhodamine model captures the key photophysical features of typical organic dye molecules. Therefore, we consider the propagation mechanism observed in our simulations, representative for organic polaritons in general. Nevertheless, to demonstrate that the mechanism is robust with respect to changes in the optical properties of the molecules, we repeated the simulations for a cavity filled with 1024~Tetracene molecules in cyclohexane, and for a cavity with 1024~Methylene Blue dyes in water. At the CIS/3-21G level of theory employed in our simulations, Tetracene has an absorption line width of 0.14~eV and a Stokes shift of 0.7~eV, both of which are significantly higher than for Rhodamine. The Rabi splitting of the Tetracene cavity was 0.24~eV. At the TD-B97/3-21G level of theory, Methylene blue has an absorption line width of 0.07~eV and a Stokes shift of 0.23~eV. The Rabi splitting of the Methylene Blue cavity was 0.175~eV.

\subsubsection{Tetracene}

\begin{figure*}[!htb]
\centering
\includegraphics[width=\textwidth]{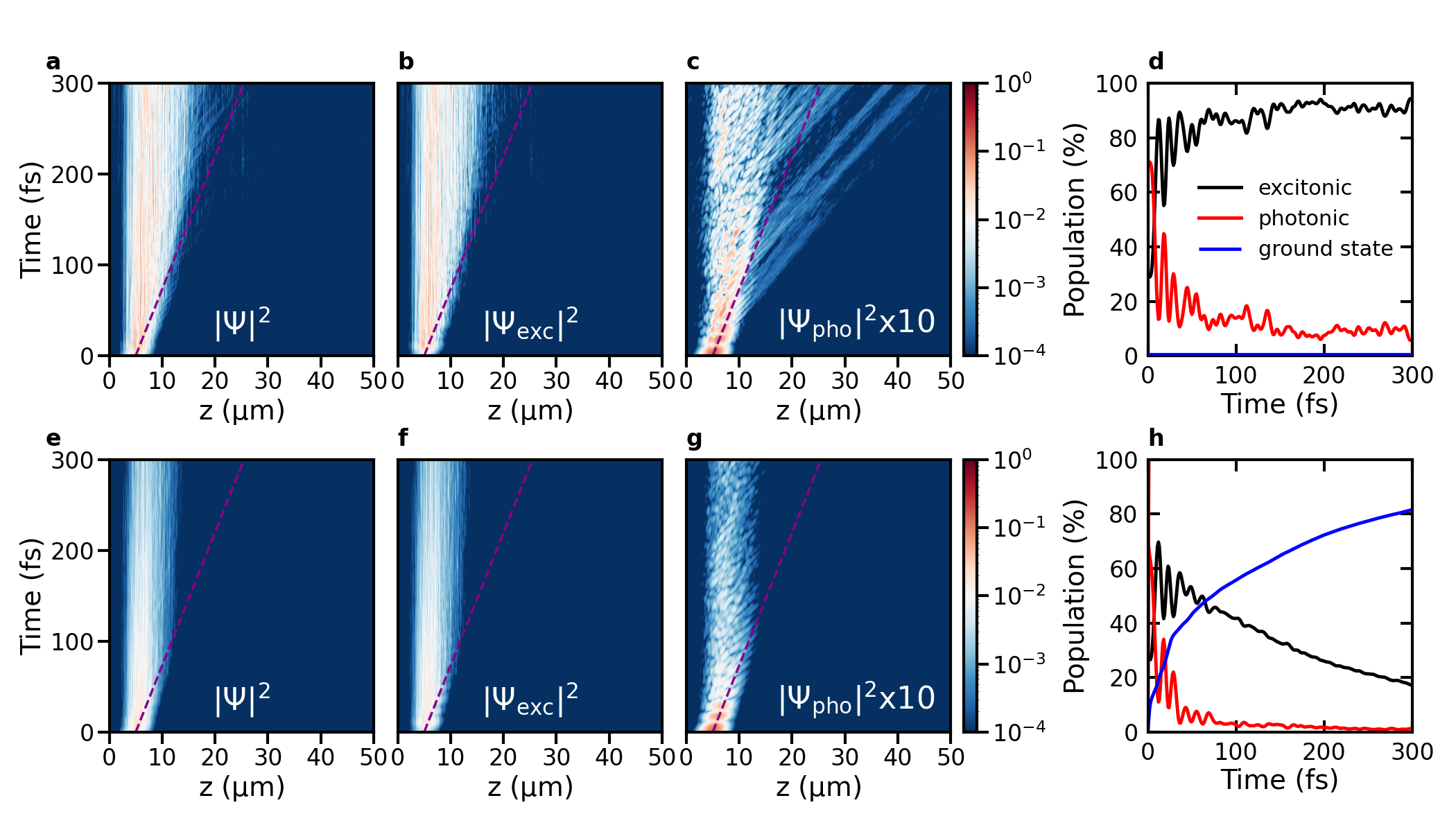}
  \caption{Polariton propagation in cavities filled with Tetracene molecules after {\bf on-resonant} excitation of a wavepacket of LP states, centered at $z=$~5~$\mu$m. Panels {\bf{a}}, {\bf{b}} and {\bf{c}}: total probability density $\vert\Psi(t)\vert^2$, probability density of the molecular excitons $\vert\Psi_{\text{exc}}(t)\vert^2$ and of the cavity mode excitations $\vert\Psi_{\text{pho}}(t)\vert^2$, respectively, as a function of distance (horizontal axis) and time (vertical axis), in a cavity with perfect mirrors ({\it i.e.}, $\gamma_{\text{cav}} = $~0~ps$^{-1}$). The magenta dashed line indicates propagation at the maximum group velocity of the LP (68~$\mu$mps$^{-1}$). Panel {\bf{d}}: Contributions of molecular excitons (black) and cavity mode excitations (red) to $\vert\Psi(t)\vert^2$ as a function of time in the perfect cavity. With no photon losses, no ground state population (blue) can build up. Panels {\bf{e}}, {\bf{f}}, and {\bf{g}}: $\vert\Psi(t)\vert^2$, $\vert\Psi_{\text{exc}}(t)\vert^2$ and $\vert\Psi_{\text{pho}}(t)\vert^2$, respectively, as a function of distance (horizontal axis) and time (vertical axis), in a lossy cavity ($\gamma_{\text{cav}}=$~66.7~ps$^{-1}$). Panel {\bf{h}}: Contributions of the molecular excitons (black) and cavity mode excitations (red) to $\vert\Psi(t)\vert^2$ as a function of time. The population in the ground state, created by radiative decay through the imperfect mirrors, is plotted in blue.}\label{fig:tetracene_wp_on_1024}
\end{figure*}

\begin{figure}[!htb]
\centering
\includegraphics[width=12cm]{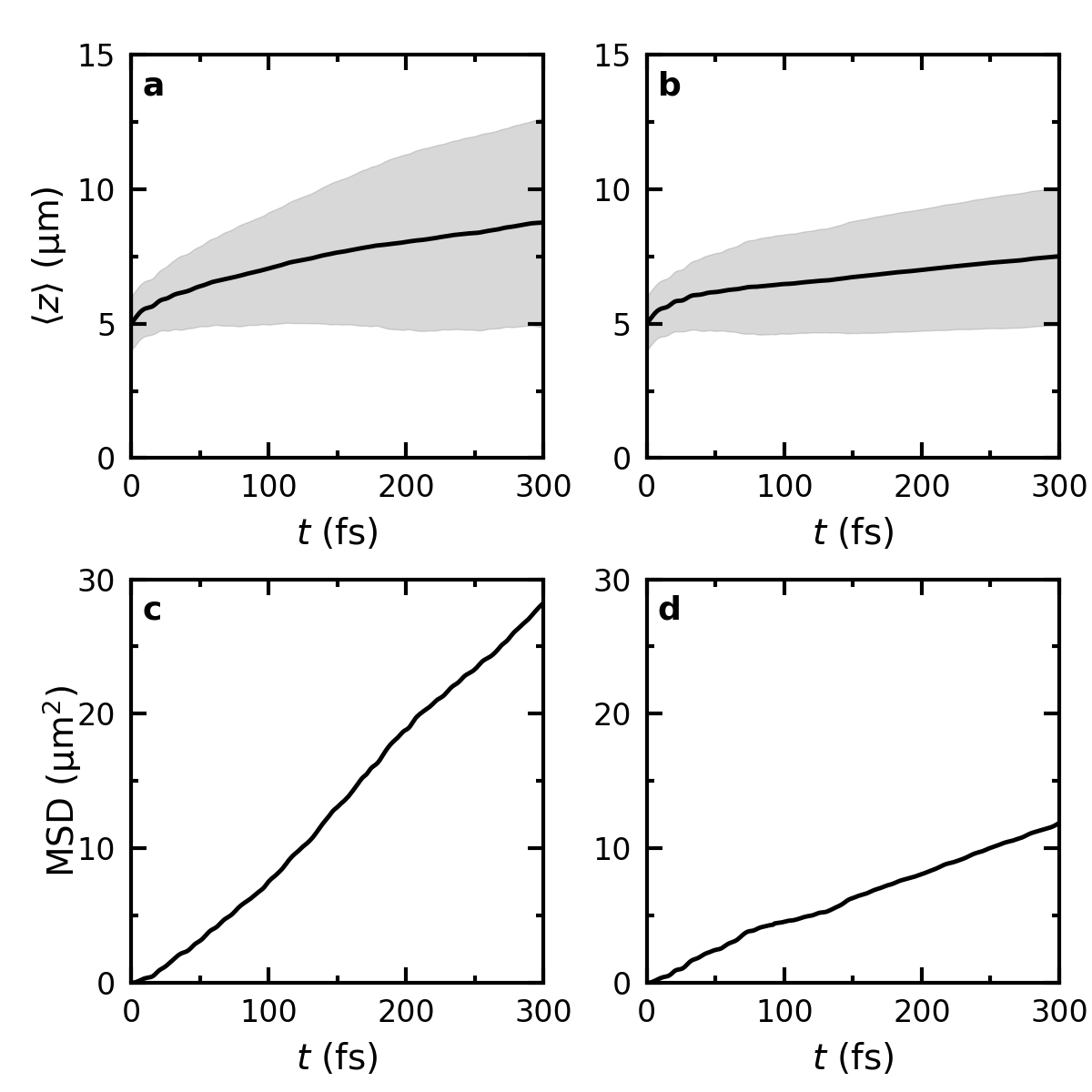}
  \caption{Top panels: Expectation value of the position of the total-time dependent wavefunction $\langle\Psi(t)\vert\hat{z}\vert\Psi(t)\rangle/\langle\Psi(t)\vert\Psi(t)\rangle$ after {\bf on-resonant} excitation of LP states in lossless ({\bf{a}}) and lossy cavities ({\bf{b}}) filled with Tetracene molecules. The black lines represent $\langle z\rangle$ while the shaded area around the lines represents the root mean squared deviation (RMSD, {\it{i.e.}}, $\sqrt{\langle (z(t)-\langle z(t)\rangle)^2\rangle}$). Bottom panels: Mean Square Displacement (MSD, {\it{i.e.}}, $\sqrt{\langle (z(t)-\langle z(0)\rangle)^2\rangle}$) in the ideal lossless cavity ({\bf{c}}) and the lossy cavity ({\bf{d}}). 
  }\label{fig:tetracene_z_on}
\end{figure}

In Figure~\ref{fig:tetracene_wp_on_1024} we show polariton propagation in a cavity filled with Tetracene in cyclohexane following an on-resonant excitation of a wavepacket of LP states for both a lossless ($\gamma_\text{cav}=$~0~ps$^{-1}$) and a lossy cavity ($\gamma_\text{cav}=$~66.7~ps$^{-1}$). After a very short ballistic phase, which is only discernible as the steep initial rise of $\langle z\rangle$ in the top panels of Figure~\ref{fig:tetracene_z_on}, propagation becomes diffusive, as evidenced by a linear time-dependence of the Mean Square Displacement  (Figure~\ref{fig:tetracene_z_on}{\bf{c}} and {\bf{d}} for the lossless and lossy cavity, respectively). The absence of a clear coherent transport regime is due to the large absorption linewidth of Tetracene. As we could show previously,\cite{Groenhof2019} the rate at which population transfers from the LP into the dark states depends on the energetic overlap between the LP and the molecular density of states, with the latter matching the absorption spectrum). Because for Tetracene the absorption spectrum is much broader than the Rabi splitting, this overlap is very large and hence the population transfer is very fast. As a consequence, the excitonic contributions start dominating the wavepacket immediately after the excitation, ending the ballistic propagation phase earlier than in cavities filled with Rhodamine, for which the overlap between the LP and the molecular dark states is much smaller. Because these transfers are reversible, the propagation continues in a diffusive manner. Adding losses does not change this picture, but because the coherent phase is much shorter than in cavities filled with Rhodamine, also the total radiative decay is smaller (bottom rows in Figures~2 and \ref{fig:tetracene_wp_on_1024}).

\begin{figure*}[!htb]
\centering
\includegraphics[width=\textwidth]{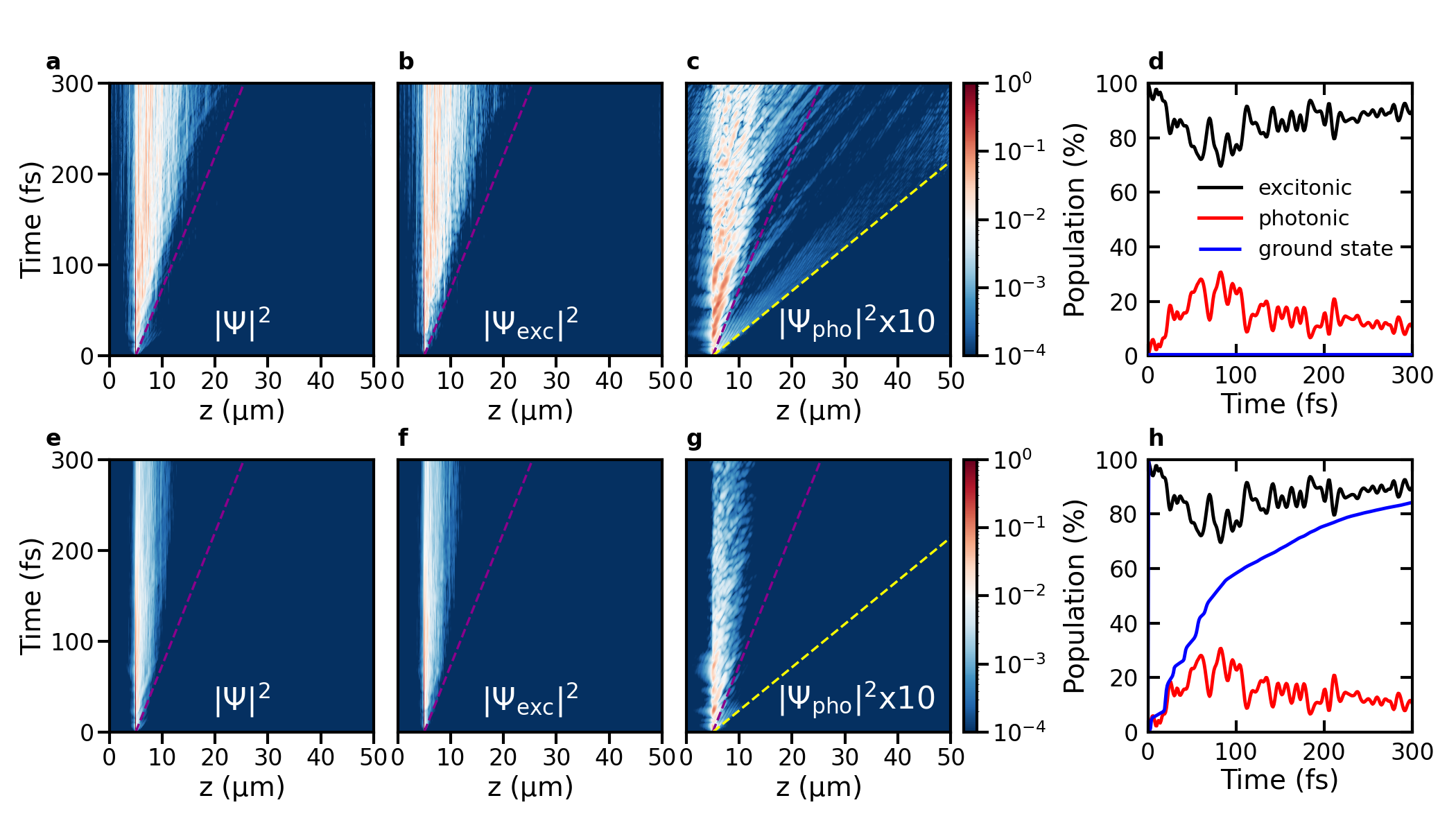}
  \caption{Polariton propagation in cavities filled with Tetracene molecules after {\bf off-resonant} excitation of a molecule located at $z=$~5~$\mu$m. Panels {\bf{a}}, {\bf{b}} and {\bf{c}}: total probability density $\vert\Psi(t)\vert^2$, probability density of the molecular excitons $\vert\Psi_{\text{exc}}(t)\vert^2$ and of the cavity mode excitations $\vert\Psi_{\text{pho}}(t)\vert^2$, respectively, as a function of distance (horizontal axis) and time (vertical axis), in a cavity with perfect mirrors ({\it i.e.}, $\gamma_{\text{cav}} = $~0~ps$^{-1}$). The magenta and yellow dashed lines indicate propagation at the maximum group velocity of the LP (68~$\mu$mps$^{-1}$) and UP (220~$\mu$mps$^{-1}$), respectively. Panel {\bf{d}}: Contributions of molecular excitons (black) and cavity mode excitations (red) to $\vert\Psi(t)\vert^2$ as a function of time in the perfect cavity. With no photon losses, no ground state population (blue) can build up. Panels {\bf{e}}, {\bf{f}}, and {\bf{g}}: $\vert\Psi(t)\vert^2$, $\vert\Psi_{\text{exc}}(t)\vert^2$ and $\vert\Psi_{\text{pho}}(t)\vert^2$, respectively, as a function of distance (horizontal axis) and time (vertical axis), in a lossy cavity ($\gamma_{\text{cav}}=$~66.7~ps$^{-1}$). Panel {\bf{h}}: Contributions of the molecular excitons (black) and cavity mode excitations (red) to $\vert\Psi(t)\vert^2$ as a function of time. The population in the ground state, created by radiative decay through the imperfect mirrors, is plotted in blue.}\label{fig:tetracene_wp_off_1024}
\end{figure*}

\begin{figure}[!htb]
\centering
\includegraphics[width=12cm]{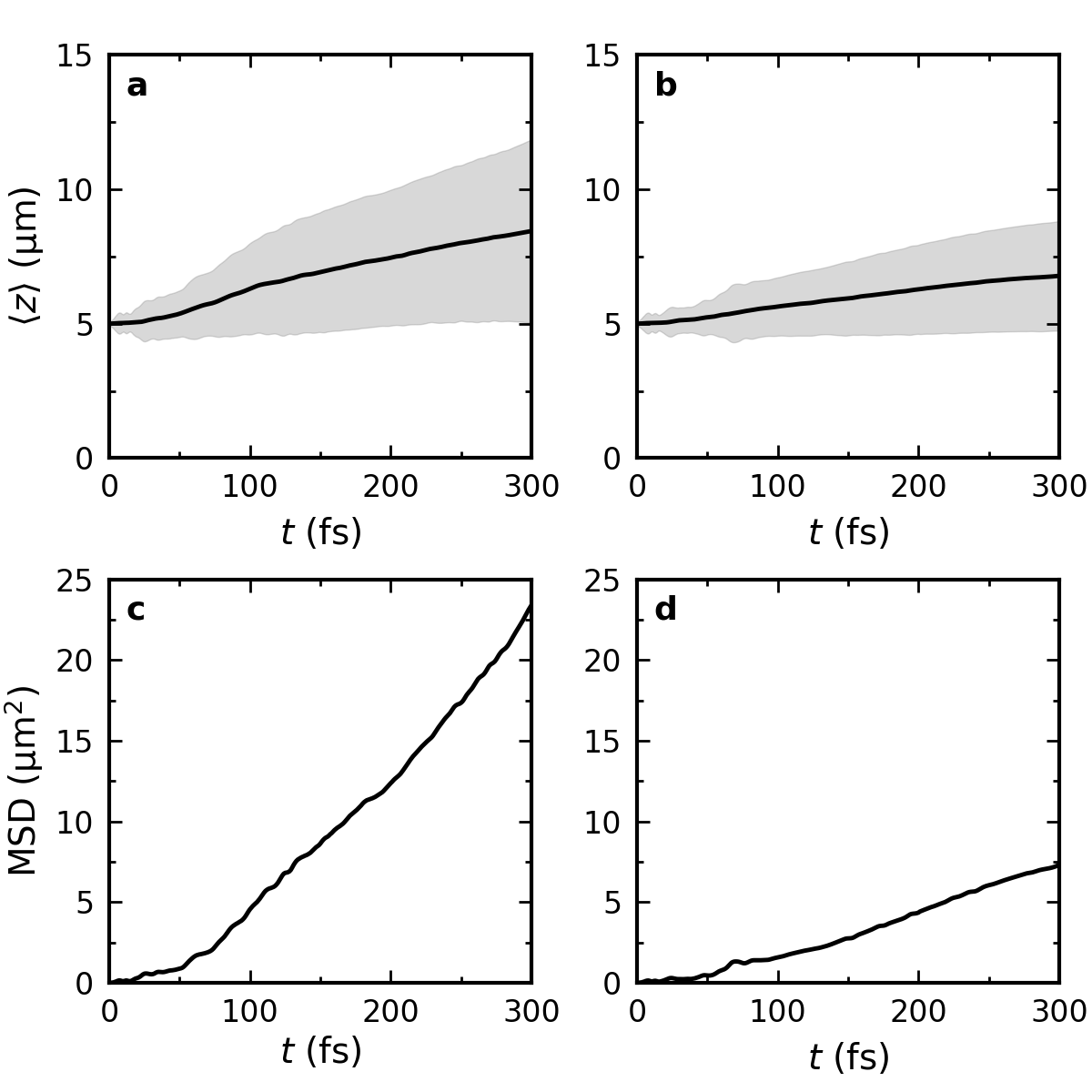}
  \caption{Top panels: Expectation value of the position of the total-time dependent wavefunction $\langle\Psi(t)\vert\hat{z}\vert\Psi(t)\rangle/\langle\Psi(t)\vert\Psi(t)\rangle$ after {\bf off-resonant} excitation of lossless ({\bf{a}}) and lossy cavities ({\bf{b}}) filled with Tetracene molecules. The black lines represent $\langle z\rangle$ while the shaded area around the lines represents the root mean squared deviation (RMSD, {\it{i.e.}}, $\sqrt{\langle (z(t)-\langle z(t)\rangle)^2\rangle}$). Bottom panels: Mean Square Displacement (MSD, {\it{i.e.}}, $\sqrt{\langle (z(t)-\langle z(0)\rangle)^2\rangle}$) in the ideal lossless cavity ({\bf{c}}) and the lossy cavity ({\bf{d}}). 
  }\label{fig:tetracene_z_off}
\end{figure}

In Figure~\ref{fig:tetracene_wp_off_1024} we show the wavepacket dynamics in both a lossless ($\gamma_\text{cav}=$~0~ps$^{-1}$) and a lossy cavity ($\gamma_\text{cav}=$~66.7~ps$^{-1}$) following off-resonant excitation of the Tetracene molecule located at 5~$\upmu$m. As in cavities filled with Rhodamine (Figure~3 in the main text), propagation is diffusive, driven by reversible exchange of population between bright states with group velocity and dark states without group velocity. Because for the on-resonant excitation, the initial coherent ballistic phases is very short due to the broad absorption of Tetracene strongly overlapping with the LP branch, the Mean Square Displacements are similar for both excitation schemes, in particular for the lossless cavity (Figures~\ref{fig:tetracene_z_on} and Figures~\ref{fig:tetracene_z_off}). This is in contrast to cavities filled with Rhodamine for which the longer lasting coherent phase with ballistic propagation, leads to a faster initial rise of the Mean Square Displacement, until the turn-over into the diffusion regime occurs (Figure~2 in the main text).

\clearpage

\subsubsection{Methylene Blue}

\begin{figure*}[!htb]
\centering
\includegraphics[width=\textwidth]{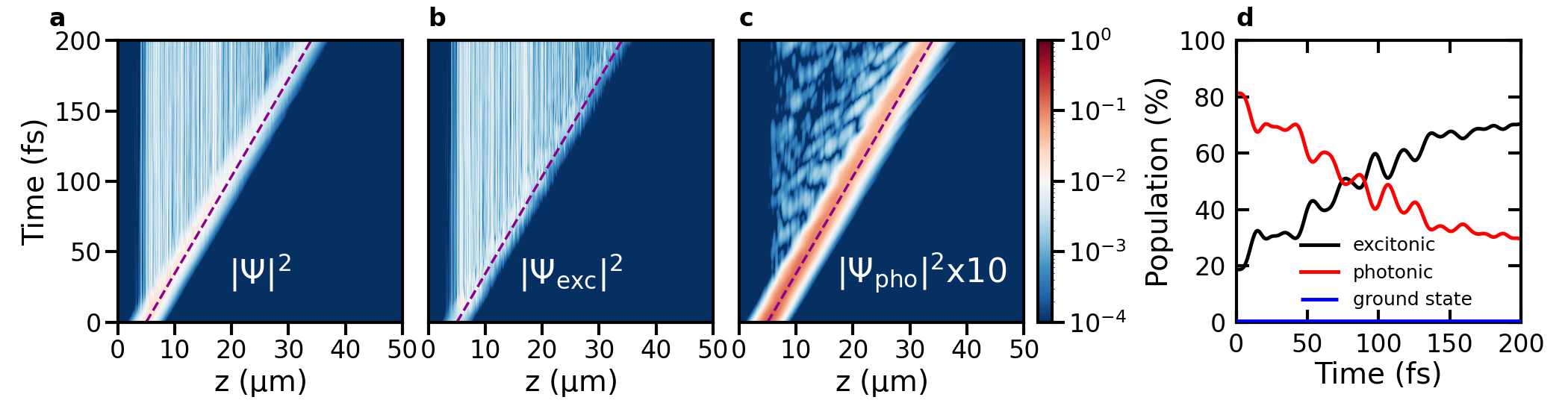}
  \caption{Polariton propagation in cavities filled with Methylene Blue molecules after {\bf on-resonant} excitation of a wavepacket of LP states centered at $z=$~5~$\mu$m. Panels {\bf{a}}, {\bf{b}} and {\bf{c}}: total probability density $\vert\Psi(t)\vert^2$, probability density of the molecular excitons $\vert\Psi_{\text{exc}}(t)\vert^2$ and of the cavity mode excitations $\vert\Psi_{\text{pho}}(t)\vert^2$, respectively, as a function of distance (horizontal axis) and time (vertical axis), in a cavity with perfect mirrors ({\it i.e.}, $\gamma_{\text{cav}} = $~0~ps$^{-1}$). The magenta dashed line indicates propagation at the maximum group velocity of the LP (145~$\mu$mps$^{-1}$). Panel {\bf{d}}: Contributions of molecular excitons (black) and cavity mode excitations (red) to $\vert\Psi(t)\vert^2$ as a function of time in the perfect cavity. With no photon losses, no ground state population (blue) can build up.}\label{fig:MeB_wp_on_1024}
\end{figure*}

\begin{figure}[!htb]
\centering
\includegraphics[width=12cm]{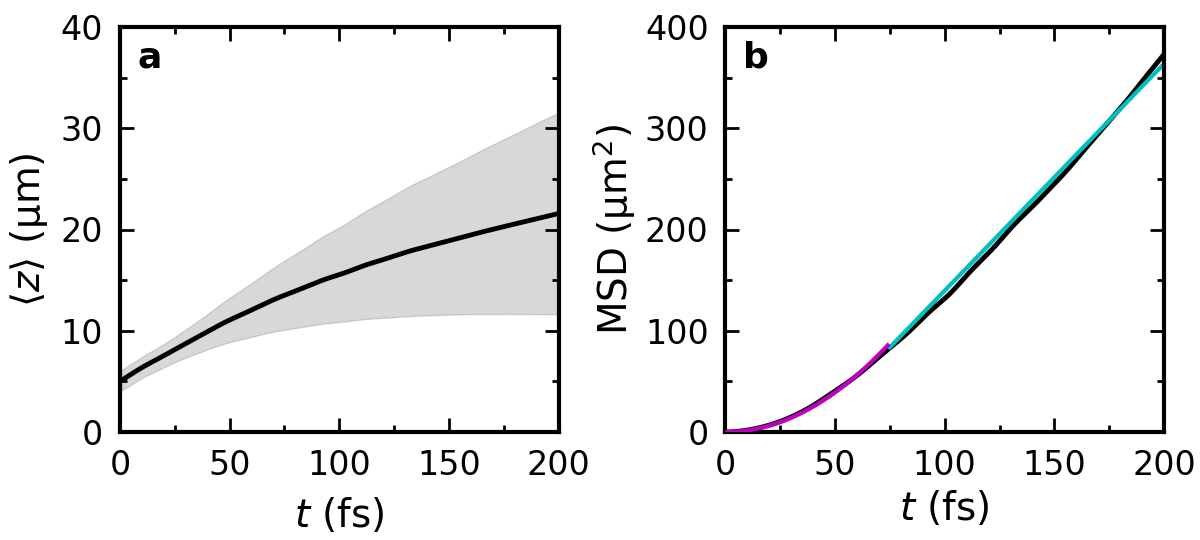}
  \caption{Panel \textbf{a}: Expectation value of the position of the total-time dependent wavefunction $\langle\Psi(t)\vert\hat{z}\vert\Psi(t)\rangle/\langle\Psi(t)\vert\Psi(t)\rangle$ after {\bf on-resonant} excitation in a lossless cavity filled with Methylene Blue molecules. The black line represents $\langle z\rangle$ while the shaded area around the line represents the root mean squared deviation (RMSD, {\it{i.e.}}, $\sqrt{\langle (z(t)-\langle z(t)\rangle)^2\rangle}$). Panel \textbf{b}: Mean square displacement (MSD, {\it{i.e.}}, $\sqrt{\langle (z(t)-\langle z(0)\rangle)^2\rangle}$) in the ideal lossless cavity. The magenta line is a quadratic fit to the Mean Square Displacement and the cyan line is a linear fit.
  }\label{fig:MeB_z_on}
\end{figure}

Because the broad absorption line-width of Tetracene 
reduces the duration of the ballistic propagation regime in Tetracene to a few fs, we also performed a simulation of Methylene Blue, which has an absorption line-width that is narrower than that of both Tetracene and Rhodamine. In Figure~\ref{fig:MeB_wp_on_1024}, we show the wavepacket dynamics after exciting a wavepacket of LP states. Indeed, with a reduced  overlap between the LP states and the molecular dark states, ballistic propagation is observed. As in Rhodamine (Figure~2 in main text), the ballistic regime ends when the excitonic contributions to the total wave function become dominant, which happens around 75~fs for the cavity filled with Methylene Blue (Figure~\ref{fig:MeB_wp_on_1024}{\bf{d}}). In line with results for Rhodamine, the Mean Square Displacement then changes from a quadratic to linear time dependence, as indicated by the quadratic (magenta) and linear (cyan) fits in Figure~\ref{fig:MeB_z_on}{\bf{b}}. Thus, after on-resonant excitation of the cavity filled with Methylene Blue, propagation proceeds as in cavities containing Rhodamine or Tetracene, with a short ballistic phase when bright states are predominantly populated, followed by a diffusion phase when the population mostly resides inside the dark state manifold.

The additional simulations thus support the generality of a polariton transport mechanism based on reversible exchange of population between propagating polaritonic states and stationary dark states.
Our results furthermore suggest that QM/MM simulations open up the possibility to systematically investigate the role of various molecular and cavity parameters on exciton transport. While further simulations are needed to disentangle these effects and formulate design rules for enhancing exciton transfer with optical resonators, the results we have obtained so far already suggest that this process can be controlled by tuning the overlap between the absorption line-width of the molecules and Rabi splitting of the strongly coupled light-matter systems.

\clearpage

\section{Animations}

Animations of the wavepackets in MPEG-4 format are available for download for the  simulation systems listed below.
        
\subsection{On-resonant excitation of a Gaussian wavepacket of LP states}
\subsubsection{Lossless cavity with 256 Rhodamines}
    \begin{itemize}
        \item total wavepacket: \verb  $wp_256_decay0.mp4$
        \item excitonic contribution to the wavepacket: \verb  $wp_mol_256_decay0.mp4$
        \item cavity mode contribution to the wavepacket: \verb  $wp_pho_256_decay0.mp4$
    \end{itemize}
    
\subsubsection{Lossless cavity with 512 Rhodamines}
    \begin{itemize}
        \item total wavepacket: \verb  $wp_512_decay0.mp4$
        \item excitonic contribution to the wavepacket: \verb  $wp_mol_512_decay0.mp4$
        \item cavity mode contribution to the wavepacket: \verb  $wp_pho_512_decay0.mp4$
    \end{itemize}
    
\subsubsection{Lossless cavity with 1024 Rhodamines}
    \begin{itemize}
        \item total wavepacket: \verb  $wp_1024_decay0.mp4$
        \item excitonic contribution to the wavepacket: \verb  $wp_mol_1024_decay0.mp4$
        \item cavity mode contribution to the wavepacket: \verb  $wp_pho_1024_decay0.mp4$
    \end{itemize}

\subsubsection{Lossless cavity with 1024 Tetracene molecules}
    \begin{itemize}
        \item total wavepacket: \verb  $wp_tc-cyclo_on_decay0.mp4$
        \item excitonic contribution to the wavepacket: \verb  $wp_mol_tc-cyclo_on_decay0.mp4$
        \item cavity mode contribution to the wavepacket: \verb  $wp_pho_tc-cyclo_on_decay0.mp4$
    \end{itemize}

\subsubsection{Lossy cavity with 256 Rhodamines}
    \begin{itemize}
        \item total wavepacket: \verb  $wp_256_decay66.7.mp4$
        \item excitonic contribution to the wavepacket: \verb  $wp_mol_256_decay66.7.mp4$
        \item cavity mode contribution to the wavepacket: \verb  $wp_pho_256_decay66.7.mp4$
    \end{itemize}
    
\subsubsection{Lossy cavity with 512 Rhodamines}
    \begin{itemize}
        \item total wavepacket: \verb  $wp_512_decay66.7.mp4$
        \item excitonic contribution to the wavepacket: \verb  $wp_mol_512_decay66.7.mp4$
        \item cavity mode contribution to the wavepacket: \verb  $wp_pho_512_decay66.7.mp4$
    \end{itemize}
    
\subsubsection{Lossy cavity with 1024 Rhodamines}
    \begin{itemize}
        \item total wavepacket: \verb  $wp_1024_decay66.7.mp4$
        \item excitonic contribution to the wavepacket: \verb  $wp_mol_1024_decay66.7.mp4$
        \item cavity mode contribution to the wavepacket: \verb  $wp_pho_1024_decay66.7.mp4$
    \end{itemize}

\subsubsection{Lossy cavity with 1024 Tetracene molecules}
    \begin{itemize}
        \item total wavepacket: \verb  $wp_tc-cyclo_on_decay66.7.mp4$
        \item excitonic contribution to the wavepacket: \verb  $wp_mol_tc-cyclo_on_decay66.7.mp4$
        \item cavity mode contribution to the wavepacket: \verb  $wp_pho_tc-cyclo_on_decay66.7.mp4$
    \end{itemize}

\subsection{Off-resonant excitation of a single molecule}
\subsubsection{Lossless cavity with 256 Rhodamines}
    \begin{itemize}
        \item total wavepacket: \verb  $wp_256_off_decay0.mp4$
        \item excitonic contribution to the wavepacket: \verb  $wp_mol_256_off_decay0.mp4$
        \item cavity mode contribution to the wavepacket: \verb  $wp_pho_256_off_decay0.mp4$
    \end{itemize}
\subsubsection{Lossless cavity with 512 Rhodamines}
    \begin{itemize}
        \item total wavepacket: \verb  $wp_512_off_decay0.mp4$
        \item excitonic contribution to the wavepacket: \verb  $wp_mol_512_off_decay0.mp4$
        \item cavity mode contribution to the wavepacket: \verb  $wp_pho_512_off_decay0.mp4$
        \end{itemize}
\subsubsection{Lossless cavity with 1024 Rhodamines}
    \begin{itemize}
        \item total wavepacket: \verb  $wp_1024_off_decay0.mp4$
        \item excitonic contribution to the wavepacket: \verb  $wp_mol_1024_off_decay0.mp4$
        \item cavity mode contribution to the wavepacket: \verb  $wp_pho_1024_off_decay0.mp4$
    \end{itemize}

\subsubsection{Lossless cavity with 1024 Tetracene molecules}
    \begin{itemize}
        \item total wavepacket: \verb  $wp_tc-cyclo_off_decay0.mp4$
        \item excitonic contribution to the wavepacket: \verb  $wp_mol_tc-cyclo_off_decay0.mp4$
        \item cavity mode contribution to the wavepacket: \verb  $wp_pho_tc-cyclo_off_decay0.mp4$
    \end{itemize}

\subsubsection{Lossy cavity with 256 Rhodamines}
    \begin{itemize}
        \item total wavepacket: \verb  $wp_256_off_decay66.7.mp4$
        \item excitonic contribution to the wavepacket: \verb  $wp_mol_256_off_decay66.7.mp4$
        \item cavity mode contribution to the wavepacket: \verb  $wp_pho_256_off_decay66.7.mp4$
    \end{itemize}
\subsubsection{Lossy cavity with 512 Rhodamines}
    \begin{itemize}
        \item total wavepacket: \verb  $wp_512_off_decay66.7.mp4$
        \item excitonic contribution to the wavepacket: \verb  $wp_mol_512_off_decay66.7.mp4$
       \item cavity mode contribution to the wavepacket: \verb  $wp_pho_512_off_decay66.7.mp4$
    \end{itemize}
\subsubsection{Lossy cavity with 1024 Rhodamines}
    \begin{itemize}
        \item total wavepacket: \verb  $wp_1024_off_decay66.7.mp4$
        \item excitonic contribution to the wavepacket: \verb  $wp_mol_1024_off_decay66.7.mp4$
        \item cavity mode contribution to the wavepacket: \verb  $wp_pho_1024_off_decay66.7.mp4$
    \end{itemize}        

\subsubsection{Lossy cavity with 1024 Tetracene molecules}
    \begin{itemize}
        \item total wavepacket: \verb  $wp_tc-cyclo_off_decay66.7.mp4$
        \item excitonic contribution to the wavepacket: \verb  $wp_mol_tc-cyclo_off_decay66.7.mp4$
        \item cavity mode contribution to the wavepacket: \verb  $wp_pho_tc-cyclo_off_decay66.7.mp4$
    \end{itemize}
    
\bibliography{supp_info}